\documentclass[%
reprint,
superscriptaddress,
 amsmath,amssymb,
 aps,
longbibliography,
floatfix,
]{revtex4-2}

\usepackage[dvipsnames,svgnames,x11names,hyperref]{xcolor}
\usepackage{siunitx}
\usepackage{tabularx}
\usepackage{color}
\usepackage{graphicx}% Include figure files
\usepackage{dcolumn}% Align table columns on decimal point
\usepackage{bm}% bold math
\usepackage{hyperref}% add hypertext capabilities
\usepackage[mathlines]{lineno}% Enable numbering of text and display math
\usepackage{mathrsfs}
\usepackage{braket}
\usepackage{placeins} % Allows use of the command \FloatBarrier to prevent figures and tables from moving past that point
\usepackage{float} % Allows the option H for force figures to appear there. Doesn't work for two-column figures placed with \begin{figure*}
\renewcommand{\sl}[1]{$_\textrm{#1}$} % `sl' = subscript letter, e.g. for writing defects where the subscripted letter corresponds to the atom being replaced

\hypersetup{colorlinks=true, 
	linkcolor={blue!75!black!80!yellow},
	citecolor={blue!75!black!80!yellow}, 
	urlcolor=black %{blue!75!black!80!yellow}
	}
\makeatletter \renewcommand\@make@capt@title[2]{%
\@ifx@empty\float@link{\@firstofone}{\expandafter\href\expandafter{\float@link}}%
\sffamily{\textbf{#1}}\@caption@fignum@sep#2 }% \makeatother

\usepackage[normalem]{ulem}
\usepackage{color}

\begin{document}

\title{Defect polaritons from first principles}

\author{Derek S. Wang}
\email{derekwang@g.harvard.edu}
\affiliation{Harvard John A. Paulson School of Engineering and Applied Sciences, Harvard University, Cambridge, MA 02138, USA}

\author{Susanne F. Yelin}
\affiliation{Department of Physics, Harvard University, Cambridge, MA 02138, USA}

\author{Johannes Flick}
\email{jflick@flatironinstitute.org}
\affiliation{Center for Computational Quantum Physics, Flatiron Institute, New York, NY 10010, USA}

\begin{abstract}
\noindent Precise control over the electronic and optical properties of defect centers in solid-state materials is necessary for their applications as quantum sensors, transducers, memories, and emitters. In this study, we demonstrate, from first principles, how to tune these properties {via} the formation of defect polaritons. Specifically, we investigate three defect types---CH\sl{B}, C\sl{B}-C\sl{B}, and C\sl{B}-V\sl{N}---in monolayer hexagonal boron nitride (hBN). The lowest-lying electronic excitation of these systems is coupled to an optical cavity where we explore the strong light-matter coupling regime. For all defect systems, we show that the polaritonic splitting that shifts the absorption energy of the lower polariton is much higher than can be expected from a Jaynes-Cummings interaction. In addition, we find that the absorption intensity of the lower polariton increases by several orders of magnitude, suggesting a possible route toward overcoming phonon-limited single photon emission from defect centers. Finally, we find that initially localized electronic transition densities can become delocalized across the entire material under strong light-matter coupling. These findings are a result of an effective continuum of electronic transitions near the lowest-lying electronic transition for both pristine hBN and hBN with defect centers that dramatically enhances the strength of the light-matter interaction. We expect our findings to spur experimental investigations of strong light-matter coupling between defect centers and cavity photons for applications in quantum technologies.
\end{abstract}
\date{\today}

\maketitle

\section*{Introduction} \label{sec:intro}
Defect emitters in solid-state materials have wide applicability in scalable and stable solid-state quantum technologies~\cite{Wrachtrup2001, Weber2010, Wrachtrup2010, Aharonovich2016, Degen2017, Childress2014, Aharonovich2016, Degen2017, Atature2018}. They are especially suitable as quantum memories ~\cite{Kurtsiefer2000, Ye2019} or as quantum transducers because they can interact with a wide range of quantum information carriers, such as phonons, magnons, and photons, across a broad spectral range \cite{Appel2015, Lemonde2018, Candido2020, Neuman2020Phononicbus, Wang2020Selection, Wang2021Perspective}. These defects---including simple substitutional or vacancy defects, as well as hybridized defect complexes~\cite{McDougall2017, MacKoit-Sinkeviciene2019, Czelej2020, Wang2020Hybridized}---can introduce spatially localized electronic states whose electronic, optical, and spin properties can be tuned by coupling them to external fields, including electric, magnetic, and strain fields, as well as to waveguides and cavity environments~\cite{Rogers2008, Momenzadeh2015, Faraon2012, Chakraborty2019, Zhang2018, Machielse2019, Neuman2020Nanomagnonics, Wang2020Selection}. Due to their flexible applications, demands to the properties of defect systems are ever-increasing, such as specific level structures for the emission of entangled photonic states~\cite{Wang2020Entangled, Trivedi2020} or implementation of multi-qubit photonic gates \cite{Dai2020}.

\begin{figure}[!tbhp]
\centering
\includegraphics[width=1.0\linewidth]{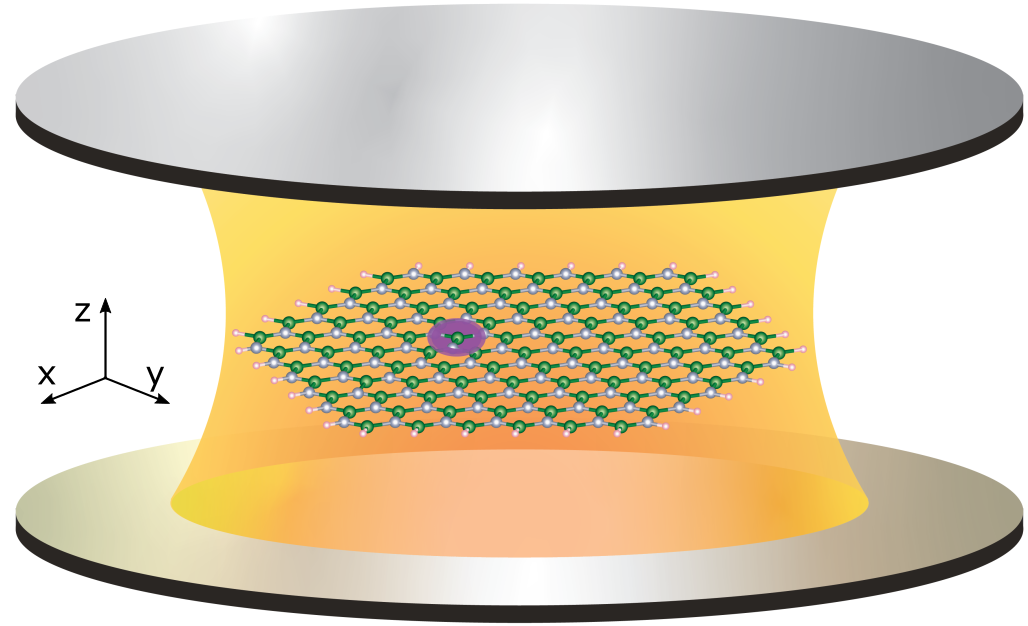}
\caption{Schematic of the 226-atom nanoflake of hexagonal boron nitride (hBN) with defect centers (in the purple region) inside an optical cavity. We simulate from first principles the coupling between the vacuum electric field (yellow-orange) of the optical cavity and the electronic transitions of defect centers.
}
\label{fig:schematic}
\end{figure}

For inspiration for additional control over the optical properties of defect emitters, we turn to recent experiments in polaritonic systems, where the light-matter interaction is strong enough to hybridize electronic excitations of molecules and materials with cavity photon modes. The light-matter coupling in polaritonic systems can range from the weak to strong to ultrastrong coupling regimes that each manifest qualitatively different phenomena. Assuming no losses due to spontaneous emission or other processes, when the cavity loss is larger than the light-matter coupling rate, the system is in the weak-coupling regime, characteristically resulting in the acceleration of excited state decay \textit{via} the Purcell effect~\cite{purcell1946spontaneous}. When the light-matter coupling rate is larger than the cavity loss rate, matter excitations hybridize with cavity photons to form polaritonic states~\cite{cwik2016excitonic, ebbesen2016strongcoupling, flick2017, Herrera2020, rivera2018, flick2018b, flick2018strong, flickexcited} that are shifted in energy from the bare electronic excitations and photon modes. This coherent, non-perturbative regime is denoted as strong light-matter coupling and results in modifications of, for instance, chemical reactivity~\cite{galego2015, galego2016, anoop2016vibreactivity, herrera2016chemistry, flick2017, galego2019, groenhof2019relaxation, ebbesenTilting}, optical properties~\cite{lidzey1999polemission, delpino2015quantum, george2015ultra, herrera2018review, zeb2018exactvibdressed, herrera2017dark, Owrutzsky2DIR}, and energy transfer~\cite{coles2014polariton, zhong2017entangled, juraschek2019cavity, DuEnergyTransfer, Wang2020LossQEDFT}.

There has been previous interest in coupling spin states of individual or ensembles of defect emitters in solid-state materials to MHz- and GHz-frequency cavities \cite{Schuster2010, Sandner2012}, demonstrating tunability of the optical emission of defect centers in diamond and hBN \cite{Wolters2010, Englund2010, Faraon2012, Schroder2017, Vogl2019, Caldwell2019} among other semiconducting host materials with wide band gaps. These experiments are generally within the weak coupling regime characterized by Purcell enhancement and enhanced emission intensity at the original emission frequency. While strong-coupling between optical excitations of a defect and a cavity mode has not been achieved yet \cite{Janitz2020}, it is a widely sought after goal that has generated much research interest. 

In this paper we study beyond the weak-coupling regime to the strong-coupling limit between a defect and cavity mode that is characterized by the shift in emission frequency due to polariton formation. We demonstrate, from first principles, how to tune the optical properties of defect centers by strongly coupling them to cavity photons and forming defect polaritons. Specifically, we investigate a flake of monolayer hexagonal boron nitride (hBN) terminated with hydrogen atoms as a host material, which has a wide band gap that is suitable for hosting spatially localized defect orbitals \cite{Wang2020Hybridized}. In addition, quantum emission of single photons from defects in monolayer hBN \cite{Grosso2017, Mendelson} has been observed, paving the way for defect centers in 2D materials to be used for local on-chip computation and state preparation.  In addition to pristine (defect-less) hBN, we study three different defect types: CH\sl{B}, C\sl{B}-C\sl{B}, and C\sl{B}-V\sl{N}. For all these systems, we couple the lowest-lying electronic excitation to a single photon mode of an optical cavity. To quantitatively study defect polaritons in the strong coupling regimes, we turn to the recently developed linear-response quantum electrodynamical density functional theory (QEDFT) method~\cite{tokatly2013,ruggenthaler2014, flick2019lmrnqe, flickexcited, Wang2020LossQEDFT}. With only the chemical structure and spectral profile of the cavity as input, QEDFT predicts the effects of non-perturbative light-matter coupling on the molecular properties, combining the power of parametric cavity quantum electrodynamics (cQED) models, theories of open-quantum systems, and electronic structure theory. 

We show that for all four systems, the polaritonic splitting that shifts the absorption energy of the lower polariton is much higher than can be expected from a Jaynes-Cummings-like Hamiltonian. In addition, we find that the oscillator strength of the lower polariton increases by several orders of magnitude while the absorption intensity of excited states in the electronic conduction band necessarily decreases due to the f-sum rule, suggesting a possible route toward overcoming loss- and decoherence-limited single photon emission from defect centers. We find that transition densities, even those that are localized on defect centers outside the cavity, can become delocalized across the entire material inside the cavity. These surprising discoveries are a result of a quasi-continuum of electronic transitions to the conduction band near the lowest-lying electronic transition for both pristine hBN and hBN with defect centers that enhances the strength of the light-matter interaction. We compare our first-principles results against a cQED and show excellent agreement. We expect our findings to spur experimental investigations into strong light-matter coupling between defect centers and cavity photons for applications in quantum technologies.

\section*{Results and Discussion} \label{sec:results}

\subsection*{Model defect systems}

\begin{figure*}[!ht]
\centering
\includegraphics[width=\linewidth]{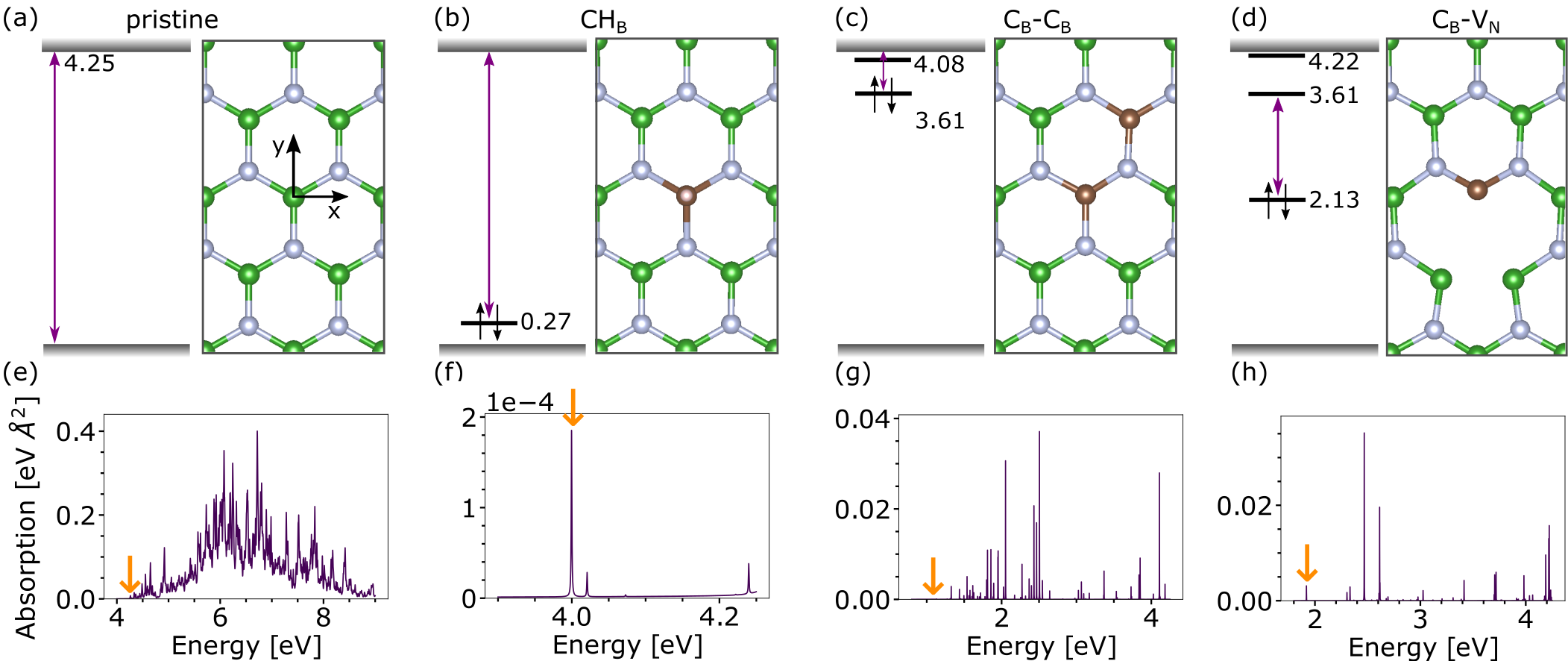}
\caption{Energetic structure, geometry, and absorption spectra of pristine and defect systems. As depicted in \textbf{(a)}, the nanoflake of pristine hBN in this study has a calculated energy gap between the highest-lying occupied (``valence band") and lowest-lying unoccupied (``conduction band") Kohn-Sham state of 4.25 eV. Within the band gap the defect systems \textbf{(b)} CH\sl{B}, \textbf{(c)} C\sl{B}-C\sl{B}, and \textbf{(d)} C\sl{B}-V\sl{N} present one, two, and three spatially localized defect orbitals, respectively, where the lowest-lying defect orbital of each is doubly occupied. B is green, N is white, C is brown, and H is pink. All energies are in eV and are relative to the valence band maximum. Chemical structures are plotted with VESTA \cite{Momma2008}. Absorption spectra of the four systems outside an optical cavity are plotted on the bottom from \textbf{(e)}-\textbf{(h)}. For the pristine flake, the entire absorption spectrum is shown from 4 eV, where the electronic excitation continuum begins, up to 9 eV, whereas for the defects CH\sl{B}, C\sl{B}-C\sl{B}, and C\sl{B}-V\sl{N}, the $x$-axis is limited to show the lowest-lying excitation up to the electronic excitation continuum. The optical cavity mode, whose energy for each system is marked by orange arrow, is tuned in resonance with lowest-lying electronic excitation whose character is dominated by the Kohn-Sham states spanning the purple arrow in the energy diagram. For visual clarity, for the pristine system, $\hbar\Gamma=10$ meV and $\Delta\omega=1$ meV (as defined in the text), and for CH\sl{B}, C\sl{B}-C\sl{B}, and C\sl{B}-V\sl{N}, $\hbar\Gamma=1$ meV and $\Delta\omega=0.1$ meV. The absorption intensity of the lowest-lying excitation for C\sl{B}-C\sl{B}, not visible on the plot, is 1.5$\cdot 10^{-5}$ eV \AA$^2$.}
\label{fig:systems}
\end{figure*}

We model the system depicted schematically in Fig. \ref{fig:schematic}, where a nanoflake of hBN is placed in an optical cavity. In addition to pristine hBN, we simulate three defect systems---CH\sl{B}, C\sl{B}-C\sl{B}, and C\sl{B}-V\sl{N} whose molecular structures are shown in Fig. \ref{fig:systems}---placed at the center of the hBN nanoflake, marked as the purple region in Fig. \ref{fig:schematic}. The subscripted letters in the names of the defect systems are the atoms of the hBN nanoflake that are being replaced, and their substituents CH, C, and V correspond to a carbon atom bonded to a hydrogen atom, a carbon atom, and a vacancy, respectively. To theoretically model the electronic structure of these defect systems, we use the pseudopotential, real-space density functional theory method Octopus~\cite{octopus1, octopus2,octopus3}, as it has been shown that real-space electronic structure calculations on hexagonal boron nitride (hBN) nanoflakes can be extrapolated onto periodic calculations~\cite{Barcza, Reimers2020}. For the ground state, we use the Perdew, Burke, and Ernzerhof (PBE) generalized gradient approximation exchange-correlation functional~\cite{Perdew1996}, and for the excited state calculation, we use a functional based on the local density approximation (LDA)~\cite{perdew1981, Perdew1992}. The 226-atom size of the nanoflake, of which 34 atoms are hydrogen, supports several unit cells of host material beyond the spatially localized defect orbitals and should be large enough to simulate bulk behavior, as discussed further in Appendix \ref{app:compmethods} and in Ref. \citenum{Wang2020Hybridized}.

As shown in Fig. \ref{fig:systems}(a), the pristine hBN nanoflake has an energy gap (``band gap") of 4.25 eV between the highest-occupied (``valence band maximum (VBM)") and the lowest-unoccupied (``conduction band minimum (CBM)") Kohn-Sham state. These Kohn-Sham states are delocalized across the nanoflake. The three defect systems CH\sl{B}, C\sl{B}-C\sl{B}, and C\sl{B}-V\sl{N} have an identical band gap within 1\% variation. In addition to these delocalized electronic states, the defect systems also present 1, 2, and 3 spatially localized defect orbitals for CH\sl{B}, C\sl{B}-C\sl{B}, and C\sl{B}-V\sl{N}, respectively, where the lowest-lying defect orbital for each defect system is fully occupied. The nature of these spatially localized defect orbitals is further described in Refs. \citenum{McDougall2017, Wang2020Hybridized}.

Using the Casida equation of linear-response time-dependent density functional theory (TDDFT)~\cite{casida1995time}, we compute the electronic excitation spectrum of the four systems. Further details on computing the absorption spectra are presented in Appendix \ref{app:compmethods}. For pristine hBN, the lowest-lying excitation has energy of 4.25 eV, which corresponds to a transition between VBM and CBM, while the energies of the lowest-lying defect systems CH\sl{B}, C\sl{B}-C\sl{B}, and C\sl{B}-V\sl{N} are lower at 4.00, 1.06, and 1.92 eV, respectively, due to states inside the band gap. For pristine hBN, the lowest-lying excitation energy corresponds closely to the difference in energy between the highest-occupied and lowest-unoccupied Kohn-Sham states. For all four systems, the lowest-lying many-body excitation is dominated in character by the transition between the highest-lying occupied and lowest-lying unoccupied Kohn-Sham states, indicated by the purple arrows in the energy level diagrams in Fig. \ref{fig:systems}. Finally, we note that throughout this manuscript for the pristine hBN and the defect systems CH\sl{B} and C\sl{B}-C\sl{B}, we plot the $x$-polarized absorption spectrum, while for C\sl{B}-V\sl{N}, we plot the $y$-polarized absorption spectrum; we simply choose the polarization direction corresponding to the higher transition dipole moment in the lowest-lying excitation. The coordinate axes are shown in Fig. \ref{fig:schematic} and Fig. \ref{fig:systems}(a), where the out-of-plane direction is given in $z$-direction.

With these four systems, based on the energy level diagrams and absorption spectra in Fig. \ref{fig:systems}, we sample electronic transitions with a representative variety of energy ranges from the near-infrared to the ultraviolet. In addition, we can investigate the difference between coupling the cavity mode to electronic transitions of different spatial character and localization, where the lowest-lying excitation in pristine hBN is dominated in character by a transition between two delocalized states and the defect systems involve transitions with localized defect orbitals. Within the defect systems, we note further distinctions: in CH\sl{B} and C\sl{B}-C\sl{B}, the electronic transition of interest is between a spatially localized defect orbital and the valence band, and in C\sl{B}-V\sl{N}, it is between two localized defect orbitals and reminiscent of a canonical two-level system given how spectrally distant it is from the quasi-continuum of electronic transitions beginning at $\sim$4.25 eV. We study both C\sl{B}-C\sl{B} and C\sl{B}-V\sl{N} to understand the impact of the additional, third defect orbital of the latter.

\subsection*{Defect polaritons}

To form defect polaritons, we strongly couple the four systems to a single, lossless cavity mode resonant with the lowest-lying transition. To understand the impact of this coupling of the optical properties on the defects, we track changes in the excitation or absorption spectrum. By inputting only the molecular structure and cavity coupling strength, we solve the Hamiltonian with QEDFT. The light-matter interaction causes the $M$ electronic excited states to hybridize with the $N$ photon modes to form $M+N$ hybrid electron-photon states experimentally observable in the excitation (or absorption) spectrum. We briefly describe the theoretical formalism of calculating the excitation spectrum based on this light-matter interaction in Appendix \ref{app:compmethods} and point the reader to more complete descriptions in Ref. \citenum{flick2017, Wang2020LossQEDFT, flick2019lmrnqe}.

\begin{figure}[!tbhp]
\centering
\includegraphics[width=0.8\linewidth]{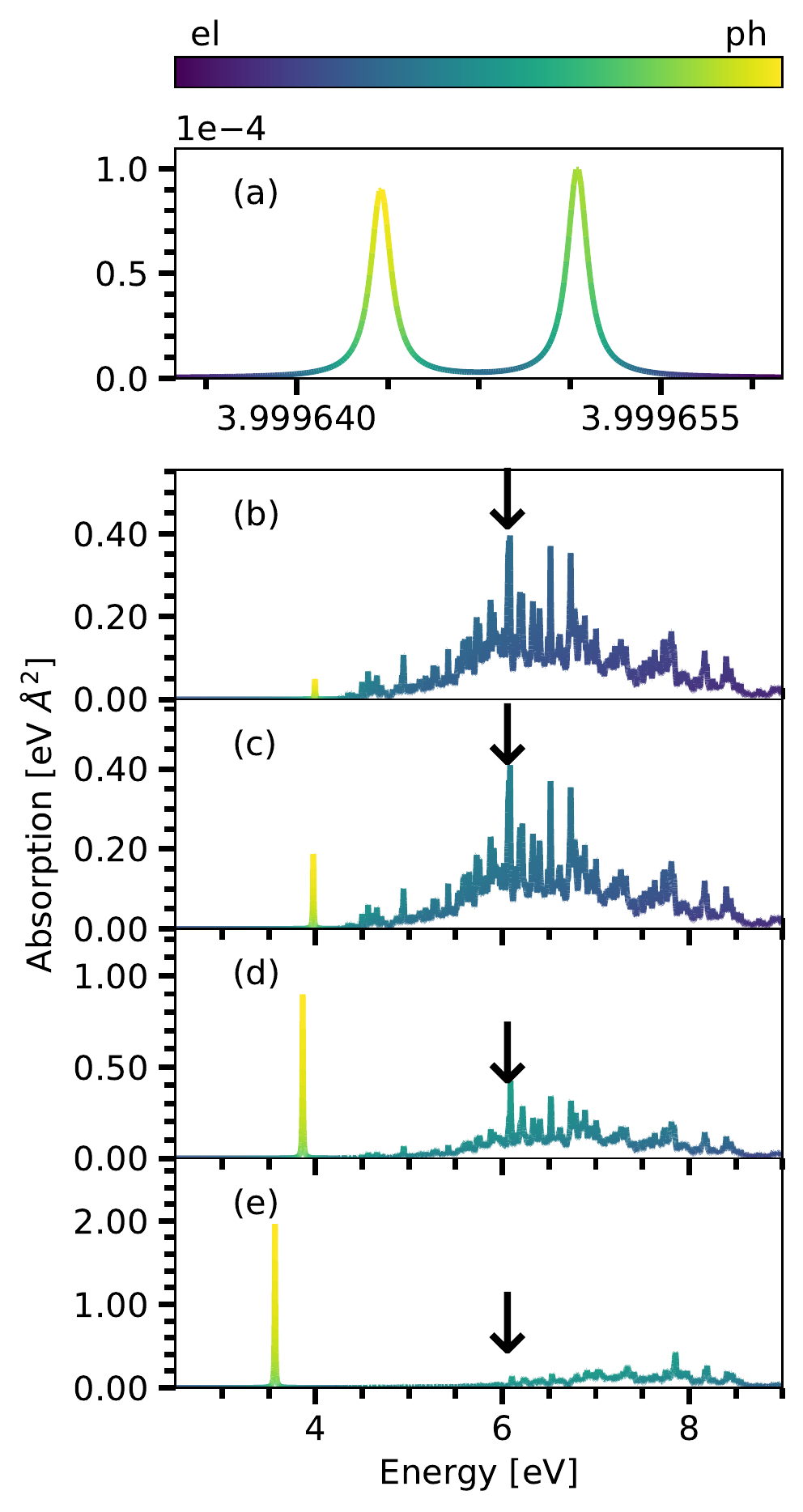}
\caption{Absorption spectra of an $x$-polarized cavity mode resonantly tuned to the lowest-lying excitation of CH\sl{B} for coupling strength $\lambda$ of \textbf{(a)} 0.001, \textbf{(b)} 0.099, \textbf{(c)} 0.197, \textbf{(d)} 0.493, and \textbf{(e)} 0.986 eV$^{1/2}$/nm. The relative electronic and photonic weight of each polariton state is overlaid on the absorption curves. Note that for the top pane, the $x$-axis range is different and the spectral broadening $\hbar\Gamma=\SI{10}{\micro\eV}$ and the energy spacing $\Delta(\hbar\omega)=\SI{1}{\micro\eV}$ for visual clarity of the upper and lower polaritons, while $\hbar\Gamma=\SI{10}{\milli\eV}$ and $\Delta(\hbar\omega)=\SI{1}{\milli\eV}$ for \textbf{(b)}-\textbf{(e)}. At the lowest coupling strength in \textbf{(a)}, the lower and upper polaritons are distinct, while the quasi-continuum of electronic excitations is largely unchanged relative to the absorption spectrum of the pristine flake in Fig. \ref{fig:schematic}(e). At higher coupling strengths, the lower polariton decreases in energy and increases in intensity as the upper polariton mixes with and quenches the optical activity of the electronic continuum starting at $\sim4$ eV. As a guide to the eye in \textbf{(b)}-\textbf{(e)}, an arrow marks the maximum of the absorption spectrum of the quasi-continuum from \textbf{(b)}.}
\label{fig:sweep}
\end{figure}

To demonstrate how the formation of defect polaritons changes the optical properties, we plot the absorption spectrum of CH\sl{B} inside an optical cavity in Fig.~\ref{fig:sweep}. In this case, we choose an $x$-polarized cavity mode resonant with the lowest-lying transition at 4.00 eV with $x$-polarized transition dipole moment of 0.027 \AA~and change the light-matter coupling strength $\lambda$ from 0.001 to 0.986 eV$^{1/2}$/nm. In addition, in the same plot we overlay the logarithm of the weight of the photonic character, defined further in Appendix \ref{app:compmethods}, in the observed polaritonic states. By tracking the photonic character, we know whether the photon field is interacting with electronic excitations at a given energy. At the lowest coupling strength $\lambda=0.001$ eV$^{1/2}$/nm in Fig. \ref{fig:sweep}(a), where the coupling energy $\hbar g_{i,k}$, defined as $-\sqrt{\frac{\hbar\omega_k}{2}}\bm{\lambda}_k\cdot \bm{d}_i$ for cavity mode $k$ and excitation $i$, is larger than the spectral broadening $\hbar\Gamma$, the upper and lower polaritons are both visible as distinct peaks with approximately equal photonic weight. As the coupling strength is increased to $\lambda=0.099$ eV$^{1/2}$/nm in Fig. \ref{fig:sweep}(b), where the spectral broadening $\hbar\Gamma$ is again set to 30 meV and much greater than the coupling strength, the lower and upper polaritons are not well resolved from each other, resulting in a single peak of relatively higher photonic weight affirming their polaritonic character. As the coupling strength increases to 0.197 eV$^{1/2}$/nm in Fig. \ref{fig:sweep}(c), the lower polariton is relatively brighter and contains more photonic character. At $\lambda=0.493$ eV$^{1/2}$/nm in Fig. \ref{fig:sweep}(d), the lower polariton becomes even brighter and shifts lower in energy, while the upper polariton mixes with the necessarily darker continuum of electronic transitions $>$4 eV due to the sum rule \cite{craig2012molecular}, increasing the weight of photonic character throughout the quasi-continuum of the conduction band states. In Fig. \ref{fig:sweep}(e), the coupling strength is further increased to 0.986 eV$^{1/2}$/nm where we see the lower polariton continue to decrease in energy while increasing in intensity as the upper polariton quenches the intensity the continuum of electronic transitions and contributes further to the photonic character. Here, the lower polariton is at its brightest with an intensity 10,000 times higher than the bare lowest-lying electronic excitation in Fig. \ref{fig:systems}(b). We note that this change of coupling strength effectively leads to a transition from a resonant coupling in (a) to an off-resonant coupling best exemplified in (e). In (a), the typical resonant coupling situation with cavity mode with energy of 4 eV tuned in resonance with the electronic excitation at 4 eV leads to distinct upper and lower polaritons split nearly symmetrically about the original excitation energy, as also observed in the two-level Rabi model \cite{JaynesCummings1963}. As the coupling strength is increased in (b)-(e), the hybridization of light and matter states now also includes the states inside the conduction band with high spectral amplitude, leading to an effective detuning. In this setup, the main intensity of the matter part is located around $\sim$6.2 eV interacting with the frequency of the photon mode at 4 eV, effectively leading to off-resonant situation. Due to the detuning in this off-resonant situation, the photonic character of the lower polariton becomes very high. Finally, we note that in Appendix \ref{app:losses}, we show how the absorption spectrum changes with non-zero cavity loss rate $\kappa$ using the formalism of Ref. \cite{Wang2020LossQEDFT} that generalizes QEDFT to dissipative electronmagnetic environments.

In summary, we observe that the character of the defect polaritons changes drastically with increasing coupling strength $\lambda$ for CH\sl{B}. The emission peak of the lower polariton remains sharp and becomes orders of magnitude brighter compared to the lowest-lying electronic transition outside of the cavity. Meanwhile, the upper polariton couples to the electronic quasi-continuum, quenching the optical activity of the latter and distributing photonic character throughout until the energy of the upper polariton exceeds the quasi-continuum.

\begin{figure*}[tbhp]
\centering
\includegraphics[width=0.8\linewidth]{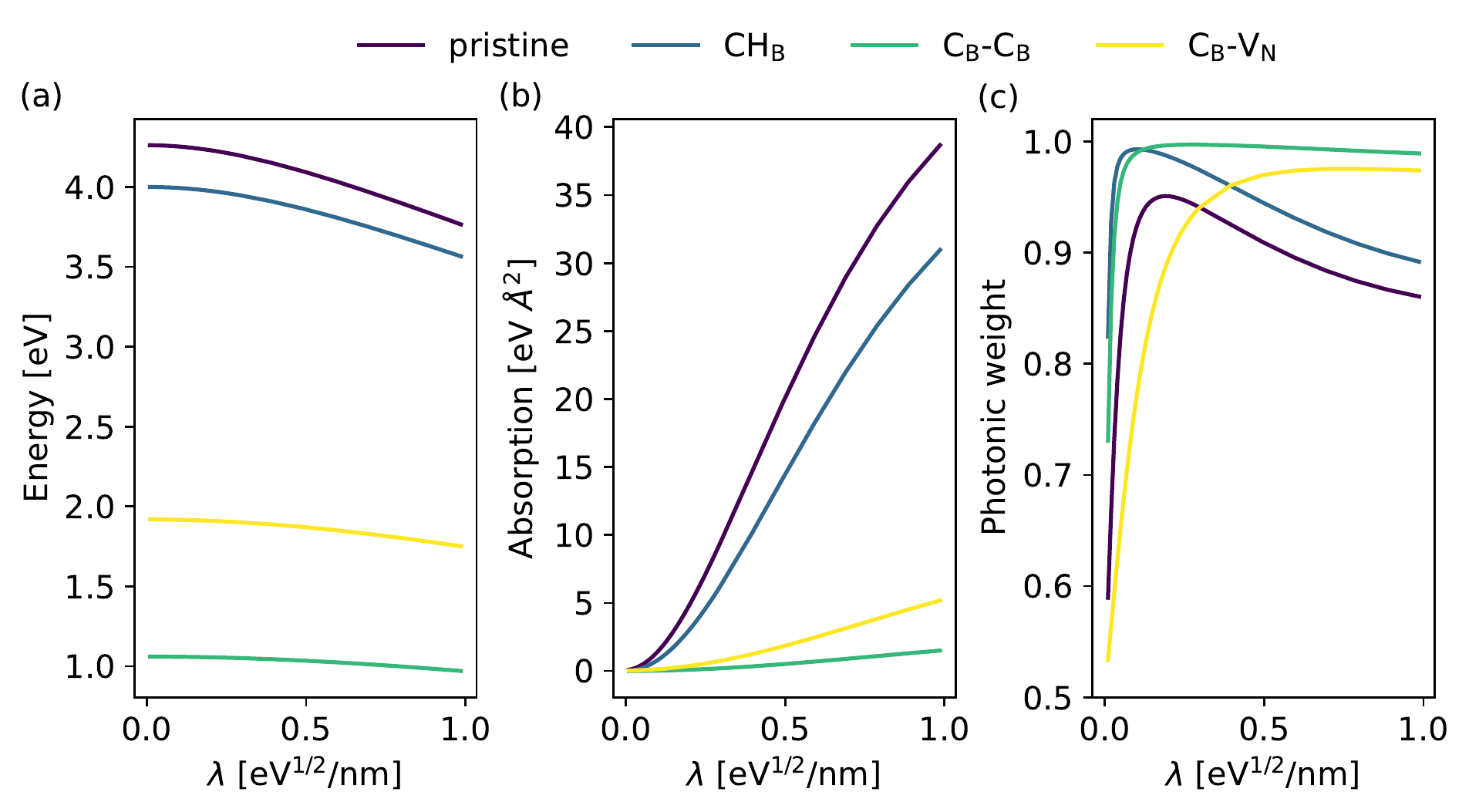}
\caption{Comparison among all four defect systems of the properties of the lowest-lying excitation, namely \textbf{(a)} excitation energy, \textbf{(b)} absorption intensity, and \textbf{(c)} photonic weight. The coupling strength $\lambda$ is swept from 0.010 to 0.986 eV$^{1/2}$/nm. All four systems exhibit similar behavior where the excitation energy decreases, absorption and thus transition dipole moment increases, and photonic weight increases from $\sim$0.5-0.7 to $\sim$0.9-1 before decreasing with increasing $\lambda$. The pristine and CH\sl{B} defect systems change most rapidly, as the cavity mode tuned in resonance with their lowest-lying excitations is closer to the energy of and interacts more strongly to the electronic quasi-continuum starting at $\sim$4 eV.
}
\label{fig:comparison}
\end{figure*}

In Fig. \ref{fig:comparison}, we compare the changes in the lower polariton upon changing the cavity strength $\lambda$ for all four defect systems resonantly tuned to a cavity mode and note the universality of the behavior described in detail for CH\sl{B} in Fig. \ref{fig:sweep}. The cavity mode is $x$-polarized for pristine hBN, CH\sl{B}, and C\sl{B}-C\sl{B} and $y$-polarized for C\sl{B}-V\sl{N} to couple with the corresponding component of the transition dipole moment of the lowest-lying excitation with the higher magnitude. Starting with Fig. \ref{fig:comparison}(a), we plot the energies of the lower polariton. As expected, at $\lambda=0.010$ eV$^{1/2}$/nm, the excitation corresponds closely to the energy of the lowest-lying electronic excitations plotted at the bottom of Fig. \ref{fig:systems}. For all four systems, as the coupling strength $\lambda$ is increased, the energy decreases. While decreasing energy of the lower polariton is expected for increasing coupling strength, the magnitude of the change far exceeds what one would naively expect from a Jaynes-Cummings model. For instance, for CH\sl{B}, the coupling energy $\hbar g$ from coupling only the lowest-lying excitation with $\hbar\omega=4.00$ eV and $d_x=0.027$ \AA~to a photon with cavity strength $\lambda=0.99$ eV$^{1/2}$/nm should be 3.7 meV, but the lower polariton is in fact 435 meV lower in energy compared to the bare electronic transition. The origin of the higher-than-expected polariton splitting energy can be understood as follows: while the cavity mode is tuned in resonance with only the lowest-lying excitation, as the cavity strength increases and the upper polariton shifts upward, the cavity mode can then interact through the upper polariton with higher energy electronic excited states, effectively enhancing the total coupling strength and shifting the lower polariton further than would be expected from coupling to a single electric transition. Notably, the excitation energies for the systems with the highest energy lowest-lying excitations, pristine hBN and CH\sl{B}, decrease more for a given coupling strength $\lambda$ compared to C\sl{B}-C\sl{B} and C\sl{B}-V\sl{N} because the lowest-lying excitations of the former are closer in energy to the bath of electronic transitions in the $>$4 eV range. Due to this polariton-mediated interaction with higher energy excited states, we restrict the coupling strength to 0.986 eV$^{1/2}$/nm because at this coupling strength, the energies of the lower polaritons have shifted 10\% or even more from the energy of the bare electronic transition, driving the system into the ultrastrong coupling regime. We validate our results up to this coupling strength in Fig. \ref{fig:cQED} with those from a parametric cQED model described in further detail in Appendix \ref{app:cQED} and note excellent agreement.

In Fig. \ref{fig:comparison}(b), we plot the absorption of the lower polariton for increasing cavity strength $\lambda$ for all four defect systems. All four systems exhibit drastic increases in absorption and, thus, dipole moment relative to their bare lowest-lying excitations. The maximum absorption for each of the four systems---pristine, CH\sl{B}, C\sl{B}-C\sl{B}, and C\sl{B}-V\sl{N}---in the range of coupling strengths considered corresponds to transition dipole moments of 3.20, 2.95, 1.24, and 1.72 \AA, respectively, compared to the transition dipole moment of their respective lowest-lying excitation of 0.16, 0.027, 0.015, and 0.16 \AA. Assuming Fermi's golden rule where the emission rate $\propto |\mathbf{d}|^2$, we can expect the emission rates of these states to increase by 2-4 orders of magnitude and potentially enabling the efficient emission of single photons by emitting faster than decoherence and loss processes; further studies incorporating effects known to hinder emission of quantum light from defects, such as electron-phonon coupling, are necessary to quantitatively predict whether the formation of defect polaritons enables such phenomena.

We rationalize the differences in how strongly the excitation energy, absorption, and photonic weight respond to increasing coupling strength among the four systems. For the two systems with lower lowest-lying excitation energies that are farther from the onset of the electronic continuum, C\sl{B}-C\sl{B} and C\sl{B}-V\sl{N}, the absorption intensities are lower than for the two systems with higher lowest-lying excitation energies, pristine hBN and CH\sl{B}. Just as the excitation energies of the lower polaritons for the pristine hBN and CH\sl{B} decrease more with increasing coupling strength than those of C\sl{B}-C\sl{B} and C\sl{B}-V\sl{N} because the former are closer in energy to the electronic continuum, lower coupling strengths are necessary to mix the upper polariton into the continuum and quench the optical activity of these electronic transitions for the higher energy lowest-lying transitions in pristine hBN and CH\sl{B}. The relatively higher interaction of the photon with the electronic continuum for pristine hBN and CH\sl{B} can also be seen in Fig. \ref{fig:comparison}(c). While the photonic weight of all four systems is $\sim$0.5-0.7 when the resonant electronic transition interacts with the cavity photon mode at the lowest coupling strength of $\lambda=0.010$ eV$^{1/2}$/nm, splitting photonic weight roughly equally between the lower and upper polariton, the photonic weight of the lower polariton decreases more quickly for increasing $\lambda$ for pristine hBN and CH\sl{B} as photonic weight is transferred to the upper polariton embedded in the electron continuum. 

\begin{figure}[!tbhp]
\centering
\includegraphics[width=1.0\linewidth]{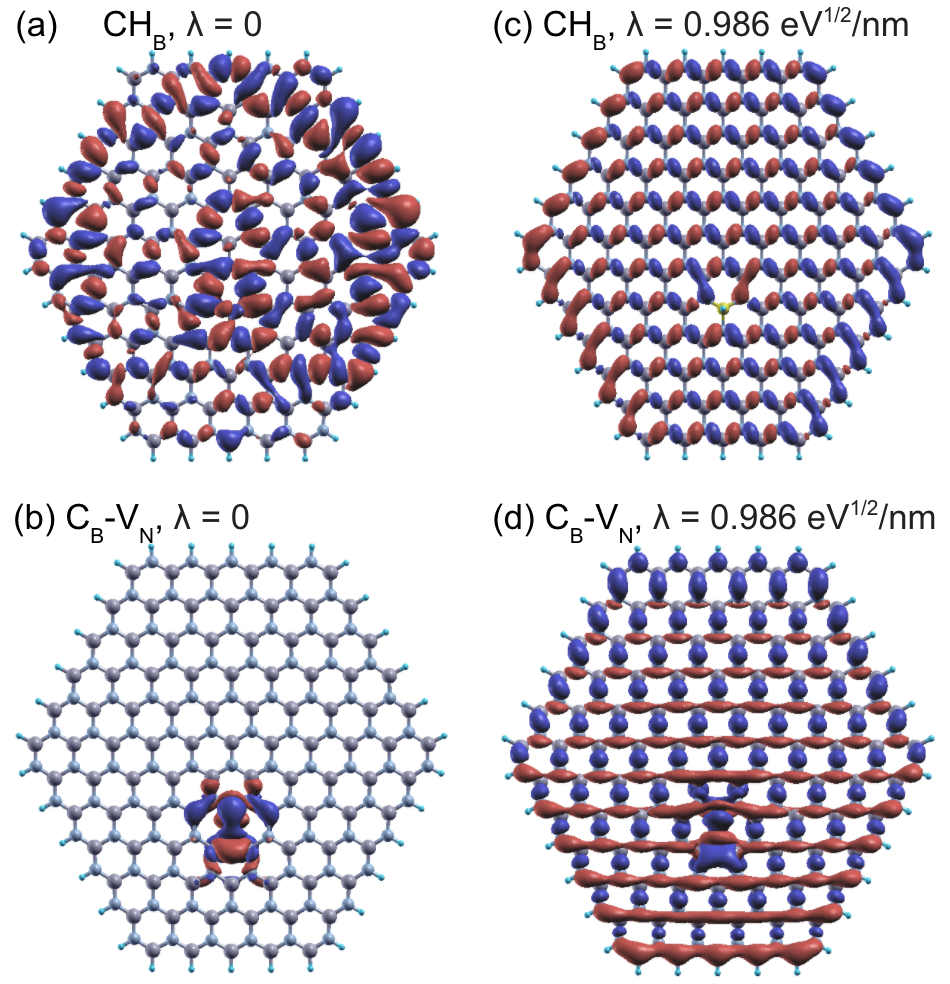}
\caption{Transition densities of the lowest-lying excitation for \textbf{(a)} CH\sl{B} outside the cavity, \textbf{(b)} CH\sl{B} inside the cavity with $\lambda=0.986$ eV$^{1/2}$/nm, \textbf{(c)} C\sl{B}-V\sl{N} outside the cavity, and \textbf{(d)} C\sl{B}-V\sl{N} inside the cavity with $\lambda=0.986$ eV$^{1/2}$/nm. The boron, nitrogen, and hydrogen atoms are gray, blue, and bright blue, respectively, and the red and purple surfaces of the transition densities correspond to opposite signs. Inside the cavity, the electronic transition densities become more uniformly delocalized across the entire hBN nanoflake, even for C\sl{B}-V\sl{N}, where the transition density is localized outside the cavity. Atoms and transition densities are visualized with XCrySDen~\cite{Kokalj1999}. 
}
\label{fig:delocalization}
\end{figure}

Fig. \ref{fig:comparison} demonstrates qualitative similarities between the experimentally observable absorption spectra of the four defect systems inside an optical cavity; in Fig. \ref{fig:delocalization}, we highlight a crucial difference that, from a methods perspective, is an understanding made possible only \textit{via} a first-principles description of light-matter interactions rooted in quantum chemistry methods. As noted in Fig. \ref{fig:systems}, the lowest-lying excitations of the four systems differ primarily in the spatial character of their transition densities. For instance, the lowest-lying excitation of CH\sl{B} is largely comprised of a transition between the spatially localized defect orbital and spatially delocalized states in the conduction band, resulting overall in a delocalized transition density outside an optical cavity as shown in Fig. \ref{fig:delocalization}(a). In contrast, the lowest-lying excitation of C\sl{B}-V\sl{N} is between two spatially localized defect orbitals, resulting in a transition density spatially localized on the defect as shown in Fig. \ref{fig:delocalization}(b). The spatial localization of the electronic transition density is generally considered a fundamental requirement of defect centers capable of quantum emission due to low electron-phonon coupling and contributes to the possibility that the C\sl{B}-V\sl{N} defect is a source of quantum emission in hBN \cite{Tawfik2017, Grosso2020}. 

Inside the cavity, the spatial localization of the transition density can change: for sufficiently high coupling strengths, the electronic character of the lowest-lying excitation becomes delocalized for both CH\sl{B} and C\sl{B}-V\sl{N}, as we show in Fig. \ref{fig:delocalization}(c) and Fig. \ref{fig:delocalization}(d), respectively. This change may be rationalized as follows: In Fig. \ref{fig:sweep}, when the defect systems are placed inside an optical cavity and the coupling strength is increased, the lower polariton increases in intensity as the lower and upper polaritons mix with and quench the oscillator strength of the electronic quasi-continuum. Because the electronic quasi-continuum generally corresponds to spatially delocalized electronic transitions from valence band and defect states to conduction bands, the electronic transition of the lower polariton also becomes spatially delocalized. 
That the transition density of the lowest-lying excitation in C\sl{B}-V\sl{N} is localized outside the cavity and delocalized inside the cavity may have important practical implications on optical properties of these defect systems. For instance, whether the emission rate of the lowest-lying excitation or the electron-phonon coupling rate increases more inside a cavity will affect whether the formation of defect polaritons is a viable method for improving quantum light emission and warrants further investigation. In addition, leveraging the delocalization of the electronic transition density may enable energy transfer among spatially separated emitters. 

\section*{Conclusions and Outlook}
The presented results demonstrate the drastic impact of the formation of defect polaritons on the optical properties of defect systems. Using the first-principles approach of linear-response QEDFT, we couple the lowest-lying excitations of three representative defect systems---CH\sl{B}, C\sl{B}-C\sl{B}, and C\sl{B}-V\sl{N}---and pristine hBN to an optical cavity mode. The four systems are chosen to represent a variety of energy ranges and transition character. We change the coupling strength up to the ultrastrong coupling regime and discover several qualitatively different regimes in the experimentally observable absorption spectra. At low coupling strengths, such that the cavity mode is coupled only to the lowest-lying excitation, we observe the emergence of lower and upper polariton states, or electron-photon hybrid states, split in energy symmetrically around the bare electronic transition. As the coupling strength is increased further, the lower polariton decreases much further in energy than can be expected from a simple Jaynes-Cummings-type model where the cavity mode is only coupled to the lowest-lying excitation. Instead, what occurs is that the upper polariton interacts with the quasi-continuum of electronic states in the conduction band, effectively enhancing the coupling strength. In addition, the absorption and, thus, magnitude of the transition dipole moment of the lower polariton increases drastically as the upper polariton quenches optical activity from this electronic quasi-continuum, which will result in emission rates that are orders of magnitude higher and potential enabling more facile emission of quantum light from defect centers. 

We find that the behavior of defect polaritons is largely universal among the four systems studied, with the principal difference being how far in energy the lowest-lying excitation is from the coupling-enhancing electronic continuum---for pristine and CH\sl{B} whose lowest-lying excitation energies are much closer to the electronic quasi-continuum than those of C\sl{B}-C\sl{B} and C\sl{B}-V\sl{N}, the magnitudes of the change in excitation energy, absorption energy, and photonic weight are much higher. We therefore expect similar phenomena to manifest in other defect systems in wide band gap semiconductors, such as diamond and silicon carbide, or other two-dimensional materials, such as the transition metal dichalcogenides (TMDs). Finally, we emphasize that the enhancement of emission intensity is a consequence of the strong coupling regime, not the weak coupling regime that results in previously observed Purcell enhancement of defect emission, although in principle the effects of such cavities could also be included in the QEDFT formalism \cite{Wang2020LossQEDFT} to further enhance the decay rate of defect polaritons, as we show in Appendix \ref{app:losses}.

We also show that the transition densities of the studied systems can be differentially impacted with an example: the transition density of the lowest-lying excitation of CH\sl{B} is spatially delocalized across the entire hBN flake both inside and outside of the cavity, while the transition density of the lowest-lying excitation of C\sl{B}-V\sl{N} is localized around the defect site outside the cavity and delocalized inside the cavity. Therefore, further work studying how the delocalization of the electronic transition density affects losses due to, for instance, electron-phonon coupling is necessary to fully predict their optical activity \textit{ab initio}. In addition, determining how to leverage delocalization of the electronic transition density to transfer energy between spatially separated emitters warrants further investigation. 

Given the large shifts in emission frequency, large increases in transition dipole moments, and delocalization of the electronic transition density due to coupling between the electronic continuum and the cavity mode, coupling defect systems to optical cavities may be a powerful control knob for tuning the optical properties of defects for quantum technological applications. We predict these properties of defect polaritons with a first-principles method that encapsulates the full complexity of the electronic structure of defects in a a solid-state material, highlighting the importance of electronic structure techniques in the development of quantum optical materials. Looking forward, we expect further theoretical and computational advances in generalizing QEDFT to the ultrastrong coupling regime and to periodic systems will enable computationally driven discoveries of further complex phenomena in solid-state systems strongly coupled to light that explore the full capability of optical cavities. In addition, we anticipate that applying these first principles-based approaches to spin-polarized systems to be especially fruitful for engineering quantum technological systems, such as NV and SiV$^-$ defect centers in diamond where logical qubits are often mapped to the spin state in the ground state manifold that operate in the GHz range and can be coupled to host lattice phonons, magnetic fields, and microwaves.

\section*{Acknowledgements}
The authors acknowledge valuable discussions with Matt Trusheim, Valentin Walther, Christopher J. Ciccarino, and Prineha Narang. D.S.W. is an NSF Graduate Research Fellow. S.F.Y. would like to thank the Department of Energy for funding under award DE-SC0020115. Calculations were performed using the computational facilities of the Flatiron Institute and the Extreme Science and Engineering Discovery Environment (XSEDE) \cite{xsede} through allocation NNT210001. The Flatiron Institute is a division of the Simons Foundation. 

\appendix
\section{QEDFT Details} \label{app:compmethods}

\begin{figure}[!tbhp]
\centering
\includegraphics[width=0.8\linewidth]{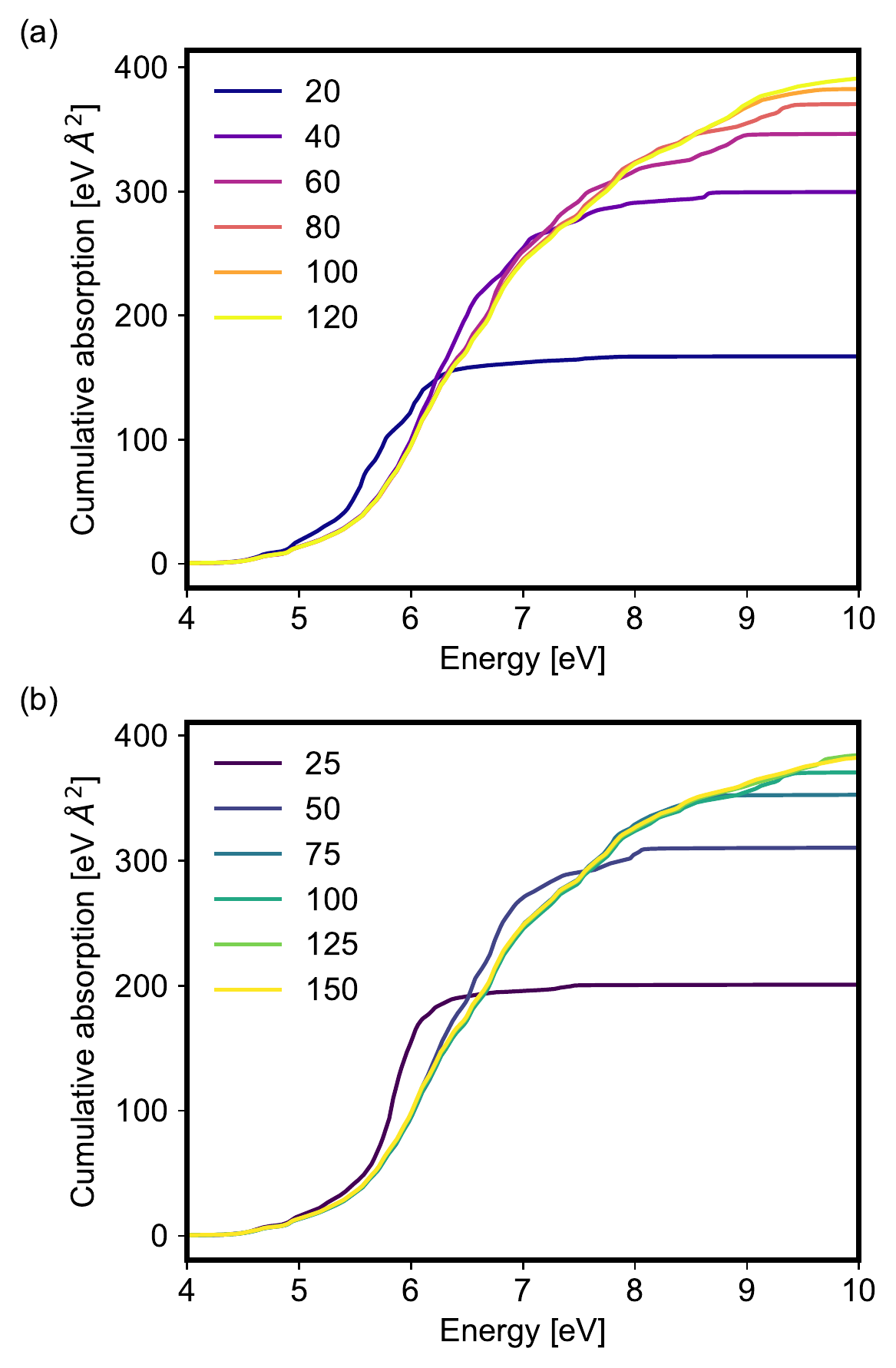}
\caption{Convergence. For all four systems, the upper polariton interacts with the near-continuum of electronic transitions starting from 4 eV to $\sim$9 eV in the coupling strength range studied. We ensure that the total light-matter coupling is converged for a given cavity mode and vacuum electric field by converging the integrated absorption relative to the number of \textbf{(a)} occupied and \textbf{(b)} unoccupied states included in the electronic component of the Casida linear-response time-dependent density functional theory for the pristine hBN nanoflake. At 9 eV, the integrated absorption is converged within 1\% for 80 occupied states and 100 unoccupied states.
}
\label{fig:converge}
\end{figure}

We briefly summarize how to compute the electronic ground state of the defect systems in the hBN flakes, as described in further detail in Ref. \cite{Wang2020Hybridized}. We use a pseudopotential, real-space density functional theory (DFT) code Octopus~\cite{octopus1, octopus2,octopus3}. To optimize chemical geometries and ground state electronic densities, we use SG15 optimized norm-conserving Vanderbilt pseudopotentials~\cite{Schlipf2015, Hamann2013} and the Perdew, Burke, and Ernzerhof (PBE) generalized gradient approximation exchange-correlation functional~\cite{Perdew1996}. The PBE functional has been used in previous studies of defects in hBN~\cite{McDougall2017, Tawfik2017, Tran2016a, Wu2017, Tancogne-dejean2018} where it has been known to underestimate the band gap and systematically mischaracterize certain optical properties~\cite{Abdi2018}, although results from PBE often qualitatively match those from the more accurate but computationally expensive Heyd–Scuseria–Ernzerhof functional (HSE) functional~\cite{Ernzerhof2003}. The real-space simulation box consists of spheres with 4~\AA~radius around each atom in a mesh with spacing of 0.20~\AA, as in Refs.~\cite{Tancogne-dejean2018, Wang2020Hybridized}. The calculated bandgap agrees with the range of bandgaps calculated for nanoflakes of hBN ~\cite{Maruyama2018} and is close to the calculated band gap of 4.50 eV with periodic DFT codes and the PBE functional~\cite{Tancogne-dejean2018}. 

To compute the excitations of the system where electronic and photonic excitations are treated on the same quantized footing ~\cite{tokatly2013,ruggenthaler2014, Ruggenthaler2018, flick2019lmrnqe}, we employ the linear-response formulation  of time-dependent QEDFT implemented in the publicly available version of Octopus, first introduced in Ref.\,\cite{flick2019lmrnqe}. This implementation has also been applied to construct potential-energy surfaces~\cite{flickexcited}, and extended to photonic losses ~\cite{Wang2020LossQEDFT, Sidler2021}. With QEDFT, we solve the light-matter Hamiltonian $H$ for a non-relativistic system of $M$ electrons interacting with the quantized light field of $N$ photon modes, in the absence of an external classical current and under the dipole approximation \cite{tokatly2013, flick2017, flickexcited}:
\begin{align} \label{eq:hamiltonian}
H = H_{\rm e} + \sum_{k=1}^{N} \frac{1}{2}\left[ p_k^2 + \omega_k^2 \left(q_k-\frac{\bm{\lambda}_k}{\omega_k}\cdot \bm{R}\right)^2\right],
\end{align}
where $H_{\rm e}$ is the electronic Hamiltonian; the $k$th quantized photon mode is given by the operators for the photon conjugate momentum $p_k={\rm i} \sqrt{\frac{\hbar\omega_k}{2}}(a_k-a_k^\dagger)$ and the photon displacement coordinate $q_k=\sqrt{\frac{\hbar}{2\omega_k}}(a_k+a_k^\dagger)$; the photon annihilation (creation) operator is $a_k$ ($a_k^\dagger$); and $\omega_k$ is the frequency of mode $k$. The photon modes couple to the electronic system through the position operator $\bm{R}=\sum_{i=1}^{M}\bm{r}_i$ of the electronic system and $q_k$ of the photonic system. The cavity strength $\bm{\lambda}_k$ determines the strength of this interaction:
\begin{align}
    \bm{\lambda}_k=\sqrt{\frac{2}{\hbar \omega_k}}\bm{E}_ke,
\end{align}  
where $\bm{E}_k$ is the amplitude of the electric field at the center of charge and $e$ is the elementary charge. This coupling strength is closely related to the commonly used coupling rate $g_{i,k}$ between the cavity mode $k$ and a electronic excitation $i$ from the electronic ground state $|{\rm g}\rangle$:
\begin{align} \label{eq:g}
    g_{i,k}=-\frac{e}{\hbar}\bm{E}_k\cdot \langle {\rm g}|\bm{R}|{\rm e}_i\rangle=-\sqrt{\frac{\omega_k}{2\hbar}}\bm{\lambda}_k\cdot \boldsymbol{d}_i.
\end{align}
The transition dipole moments $\bm{d}_i=\langle {\rm g}|\bm{R}|{\rm e}_i\rangle$ and energies of electronic excitations $\hbar\omega_i$ can also be determined with a standard linear-response time-dependent density-functional theory (TDDFT)~\cite{casida1995time}. In the strong-coupling limit shown in the main text, we include a single, lossless cavity mode ($N=1$), or equivalently $\kappa=0$, such that we are in the strong-coupling regime for $g>0$. We can also incorporate cavity losses by inputting the spectral profile of a lossy cavity, as described in Ref. \citenum{Wang2020LossQEDFT} and shown for CH\sl{B} in Appendix \ref{app:losses}, to model Purcell enhancements simultaneously with the formation of defect polaritons.

To solve the Hamiltonian in Eq. \eqref{eq:hamiltonian}, we apply the generalized Casida equation~\cite{flick2019lmrnqe,casida1995time}, where the electron-electron interactions included in TDDFT and the electron-photon interactions are solved simultaneously. We diagonalize the Casida matrix of size $N$\sl{o}$N$\sl{u}$+N=MN$, where $N$\sl{o} is the number of occupied orbitals, $N$\sl{u} is the number of unoccupied orbitals, $N$\sl{o}$N$\sl{u}$=M$ is the number of electronic excitations, and $N$ the number of photon modes, to obtain excitation energies and transition dipole moments that generate the absorption spectrum: 
\begin{align}\label{eq:polaritonicAbsorption}
     A_{j}(\hbar\omega)= \mathcal{C}\sum_{l=1}^{M+N}\delta(\omega-\omega_l)\hbar\omega_l |\sum_{i=1}^{M} C_{il}^{\rm el} {d}_{i,j}|^2,
 \end{align}
where the absorption of $j$-polarized light (with $j\in \{x,y,z\}$) is a function of a frequency-independent pre-factor $\mathcal{C}=2m_{\rm e}/(3\hbar^2)$ (with $m_{\rm e}$ the electron mass); the mode energy is $\hbar\omega_i$; $C_{il}^{\rm el}$ ($C_{kl}^{\rm ph}$) is the projection of an original, unmixed electronic (photonic) state $|{\rm e}_i,0\rangle$ ($|{\rm g},1_k\rangle$) to a resulting polaritonic state $|v_l\rangle$; and $d_{i,j}$ is the $j$-component of the transition dipole moment $\bm{d}_i$ of electronic excitation $i\in$ $M$ excited electronic states. For presentation purposes, we broaden the delta function $\delta(\omega-\omega_l)$ with a discrete Lorentzian: $\delta(\omega-\omega_l)\rightarrow\Gamma\Delta\omega\big(2\pi[(\omega-\omega_l)^2+(\Gamma/2)^2]\big)^{-1}$, where $\Delta(\hbar\omega)$ is the energy spacing and $\hbar\Gamma$ is the spectral broadening. We drop $\mathcal{C}$ for convenience when plotting the absorption. Using the projections $C_{il}^{\rm el}$ and $C_{kl}^{\rm ph}$, we can additionally calculate the logarithm of the weight of photon character $w_l^{\rm ph}= \sum_{k=1}^{N} W_{kl}^{\rm ph}=1-w_l^{\rm el}$ where $W_{kl}^{\rm ph}=|C_{kl}^{\rm ph}|^2$. In the absence of cavity photons, Eq. \eqref{eq:polaritonicAbsorption} reduces to the standard form of absorption of an all-electronic system:
\begin{align} \label{eq:elecabsorption}
     A_{j}(\hbar\omega)= \mathcal{C}\sum_{i=1}^{M}\delta(\omega-\omega_i)\hbar\omega_i |{d}_{i,j}|^2.
 \end{align}

Since the upper polariton of the defect polariton of all four systems interacts with the near-continuum of electronic transitions starting from $\sim$4 to $\sim$9 eV within the cavity strength range of $\lambda$ considered, we ensure that the total light-matter coupling is converged for a given cavity mode and vacuum electric field by converging the integrated absorption relative to the number of occupied and unoccupied states included in the electronic component of the Casida calculation for the pristine hBN nanoflake, as shown in Fig. \ref{fig:converge}(a) and (b), respectively. At 9 eV, the integrated absorption is converged within 1\% for 80 occupied states and 100 unoccupied states. We use this number of occupied and unoccupied states for all four defect systems considered. 

\section{Cavity losses} \label{app:losses}

\begin{figure}[!tbhp]
\centering
\includegraphics[width=0.8\linewidth]{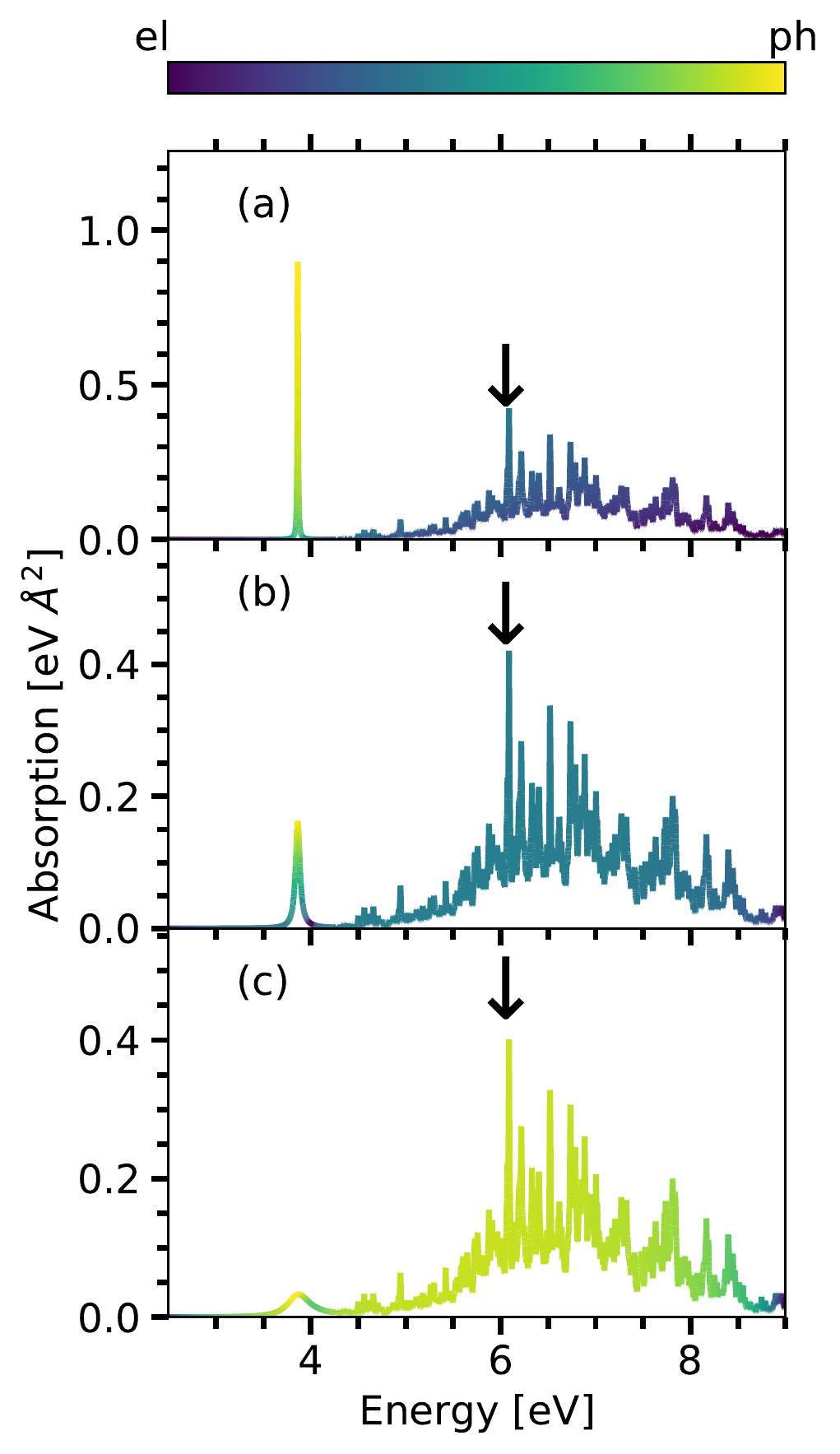}
\caption{Cavity losses. Absorption spectra of an $x$-polarized cavity mode resonantly tuned to the lowest-lying excitation of CH\sl{B} for coupling strength $\lambda$ of 0.493 eV$^{1/2}$/nm and cavity loss rate $\kappa$ of \textbf{(a)} 0.00, \textbf{(b)} 0.05, and \textbf{(c)} 0.30 eV. Therefore, Fig. \ref{fig:losses}(a) corresponds to Fig. \ref{fig:sweep}(d). The relative electronic and photonic weight of each polariton state is overlaid on the absorption curves. We set $\hbar\Gamma=\SI{10}{\milli\eV}$ and $\Delta(\hbar\omega)=\SI{1}{\milli\eV}$. For increasing coupling strength, for the peak corresponding to the lower polariton, the linewidth increases, the maximum intensity decreases, and the central frequency does not shift. Meanwhile, the shape and intensity of the quasi-continuum absorption do not change significantly, as indicated by the black arrow indicating the peak absorption of the quasi-continuum for $\kappa=0$ in (a), although the relative photonic character increases.
}
\label{fig:losses}
\end{figure}

As in Ref. \cite{Wang2020LossQEDFT}, we model the effect of non-zero cavity losses on the absorption spectrum of defect polaritons in Fig. \ref{fig:losses} by replacing the cavity strength $\lambda_c$ of a cavity with energy $\hbar\omega_c$ with many photon modes whose spectral density follows a Lorentzian profile:
\begin{align} \label{eq:lorentz}
    |\bm{\lambda}_k|^2=|\bm{\lambda}_{\rm c}|^2L(\Delta\omega,\kappa,\omega_{k{\rm c}}),
\end{align}
where 
\begin{align}
    L(\Delta\omega,\kappa,\omega_{k{\rm c}})=\Delta\omega\frac{1}{2\pi}\frac{\kappa}{(\omega_k-\omega_c)^2+(\kappa/2)^2},
\end{align}
where $\kappa$ is the cavity loss rate. Notably, the peak intensity of the lower polariton decreases and the linewidth increases, while the quasi-continuum remains largely unchanged. 

\FloatBarrier

\section{Comparison with a cQED model} \label{app:cQED}

\begin{figure*}[tbhp]
\centering
\includegraphics[width=0.8\linewidth]{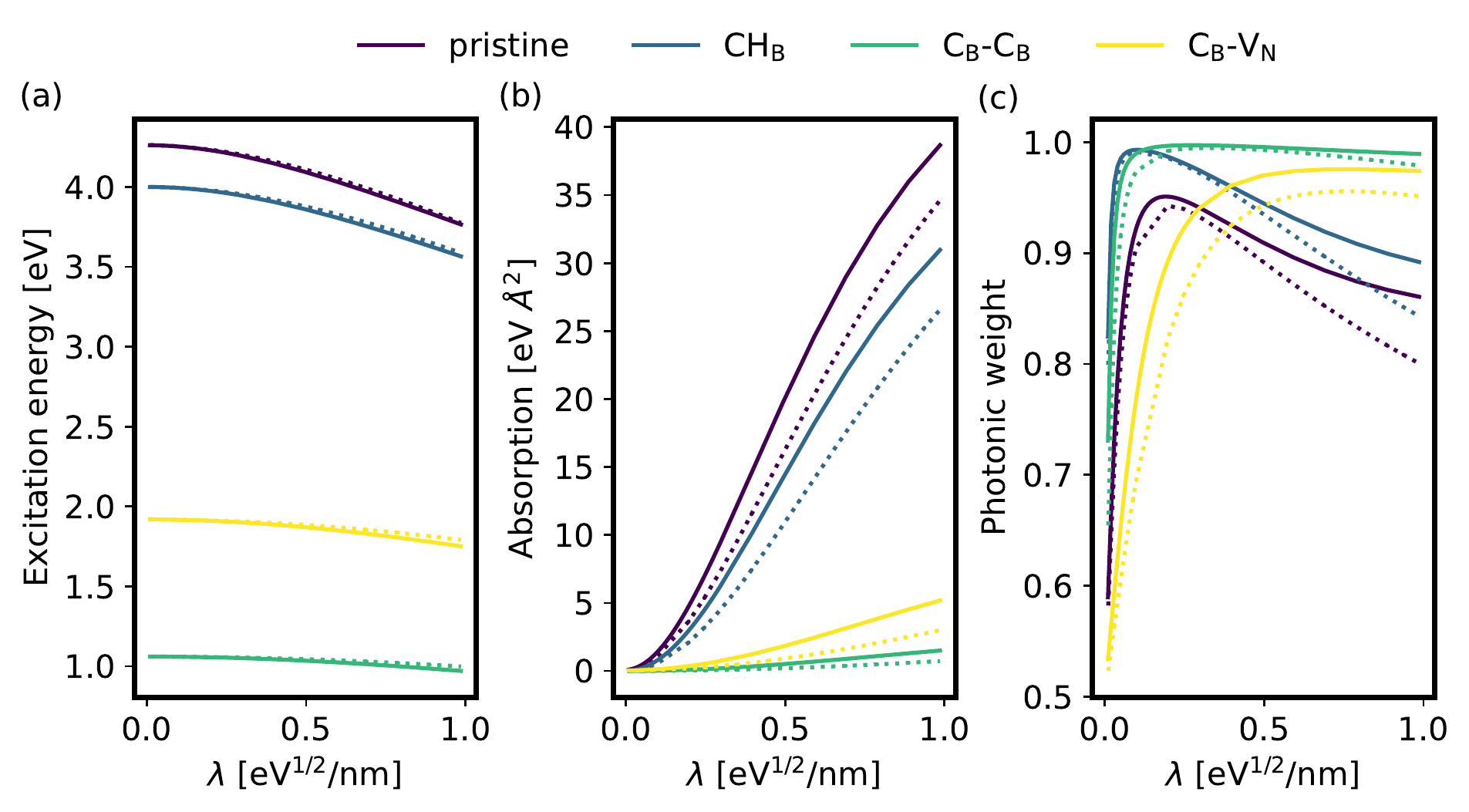}
\caption{Comparison of QEDFT (solid lines) and cQED model (dotted lines). As in Fig. \ref{fig:comparison}, for all four defect systems, we show the \textbf{(a)} excitation energy, \textbf{(b)} absorption with zero spectral broadening, and \textbf{(c)} photonic weight of the lowest-lying excitation, or the lower polariton for coupling strength $\lambda$ from 0.010 to 0.986 eV$^{1/2}$/nm. QEDFT and cQED model present similar results that differ more strongly at higher coupling strengths evidently due to the inclusion of the dipole self-energy and counter-rotating terms in QEDFT.
}
\label{fig:cQED}
\end{figure*}

We compare the results in Fig. \ref{fig:comparison} calculated with first-principles QEDFT with those from a cQED model briefly summarized here and described in further detail in Ref. \cite{Wang2020LossQEDFT}. The Hamiltonian $H_{\rm cQED}$ describing the light-matter interaction between electronic excitations and cavity photon modes, in the rotating wave approximation, is given by
\begin{multline} \label{app:eq1}
H_{\rm cQED}=\sum_{i=1}^{M}\hbar\omega^{\rm el}_{i}\sigma_i^\dagger\sigma_i + \sum_{k=1}^{N}\hbar \omega_k a^\dagger_k a_k \\
+ \hbar\sum_{i,k=1}^{M,N} (g_{i,k} \sigma_i^\dagger a_k+{\rm H.c.}),
\end{multline}
where $\sigma_i^\dagger$ ($\sigma_i$), $a_k^\dagger$ ($a_k$) are creation (annihilation) operators for the $i$th of $M$ electronic excitations and $k$th of $N$ photon modes, respectively, H.c. is the Hermitian conjugate, and $\hbar\omega^{\rm el}_{i}$ and $\hbar\omega_{k}$ are the mode energies.

The state of the system in the single excitation subspace is
\begin{align} \label{app:eqFano1}
    |\psi\rangle=& c_0|{\rm g},\{0_k\}\rangle+\sum_{i=1}^{M}c_{i}^{\rm el}|{{\rm e}_i}, \{0_k\}\rangle+\sum_{k=1}^{N}c_{k}^{\rm ph}|{\rm g}, \{1_k\}\rangle,
\end{align}
where the coefficients $c_0$, $c_i^{\rm el}$, $c_k^{\rm ph}$  are time-dependent. We plug this \textit{ansatz} into the Schr\"{o}dinger equation to obtain a system of linear differential equations:
\begin{align} \label{app:eqFano2}
    \dot c_{i}^{\rm el} &=-{\rm i}\omega^{\rm el}_i c_{i}^{\rm el}-{\rm i}\sum_{k=1}^{N} g_{i,k} c_k^{\rm ph},\\
    \dot c_k^{\rm ph} &=-{\rm i}\sum_{i=1}^{M}g^\ast_{i,k} c_{i}^{\rm el}-{\rm i}\omega_k c_k .
\end{align}
This system of equations can be solved to obtain the polariton energies $\hbar\omega_l$ and polaritonic states as projections onto the bare electronic excitations and photons, which can be used to calculate the electronic absorption spectrum and weight of the original, unmixed electronic and photonic states as described in Appendix \ref{app:compmethods}.

Importantly, both the cQED and QEDFT models are within the point dipole approximation and single excitation subspace. However, the QEDFT model has the advantage of including the dipole self-energy term $\bm{R}^2$, necessarily includes all electronic and photonic degrees of freedom for good quantitative agreement with experimental data up to the accuracy of the density functional used, and includes counter-rotating terms typically neglected in the rotating wave approximation. In contrast, calculations based on the cQED model generally include only a few, judiciously chosen electronic excitations to minimize computational complexity, although in principle, many can be included, as we do in Fig. \ref{fig:comparison} where we compare the results from the cQED and QEDFT models. Despite the aforementioned differences in these models, we note excellent qualitative agreement, although quantitatively the results differ more strongly at higher coupling strengths due to the inclusion of the dipole self-energy and counter-rotating terms in QEDFT.

\newcommand{\noopsort}[1]{} \newcommand{\printfirst}[2]{#1}
  \newcommand{\singleletter}[1]{#1} \newcommand{\switchargs}[2]{#2#1}


\begin{thebibliography}{96}%
\makeatletter
\providecommand \@ifxundefined [1]{%
 \@ifx{#1\undefined}
}%
\providecommand \@ifnum [1]{%
 \ifnum #1\expandafter \@firstoftwo
 \else \expandafter \@secondoftwo
 \fi
}%
\providecommand \@ifx [1]{%
 \ifx #1\expandafter \@firstoftwo
 \else \expandafter \@secondoftwo
 \fi
}%
\providecommand \natexlab [1]{#1}%
\providecommand \enquote  [1]{``#1''}%
\providecommand \bibnamefont  [1]{#1}%
\providecommand \bibfnamefont [1]{#1}%
\providecommand \citenamefont [1]{#1}%
\providecommand \href@noop [0]{\@secondoftwo}%
\providecommand \href [0]{\begingroup \@sanitize@url \@href}%
\providecommand \@href[1]{\@@startlink{#1}\@@href}%
\providecommand \@@href[1]{\endgroup#1\@@endlink}%
\providecommand \@sanitize@url [0]{\catcode `\\12\catcode `\$12\catcode
  `\&12\catcode `\#12\catcode `\^12\catcode `\_12\catcode `\%12\relax}%
\providecommand \@@startlink[1]{}%
\providecommand \@@endlink[0]{}%
\providecommand \url  [0]{\begingroup\@sanitize@url \@url }%
\providecommand \@url [1]{\endgroup\@href {#1}{\urlprefix }}%
\providecommand \urlprefix  [0]{URL }%
\providecommand \Eprint [0]{\href }%
\providecommand \doibase [0]{https://doi.org/}%
\providecommand \selectlanguage [0]{\@gobble}%
\providecommand \bibinfo  [0]{\@secondoftwo}%
\providecommand \bibfield  [0]{\@secondoftwo}%
\providecommand \translation [1]{[#1]}%
\providecommand \BibitemOpen [0]{}%
\providecommand \bibitemStop [0]{}%
\providecommand \bibitemNoStop [0]{.\EOS\space}%
\providecommand \EOS [0]{\spacefactor3000\relax}%
\providecommand \BibitemShut  [1]{\csname bibitem#1\endcsname}%
\let\auto@bib@innerbib\@empty
%</preamble>
\bibitem [{\citenamefont {Wrachtrup}\ \emph {et~al.}(2001)\citenamefont
  {Wrachtrup}, \citenamefont {Kilin},\ and\ \citenamefont
  {Nizovtsev}}]{Wrachtrup2001}%
  \BibitemOpen
  \bibfield  {author} {\bibinfo {author} {\bibfnamefont {J.}~\bibnamefont
  {Wrachtrup}}, \bibinfo {author} {\bibfnamefont {S.~Y.}\ \bibnamefont
  {Kilin}},\ and\ \bibinfo {author} {\bibfnamefont {A.~P.}\ \bibnamefont
  {Nizovtsev}},\ }\bibfield  {title} {\bibinfo {title} {{Quantum Qomputation
  Using the 13C Nuclear Spins Nearby the Single NV Defect Center in Diamond}},\
  }\href {https://link.springer.com/article/10.1134/1.1405224} {\bibfield
  {journal} {\bibinfo  {journal} {Opt. Spectrosc.}\ }\textbf {\bibinfo {volume}
  {91}},\ \bibinfo {pages} {429} (\bibinfo {year} {2001})}\BibitemShut
  {NoStop}%
\bibitem [{\citenamefont {Weber}\ \emph {et~al.}(2010)\citenamefont {Weber},
  \citenamefont {Koehl}, \citenamefont {Varley}, \citenamefont {Janotti},
  \citenamefont {Buckley}, \citenamefont {{Van de Walle}},\ and\ \citenamefont
  {Awschalom}}]{Weber2010}%
  \BibitemOpen
  \bibfield  {author} {\bibinfo {author} {\bibfnamefont {J.~R.}\ \bibnamefont
  {Weber}}, \bibinfo {author} {\bibfnamefont {W.~F.}\ \bibnamefont {Koehl}},
  \bibinfo {author} {\bibfnamefont {J.~B.}\ \bibnamefont {Varley}}, \bibinfo
  {author} {\bibfnamefont {A.}~\bibnamefont {Janotti}}, \bibinfo {author}
  {\bibfnamefont {B.~B.}\ \bibnamefont {Buckley}}, \bibinfo {author}
  {\bibfnamefont {C.~G.}\ \bibnamefont {{Van de Walle}}},\ and\ \bibinfo
  {author} {\bibfnamefont {D.~D.}\ \bibnamefont {Awschalom}},\ }\bibfield
  {title} {\bibinfo {title} {Quantum computing with defects.},\ }\href
  {https://doi.org/10.1073/pnas.1003052107} {\bibfield  {journal} {\bibinfo
  {journal} {Proc. Natl. Acad. Sci. U.S.A}\ }\textbf {\bibinfo {volume}
  {107}},\ \bibinfo {pages} {8513} (\bibinfo {year} {2010})}\BibitemShut
  {NoStop}%
\bibitem [{\citenamefont {Wrachtrup}(2010)}]{Wrachtrup2010}%
  \BibitemOpen
  \bibfield  {author} {\bibinfo {author} {\bibfnamefont {J.}~\bibnamefont
  {Wrachtrup}},\ }\bibfield  {title} {\bibinfo {title} {{Defect Center
  Room-Temperature Quantum Processors}},\ }\href
  {https://doi.org/10.1073/pnas.1004033107} {\bibfield  {journal} {\bibinfo
  {journal} {Proc. Natl. Acad. Sci. U.S.A.}\ }\textbf {\bibinfo {volume}
  {107}},\ \bibinfo {pages} {9479} (\bibinfo {year} {2010})}\BibitemShut
  {NoStop}%
\bibitem [{\citenamefont {Aharonovich}\ \emph {et~al.}(2016)\citenamefont
  {Aharonovich}, \citenamefont {Englund},\ and\ \citenamefont
  {Toth}}]{Aharonovich2016}%
  \BibitemOpen
  \bibfield  {author} {\bibinfo {author} {\bibfnamefont {I.}~\bibnamefont
  {Aharonovich}}, \bibinfo {author} {\bibfnamefont {D.}~\bibnamefont
  {Englund}},\ and\ \bibinfo {author} {\bibfnamefont {M.}~\bibnamefont
  {Toth}},\ }\bibfield  {title} {\bibinfo {title} {{Solid-State Single-Photon
  Emitters}},\ }\href {https://doi.org/10.1038/nphoton.2016.186} {\bibfield
  {journal} {\bibinfo  {journal} {Nat. Photonics}\ }\textbf {\bibinfo {volume}
  {10}},\ \bibinfo {pages} {631} (\bibinfo {year} {2016})}\BibitemShut
  {NoStop}%
\bibitem [{\citenamefont {Degen}\ \emph {et~al.}(2017)\citenamefont {Degen},
  \citenamefont {Reinhard},\ and\ \citenamefont {Cappellaro}}]{Degen2017}%
  \BibitemOpen
  \bibfield  {author} {\bibinfo {author} {\bibfnamefont {C.~L.}\ \bibnamefont
  {Degen}}, \bibinfo {author} {\bibfnamefont {F.}~\bibnamefont {Reinhard}},\
  and\ \bibinfo {author} {\bibfnamefont {P.}~\bibnamefont {Cappellaro}},\
  }\bibfield  {title} {\bibinfo {title} {{Quantum Sensing}},\ }\href
  {https://doi.org/10.1103/RevModPhys.89.035002} {\bibfield  {journal}
  {\bibinfo  {journal} {Rev. Mod. Phys.}\ }\textbf {\bibinfo {volume} {89}},\
  \bibinfo {pages} {035002} (\bibinfo {year} {2017})}\BibitemShut {NoStop}%
\bibitem [{\citenamefont {Childress}\ \emph {et~al.}(2014)\citenamefont
  {Childress}, \citenamefont {Walsworth},\ and\ \citenamefont
  {Lukin}}]{Childress2014}%
  \BibitemOpen
  \bibfield  {author} {\bibinfo {author} {\bibfnamefont {L.}~\bibnamefont
  {Childress}}, \bibinfo {author} {\bibfnamefont {R.}~\bibnamefont
  {Walsworth}},\ and\ \bibinfo {author} {\bibfnamefont {M.}~\bibnamefont
  {Lukin}},\ }\bibfield  {title} {\bibinfo {title} {{Atom-Like Crystal Defects:
  From Quantum Computers to Biological Sensors}},\ }\href
  {https://doi.org/10.1063/PT.3.2549} {\bibfield  {journal} {\bibinfo
  {journal} {Phys. Today}\ }\textbf {\bibinfo {volume} {67}},\ \bibinfo {pages}
  {38} (\bibinfo {year} {2014})}\BibitemShut {NoStop}%
\bibitem [{\citenamefont {Atat{\"{u}}re}\ \emph {et~al.}(2018)\citenamefont
  {Atat{\"{u}}re}, \citenamefont {Englund}, \citenamefont {Vamivakas},
  \citenamefont {Lee},\ and\ \citenamefont {Wrachtrup}}]{Atature2018}%
  \BibitemOpen
  \bibfield  {author} {\bibinfo {author} {\bibfnamefont {M.}~\bibnamefont
  {Atat{\"{u}}re}}, \bibinfo {author} {\bibfnamefont {D.}~\bibnamefont
  {Englund}}, \bibinfo {author} {\bibfnamefont {N.}~\bibnamefont {Vamivakas}},
  \bibinfo {author} {\bibfnamefont {S.~Y.}\ \bibnamefont {Lee}},\ and\ \bibinfo
  {author} {\bibfnamefont {J.}~\bibnamefont {Wrachtrup}},\ }\bibfield  {title}
  {\bibinfo {title} {{Material Platforms for Spin-based Photonic Quantum
  Technologies}},\ }\href {https://doi.org/10.1038/s41578-018-0008-9}
  {\bibfield  {journal} {\bibinfo  {journal} {Nat. Rev. Mater.}\ }\textbf
  {\bibinfo {volume} {3}},\ \bibinfo {pages} {38} (\bibinfo {year}
  {2018})}\BibitemShut {NoStop}%
\bibitem [{\citenamefont {Kurtsiefer}\ \emph {et~al.}(2000)\citenamefont
  {Kurtsiefer}, \citenamefont {Mayer}, \citenamefont {Zarda},\ and\
  \citenamefont {Weinfurter}}]{Kurtsiefer2000}%
  \BibitemOpen
  \bibfield  {author} {\bibinfo {author} {\bibfnamefont {C.}~\bibnamefont
  {Kurtsiefer}}, \bibinfo {author} {\bibfnamefont {S.}~\bibnamefont {Mayer}},
  \bibinfo {author} {\bibfnamefont {P.}~\bibnamefont {Zarda}},\ and\ \bibinfo
  {author} {\bibfnamefont {H.}~\bibnamefont {Weinfurter}},\ }\bibfield  {title}
  {\bibinfo {title} {{Stable Solid-State Source of Single Photons}},\ }\href
  {https://doi.org/10.1103/PhysRevLett.85.290} {\bibfield  {journal} {\bibinfo
  {journal} {Phys. Rev. Lett.}\ }\textbf {\bibinfo {volume} {85}},\ \bibinfo
  {pages} {290} (\bibinfo {year} {2000})}\BibitemShut {NoStop}%
\bibitem [{\citenamefont {Ye}\ \emph {et~al.}(2019)\citenamefont {Ye},
  \citenamefont {Seo},\ and\ \citenamefont {Galli}}]{Ye2019}%
  \BibitemOpen
  \bibfield  {author} {\bibinfo {author} {\bibfnamefont {M.}~\bibnamefont
  {Ye}}, \bibinfo {author} {\bibfnamefont {H.}~\bibnamefont {Seo}},\ and\
  \bibinfo {author} {\bibfnamefont {G.}~\bibnamefont {Galli}},\ }\bibfield
  {title} {\bibinfo {title} {{Spin Coherence in Two-Dimensional Materials}},\
  }\href {https://doi.org/10.1038/s41524-019-0182-3} {\bibfield  {journal}
  {\bibinfo  {journal} {npj Comput. Mater.}\ }\textbf {\bibinfo {volume} {5}},\
  \bibinfo {pages} {1} (\bibinfo {year} {2019})}\BibitemShut {NoStop}%
\bibitem [{\citenamefont {Appel}\ \emph {et~al.}(2015)\citenamefont {Appel},
  \citenamefont {Ganzhorn}, \citenamefont {Neu},\ and\ \citenamefont
  {Maletinsky}}]{Appel2015}%
  \BibitemOpen
  \bibfield  {author} {\bibinfo {author} {\bibfnamefont {P.}~\bibnamefont
  {Appel}}, \bibinfo {author} {\bibfnamefont {M.}~\bibnamefont {Ganzhorn}},
  \bibinfo {author} {\bibfnamefont {E.}~\bibnamefont {Neu}},\ and\ \bibinfo
  {author} {\bibfnamefont {P.}~\bibnamefont {Maletinsky}},\ }\bibfield  {title}
  {\bibinfo {title} {{Nanoscale microwave imaging with a single electron spin
  in diamond}},\ }\href {https://doi.org/10.1088/1367-2630/17/11/112001}
  {\bibfield  {journal} {\bibinfo  {journal} {New J. Phys.}\ }\textbf {\bibinfo
  {volume} {17}},\ \bibinfo {pages} {112001} (\bibinfo {year}
  {2015})}\BibitemShut {NoStop}%
\bibitem [{\citenamefont {Lemonde}\ \emph {et~al.}(2018)\citenamefont
  {Lemonde}, \citenamefont {Meesala}, \citenamefont {Sipahigil}, \citenamefont
  {Schuetz}, \citenamefont {Lukin}, \citenamefont {Loncar},\ and\ \citenamefont
  {Rabl}}]{Lemonde2018}%
  \BibitemOpen
  \bibfield  {author} {\bibinfo {author} {\bibfnamefont {M.~A.}\ \bibnamefont
  {Lemonde}}, \bibinfo {author} {\bibfnamefont {S.}~\bibnamefont {Meesala}},
  \bibinfo {author} {\bibfnamefont {A.}~\bibnamefont {Sipahigil}}, \bibinfo
  {author} {\bibfnamefont {M.~J.}\ \bibnamefont {Schuetz}}, \bibinfo {author}
  {\bibfnamefont {M.~D.}\ \bibnamefont {Lukin}}, \bibinfo {author}
  {\bibfnamefont {M.}~\bibnamefont {Loncar}},\ and\ \bibinfo {author}
  {\bibfnamefont {P.}~\bibnamefont {Rabl}},\ }\bibfield  {title} {\bibinfo
  {title} {{Phonon Networks with Silicon-Vacancy Centers in Diamond
  Waveguides}},\ }\href {https://doi.org/10.1103/PhysRevLett.120.213603}
  {\bibfield  {journal} {\bibinfo  {journal} {Phys. Rev. Lett.}\ }\textbf
  {\bibinfo {volume} {120}},\ \bibinfo {pages} {213603} (\bibinfo {year}
  {2018})}\BibitemShut {NoStop}%
\bibitem [{\citenamefont {Candido}\ \emph {et~al.}(2020)\citenamefont
  {Candido}, \citenamefont {Fuchs}, \citenamefont {Johnston-Halperin},\ and\
  \citenamefont {Flatte}}]{Candido2020}%
  \BibitemOpen
  \bibfield  {author} {\bibinfo {author} {\bibfnamefont {D.~R.}\ \bibnamefont
  {Candido}}, \bibinfo {author} {\bibfnamefont {G.~D.}\ \bibnamefont {Fuchs}},
  \bibinfo {author} {\bibfnamefont {E.}~\bibnamefont {Johnston-Halperin}},\
  and\ \bibinfo {author} {\bibfnamefont {M.~E.}\ \bibnamefont {Flatte}},\
  }\bibfield  {title} {\bibinfo {title} {{Predicted strong coupling of
  solid-state spins via a single magnon mode}},\ }\href
  {https://iopscience.iop.org/article/10.1088/2633-4356/ab9a55} {\bibfield
  {journal} {\bibinfo  {journal} {Mater. Quantum. Technol}\ } (\bibinfo {year}
  {2020})}\BibitemShut {NoStop}%
\bibitem [{\citenamefont {Neuman}\ \emph
  {et~al.}(2020{\natexlab{a}})\citenamefont {Neuman}, \citenamefont
  {Eichenfield}, \citenamefont {Trusheim}, \citenamefont {Hackett},
  \citenamefont {Narang},\ and\ \citenamefont
  {Englund}}]{Neuman2020Phononicbus}%
  \BibitemOpen
  \bibfield  {author} {\bibinfo {author} {\bibfnamefont {T.}~\bibnamefont
  {Neuman}}, \bibinfo {author} {\bibfnamefont {M.}~\bibnamefont {Eichenfield}},
  \bibinfo {author} {\bibfnamefont {M.}~\bibnamefont {Trusheim}}, \bibinfo
  {author} {\bibfnamefont {L.}~\bibnamefont {Hackett}}, \bibinfo {author}
  {\bibfnamefont {P.}~\bibnamefont {Narang}},\ and\ \bibinfo {author}
  {\bibfnamefont {D.}~\bibnamefont {Englund}},\ }\bibfield  {title} {\bibinfo
  {title} {{A Phononic Bus for Coherent Interfaces Between a Superconducting
  Quantum Processor, Spin Memory, and Photonic Quantum Networks}},\ }\href
  {http://arxiv.org/abs/2003.08383} {\bibfield  {journal} {\bibinfo  {journal}
  {arXiv:2003.08383}\ } (\bibinfo {year} {2020}{\natexlab{a}})}\BibitemShut
  {NoStop}%
\bibitem [{\citenamefont {Wang}\ \emph
  {et~al.}(2020{\natexlab{a}})\citenamefont {Wang}, \citenamefont {Neuman},\
  and\ \citenamefont {Narang}}]{Wang2020Selection}%
  \BibitemOpen
  \bibfield  {author} {\bibinfo {author} {\bibfnamefont {D.~S.}\ \bibnamefont
  {Wang}}, \bibinfo {author} {\bibfnamefont {T.}~\bibnamefont {Neuman}},\ and\
  \bibinfo {author} {\bibfnamefont {P.}~\bibnamefont {Narang}},\ }\bibfield
  {title} {\bibinfo {title} {Spin emitters beyond the point dipole
  approximation in nanomagnonic cavities},\ }\href
  {https://arxiv.org/abs/2012.04662} {\bibfield  {journal} {\bibinfo  {journal}
  {arXiv:2012.04662}\ } (\bibinfo {year} {2020}{\natexlab{a}})}\BibitemShut
  {NoStop}%
\bibitem [{\citenamefont {Wang}\ \emph
  {et~al.}(2021{\natexlab{a}})\citenamefont {Wang}, \citenamefont {Haas},\ and\
  \citenamefont {Narang}}]{Wang2021Perspective}%
  \BibitemOpen
  \bibfield  {author} {\bibinfo {author} {\bibfnamefont {D.~S.}\ \bibnamefont
  {Wang}}, \bibinfo {author} {\bibfnamefont {M.}~\bibnamefont {Haas}},\ and\
  \bibinfo {author} {\bibfnamefont {P.}~\bibnamefont {Narang}},\ }\bibfield
  {title} {\bibinfo {title} {Quantum interfaces to the nanoscale},\ }\href@noop
  {} {\bibfield  {journal} {\bibinfo  {journal} {ACS Nano}\ } (\bibinfo {year}
  {2021}{\natexlab{a}})}\BibitemShut {NoStop}%
\bibitem [{\citenamefont {McDougall}\ \emph {et~al.}(2017)\citenamefont
  {McDougall}, \citenamefont {Partridge}, \citenamefont {Nicholls},
  \citenamefont {Russo},\ and\ \citenamefont {McCulloch}}]{McDougall2017}%
  \BibitemOpen
  \bibfield  {author} {\bibinfo {author} {\bibfnamefont {N.~L.}\ \bibnamefont
  {McDougall}}, \bibinfo {author} {\bibfnamefont {J.~G.}\ \bibnamefont
  {Partridge}}, \bibinfo {author} {\bibfnamefont {R.~J.}\ \bibnamefont
  {Nicholls}}, \bibinfo {author} {\bibfnamefont {S.~P.}\ \bibnamefont
  {Russo}},\ and\ \bibinfo {author} {\bibfnamefont {D.~G.}\ \bibnamefont
  {McCulloch}},\ }\bibfield  {title} {\bibinfo {title} {{Influence of Point
  Defects on the Near Edge Structure of Hexagonal Boron Nitride}},\ }\href
  {https://doi.org/10.1103/PhysRevB.96.144106} {\bibfield  {journal} {\bibinfo
  {journal} {Phys. Rev. B}\ }\textbf {\bibinfo {volume} {96}},\ \bibinfo
  {pages} {144106} (\bibinfo {year} {2017})}\BibitemShut {NoStop}%
\bibitem [{\citenamefont {MacKoit-Sinkevi{\v{c}}iene}\ \emph
  {et~al.}(2019)\citenamefont {MacKoit-Sinkevi{\v{c}}iene}, \citenamefont
  {MacIaszek}, \citenamefont {{Van De Walle}},\ and\ \citenamefont
  {Alkauskas}}]{MacKoit-Sinkeviciene2019}%
  \BibitemOpen
  \bibfield  {author} {\bibinfo {author} {\bibfnamefont {M.}~\bibnamefont
  {MacKoit-Sinkevi{\v{c}}iene}}, \bibinfo {author} {\bibfnamefont
  {M.}~\bibnamefont {MacIaszek}}, \bibinfo {author} {\bibfnamefont {C.~G.}\
  \bibnamefont {{Van De Walle}}},\ and\ \bibinfo {author} {\bibfnamefont
  {A.}~\bibnamefont {Alkauskas}},\ }\bibfield  {title} {\bibinfo {title}
  {{Carbon Dimer Defect as a Source of the 4.1 eV Luminescence in Hexagonal
  Boron Nitride}},\ }\href {https://doi.org/10.1063/1.5124153} {\bibfield
  {journal} {\bibinfo  {journal} {Appl. Phys. Lett.}\ }\textbf {\bibinfo
  {volume} {115}},\ \bibinfo {pages} {212101} (\bibinfo {year}
  {2019})}\BibitemShut {NoStop}%
\bibitem [{\citenamefont {Czelej}\ and\ \citenamefont
  {Majewski}(2020)}]{Czelej2020}%
  \BibitemOpen
  \bibfield  {author} {\bibinfo {author} {\bibfnamefont {K.}~\bibnamefont
  {Czelej}}\ and\ \bibinfo {author} {\bibfnamefont {J.~A.}\ \bibnamefont
  {Majewski}},\ }\bibfield  {title} {\bibinfo {title} {{Electronic Structure
  and Magneto-Optical Properties of Silicon-Nitrogen-Vacancy Complexes in
  Diamond}},\ }\href
  {https://journals.aps.org/prb/abstract/10.1103/PhysRevB.102.115102}
  {\bibfield  {journal} {\bibinfo  {journal} {Phys. Rev. B}\ }\textbf {\bibinfo
  {volume} {102}},\ \bibinfo {pages} {115102} (\bibinfo {year}
  {2020})}\BibitemShut {NoStop}%
\bibitem [{\citenamefont {Wang}\ \emph
  {et~al.}(2021{\natexlab{b}})\citenamefont {Wang}, \citenamefont {Ciccarino},
  \citenamefont {Flick},\ and\ \citenamefont {Narang}}]{Wang2020Hybridized}%
  \BibitemOpen
  \bibfield  {author} {\bibinfo {author} {\bibfnamefont {D.~S.}\ \bibnamefont
  {Wang}}, \bibinfo {author} {\bibfnamefont {C.~J.}\ \bibnamefont {Ciccarino}},
  \bibinfo {author} {\bibfnamefont {J.}~\bibnamefont {Flick}},\ and\ \bibinfo
  {author} {\bibfnamefont {P.}~\bibnamefont {Narang}},\ }\bibfield  {title}
  {\bibinfo {title} {Hybridized defects in solid-state materials as artificial
  molecules},\ }\href@noop {} {\bibfield  {journal} {\bibinfo  {journal} {ACS
  Nano}\ } (\bibinfo {year} {2021}{\natexlab{b}})}\BibitemShut {NoStop}%
\bibitem [{\citenamefont {Rogers}\ \emph {et~al.}(2008)\citenamefont {Rogers},
  \citenamefont {Armstrong}, \citenamefont {Sellars},\ and\ \citenamefont
  {Manson}}]{Rogers2008}%
  \BibitemOpen
  \bibfield  {author} {\bibinfo {author} {\bibfnamefont {L.~J.}\ \bibnamefont
  {Rogers}}, \bibinfo {author} {\bibfnamefont {S.}~\bibnamefont {Armstrong}},
  \bibinfo {author} {\bibfnamefont {M.~J.}\ \bibnamefont {Sellars}},\ and\
  \bibinfo {author} {\bibfnamefont {N.~B.}\ \bibnamefont {Manson}},\ }\bibfield
   {title} {\bibinfo {title} {{Infrared Emission of the NV Centre in Diamond:
  Zeeman and Uniaxial Stress Studies}},\ }\href
  {https://doi.org/10.1088/1367-2630/10/10/103024} {\bibfield  {journal}
  {\bibinfo  {journal} {New. J. Phys.}\ }\textbf {\bibinfo {volume} {10}},\
  \bibinfo {pages} {103024} (\bibinfo {year} {2008})}\BibitemShut {NoStop}%
\bibitem [{\citenamefont {Momenzadeh}\ \emph {et~al.}(2015)\citenamefont
  {Momenzadeh}, \citenamefont {St{\"{o}}hr}, \citenamefont {{De Oliveira}},
  \citenamefont {Brunner}, \citenamefont {Denisenko}, \citenamefont {Yang},
  \citenamefont {Reinhard},\ and\ \citenamefont {Wrachtrup}}]{Momenzadeh2015}%
  \BibitemOpen
  \bibfield  {author} {\bibinfo {author} {\bibfnamefont {S.~A.}\ \bibnamefont
  {Momenzadeh}}, \bibinfo {author} {\bibfnamefont {R.~J.}\ \bibnamefont
  {St{\"{o}}hr}}, \bibinfo {author} {\bibfnamefont {F.~F.}\ \bibnamefont {{De
  Oliveira}}}, \bibinfo {author} {\bibfnamefont {A.}~\bibnamefont {Brunner}},
  \bibinfo {author} {\bibfnamefont {A.}~\bibnamefont {Denisenko}}, \bibinfo
  {author} {\bibfnamefont {S.}~\bibnamefont {Yang}}, \bibinfo {author}
  {\bibfnamefont {F.}~\bibnamefont {Reinhard}},\ and\ \bibinfo {author}
  {\bibfnamefont {J.}~\bibnamefont {Wrachtrup}},\ }\bibfield  {title} {\bibinfo
  {title} {{Nanoengineered Diamond Waveguide as a Robust Bright Platform for
  Nanomagnetometry Using Shallow Nitrogen Vacancy Centers}},\ }\href
  {https://doi.org/10.1021/nl503326t} {\bibfield  {journal} {\bibinfo
  {journal} {Nano Lett.}\ }\textbf {\bibinfo {volume} {15}},\ \bibinfo {pages}
  {165} (\bibinfo {year} {2015})}\BibitemShut {NoStop}%
\bibitem [{\citenamefont {Faraon}\ \emph {et~al.}(2012)\citenamefont {Faraon},
  \citenamefont {Santori}, \citenamefont {Huang}, \citenamefont {Acosta},\ and\
  \citenamefont {Beausoleil}}]{Faraon2012}%
  \BibitemOpen
  \bibfield  {author} {\bibinfo {author} {\bibfnamefont {A.}~\bibnamefont
  {Faraon}}, \bibinfo {author} {\bibfnamefont {C.}~\bibnamefont {Santori}},
  \bibinfo {author} {\bibfnamefont {Z.}~\bibnamefont {Huang}}, \bibinfo
  {author} {\bibfnamefont {V.~M.}\ \bibnamefont {Acosta}},\ and\ \bibinfo
  {author} {\bibfnamefont {R.~G.}\ \bibnamefont {Beausoleil}},\ }\bibfield
  {title} {\bibinfo {title} {{Coupling of nitrogen-vacancy centers to photonic
  crystal cavities in monocrystalline diamond}},\ }\href
  {https://doi.org/10.1103/PhysRevLett.109.033604} {\bibfield  {journal}
  {\bibinfo  {journal} {Phys. Rev. Lett.}\ }\textbf {\bibinfo {volume} {109}},\
  \bibinfo {pages} {033604} (\bibinfo {year} {2012})}\BibitemShut {NoStop}%
\bibitem [{\citenamefont {Chakraborty}\ \emph {et~al.}(2019)\citenamefont
  {Chakraborty}, \citenamefont {Jungwirth}, \citenamefont {Fuchs},\ and\
  \citenamefont {Vamivakas}}]{Chakraborty2019}%
  \BibitemOpen
  \bibfield  {author} {\bibinfo {author} {\bibfnamefont {C.}~\bibnamefont
  {Chakraborty}}, \bibinfo {author} {\bibfnamefont {N.~R.}\ \bibnamefont
  {Jungwirth}}, \bibinfo {author} {\bibfnamefont {G.~D.}\ \bibnamefont
  {Fuchs}},\ and\ \bibinfo {author} {\bibfnamefont {A.~N.}\ \bibnamefont
  {Vamivakas}},\ }\bibfield  {title} {\bibinfo {title} {{Electrical
  Manipulation of the Fine-Structure Splitting of WSe2 Quantum Emitters}},\
  }\href {https://doi.org/10.1103/PhysRevB.99.045308} {\bibfield  {journal}
  {\bibinfo  {journal} {Phys. Rev. B}\ }\textbf {\bibinfo {volume} {99}},\
  \bibinfo {pages} {045308} (\bibinfo {year} {2019})}\BibitemShut {NoStop}%
\bibitem [{\citenamefont {Zhang}\ \emph {et~al.}(2018)\citenamefont {Zhang},
  \citenamefont {Sun}, \citenamefont {Burek}, \citenamefont {Dory},
  \citenamefont {Tzeng}, \citenamefont {Fischer}, \citenamefont {Kelaita},
  \citenamefont {Lagoudakis}, \citenamefont {Radulaski}, \citenamefont {Shen},
  \citenamefont {Melosh}, \citenamefont {Chu}, \citenamefont {Lon{\v{c}}ar},\
  and\ \citenamefont {Vu{\v{c}}kovi{\'{c}}}}]{Zhang2018}%
  \BibitemOpen
  \bibfield  {author} {\bibinfo {author} {\bibfnamefont {J.~L.}\ \bibnamefont
  {Zhang}}, \bibinfo {author} {\bibfnamefont {S.}~\bibnamefont {Sun}}, \bibinfo
  {author} {\bibfnamefont {M.~J.}\ \bibnamefont {Burek}}, \bibinfo {author}
  {\bibfnamefont {C.}~\bibnamefont {Dory}}, \bibinfo {author} {\bibfnamefont
  {Y.~K.}\ \bibnamefont {Tzeng}}, \bibinfo {author} {\bibfnamefont {K.~A.}\
  \bibnamefont {Fischer}}, \bibinfo {author} {\bibfnamefont {Y.}~\bibnamefont
  {Kelaita}}, \bibinfo {author} {\bibfnamefont {K.~G.}\ \bibnamefont
  {Lagoudakis}}, \bibinfo {author} {\bibfnamefont {M.}~\bibnamefont
  {Radulaski}}, \bibinfo {author} {\bibfnamefont {Z.~X.}\ \bibnamefont {Shen}},
  \bibinfo {author} {\bibfnamefont {N.~A.}\ \bibnamefont {Melosh}}, \bibinfo
  {author} {\bibfnamefont {S.}~\bibnamefont {Chu}}, \bibinfo {author}
  {\bibfnamefont {M.}~\bibnamefont {Lon{\v{c}}ar}},\ and\ \bibinfo {author}
  {\bibfnamefont {J.}~\bibnamefont {Vu{\v{c}}kovi{\'{c}}}},\ }\bibfield
  {title} {\bibinfo {title} {{Strongly Cavity-Enhanced Spontaneous Emission
  from Silicon-Vacancy Centers in Diamond}},\ }\href
  {https://doi.org/10.1021/acs.nanolett.7b05075} {\bibfield  {journal}
  {\bibinfo  {journal} {Nano Lett.}\ }\textbf {\bibinfo {volume} {18}},\
  \bibinfo {pages} {1360} (\bibinfo {year} {2018})}\BibitemShut {NoStop}%
\bibitem [{\citenamefont {Machielse}\ \emph {et~al.}(2019)\citenamefont
  {Machielse}, \citenamefont {Bogdanovic}, \citenamefont {Meesala},
  \citenamefont {Gauthier}, \citenamefont {Burek}, \citenamefont {Joe},
  \citenamefont {Chalupnik}, \citenamefont {Sohn}, \citenamefont {Holzgrafe},
  \citenamefont {Evans}, \citenamefont {Chia}, \citenamefont {Atikian},
  \citenamefont {Bhaskar}, \citenamefont {Sukachev}, \citenamefont {Shao},
  \citenamefont {Maity}, \citenamefont {Lukin},\ and\ \citenamefont
  {Lon{\v{c}}ar}}]{Machielse2019}%
  \BibitemOpen
  \bibfield  {author} {\bibinfo {author} {\bibfnamefont {B.}~\bibnamefont
  {Machielse}}, \bibinfo {author} {\bibfnamefont {S.}~\bibnamefont
  {Bogdanovic}}, \bibinfo {author} {\bibfnamefont {S.}~\bibnamefont {Meesala}},
  \bibinfo {author} {\bibfnamefont {S.}~\bibnamefont {Gauthier}}, \bibinfo
  {author} {\bibfnamefont {M.~J.}\ \bibnamefont {Burek}}, \bibinfo {author}
  {\bibfnamefont {G.}~\bibnamefont {Joe}}, \bibinfo {author} {\bibfnamefont
  {M.}~\bibnamefont {Chalupnik}}, \bibinfo {author} {\bibfnamefont {Y.~I.}\
  \bibnamefont {Sohn}}, \bibinfo {author} {\bibfnamefont {J.}~\bibnamefont
  {Holzgrafe}}, \bibinfo {author} {\bibfnamefont {R.~E.}\ \bibnamefont
  {Evans}}, \bibinfo {author} {\bibfnamefont {C.}~\bibnamefont {Chia}},
  \bibinfo {author} {\bibfnamefont {H.}~\bibnamefont {Atikian}}, \bibinfo
  {author} {\bibfnamefont {M.~K.}\ \bibnamefont {Bhaskar}}, \bibinfo {author}
  {\bibfnamefont {D.~D.}\ \bibnamefont {Sukachev}}, \bibinfo {author}
  {\bibfnamefont {L.}~\bibnamefont {Shao}}, \bibinfo {author} {\bibfnamefont
  {S.}~\bibnamefont {Maity}}, \bibinfo {author} {\bibfnamefont {M.~D.}\
  \bibnamefont {Lukin}},\ and\ \bibinfo {author} {\bibfnamefont
  {M.}~\bibnamefont {Lon{\v{c}}ar}},\ }\bibfield  {title} {\bibinfo {title}
  {{Quantum Interference of Electromechanically Stabilized Emitters in
  Nanophotonic Devices}},\ }\href {https://doi.org/10.1103/physrevx.9.031022}
  {\bibfield  {journal} {\bibinfo  {journal} {Phys. Rev. X}\ }\textbf {\bibinfo
  {volume} {9}},\ \bibinfo {pages} {031022} (\bibinfo {year}
  {2019})}\BibitemShut {NoStop}%
\bibitem [{\citenamefont {Neuman}\ \emph
  {et~al.}(2020{\natexlab{b}})\citenamefont {Neuman}, \citenamefont {Wang},\
  and\ \citenamefont {Narang}}]{Neuman2020Nanomagnonics}%
  \BibitemOpen
  \bibfield  {author} {\bibinfo {author} {\bibfnamefont {T.}~\bibnamefont
  {Neuman}}, \bibinfo {author} {\bibfnamefont {D.~S.}\ \bibnamefont {Wang}},\
  and\ \bibinfo {author} {\bibfnamefont {P.}~\bibnamefont {Narang}},\
  }\bibfield  {title} {\bibinfo {title} {Nanomagnonic cavities for strong
  spin-magnon coupling},\ }\href
  {https://journals.aps.org/prl/abstract/10.1103/PhysRevLett.125.247702}
  {\bibfield  {journal} {\bibinfo  {journal} {Phys. Rev. Lett.}\ }\textbf
  {\bibinfo {volume} {125}},\ \bibinfo {pages} {247702} (\bibinfo {year}
  {2020}{\natexlab{b}})}\BibitemShut {NoStop}%
\bibitem [{\citenamefont {Wang}\ \emph
  {et~al.}(2020{\natexlab{b}})\citenamefont {Wang}, \citenamefont {Neuman},\
  and\ \citenamefont {Narang}}]{Wang2020Entangled}%
  \BibitemOpen
  \bibfield  {author} {\bibinfo {author} {\bibfnamefont {D.~S.}\ \bibnamefont
  {Wang}}, \bibinfo {author} {\bibfnamefont {T.}~\bibnamefont {Neuman}},\ and\
  \bibinfo {author} {\bibfnamefont {P.}~\bibnamefont {Narang}},\ }\bibfield
  {title} {\bibinfo {title} {{Dipole-coupled emitters as deterministic
  entangled photon-pair sources}},\ }\href
  {https://journals.aps.org/prresearch/abstract/10.1103/PhysRevResearch.2.043328}
  {\bibfield  {journal} {\bibinfo  {journal} {Phys. Rev. Res.}\ }\textbf
  {\bibinfo {volume} {2}},\ \bibinfo {pages} {043328} (\bibinfo {year}
  {2020}{\natexlab{b}})}\BibitemShut {NoStop}%
\bibitem [{\citenamefont {Trivedi}\ \emph {et~al.}(2020)\citenamefont
  {Trivedi}, \citenamefont {Fischer}, \citenamefont {Vu{\v{c}}kovi{\'{c}}},\
  and\ \citenamefont {M{\"{u}}ller}}]{Trivedi2020}%
  \BibitemOpen
  \bibfield  {author} {\bibinfo {author} {\bibfnamefont {R.}~\bibnamefont
  {Trivedi}}, \bibinfo {author} {\bibfnamefont {K.~A.}\ \bibnamefont
  {Fischer}}, \bibinfo {author} {\bibfnamefont {J.}~\bibnamefont
  {Vu{\v{c}}kovi{\'{c}}}},\ and\ \bibinfo {author} {\bibfnamefont
  {K.}~\bibnamefont {M{\"{u}}ller}},\ }\bibfield  {title} {\bibinfo {title}
  {{Generation of Non‐Classical Light Using Semiconductor Quantum Dots}},\
  }\href {https://doi.org/10.1002/qute.201900007} {\bibfield  {journal}
  {\bibinfo  {journal} {Adv. Quantum Technol.}\ }\textbf {\bibinfo {volume}
  {3}},\ \bibinfo {pages} {1900007} (\bibinfo {year} {2020})}\BibitemShut
  {NoStop}%
\bibitem [{\citenamefont {Dai}\ \emph {et~al.}(2020)\citenamefont {Dai},
  \citenamefont {Wang},\ and\ \citenamefont {Narang}}]{Dai2020}%
  \BibitemOpen
  \bibfield  {author} {\bibinfo {author} {\bibfnamefont {D.~D.}\ \bibnamefont
  {Dai}}, \bibinfo {author} {\bibfnamefont {D.~S.}\ \bibnamefont {Wang}},\ and\
  \bibinfo {author} {\bibfnamefont {P.}~\bibnamefont {Narang}},\ }\bibfield
  {title} {\bibinfo {title} {Passive controlled-variable phase gate on photonic
  qubits via cascade emitter},\ }\href@noop {} {\bibfield  {journal} {\bibinfo
  {journal} {arXiv:2011.09302}\ } (\bibinfo {year} {2020})}\BibitemShut
  {NoStop}%
\bibitem [{\citenamefont {Purcell}\ \emph {et~al.}(1946)\citenamefont
  {Purcell}, \citenamefont {Torrey},\ and\ \citenamefont
  {Pound}}]{purcell1946spontaneous}%
  \BibitemOpen
  \bibfield  {author} {\bibinfo {author} {\bibfnamefont {E.~M.}\ \bibnamefont
  {Purcell}}, \bibinfo {author} {\bibfnamefont {H.~C.}\ \bibnamefont
  {Torrey}},\ and\ \bibinfo {author} {\bibfnamefont {R.~V.}\ \bibnamefont
  {Pound}},\ }\bibfield  {title} {\bibinfo {title} {Resonance absorption by
  nuclear magnetic moments in a solid},\ }\href
  {https://doi.org/10.1103/PhysRev.69.37} {\bibfield  {journal} {\bibinfo
  {journal} {Phys. Rev.}\ }\textbf {\bibinfo {volume} {69}},\ \bibinfo {pages}
  {37} (\bibinfo {year} {1946})}\BibitemShut {NoStop}%
\bibitem [{\citenamefont {\ifmmode~\acute{C}\else \'{C}\fi{}wik}\ \emph
  {et~al.}(2016)\citenamefont {\ifmmode~\acute{C}\else \'{C}\fi{}wik},
  \citenamefont {Kirton}, \citenamefont {De~Liberato},\ and\ \citenamefont
  {Keeling}}]{cwik2016excitonic}%
  \BibitemOpen
  \bibfield  {author} {\bibinfo {author} {\bibfnamefont {J.~A.}\ \bibnamefont
  {\ifmmode~\acute{C}\else \'{C}\fi{}wik}}, \bibinfo {author} {\bibfnamefont
  {P.}~\bibnamefont {Kirton}}, \bibinfo {author} {\bibfnamefont
  {S.}~\bibnamefont {De~Liberato}},\ and\ \bibinfo {author} {\bibfnamefont
  {J.}~\bibnamefont {Keeling}},\ }\bibfield  {title} {\bibinfo {title}
  {Excitonic spectral features in strongly coupled organic polaritons},\ }\href
  {https://doi.org/10.1103/PhysRevA.93.033840} {\bibfield  {journal} {\bibinfo
  {journal} {Phys. Rev. A}\ }\textbf {\bibinfo {volume} {93}},\ \bibinfo
  {pages} {033840} (\bibinfo {year} {2016})}\BibitemShut {NoStop}%
\bibitem [{\citenamefont {Ebbesen}(2016)}]{ebbesen2016strongcoupling}%
  \BibitemOpen
  \bibfield  {author} {\bibinfo {author} {\bibfnamefont {T.~W.}\ \bibnamefont
  {Ebbesen}},\ }\bibfield  {title} {\bibinfo {title} {Hybrid light–matter
  states in a molecular and material science perspective},\ }\href
  {https://doi.org/10.1021/acs.accounts.6b00295} {\bibfield  {journal}
  {\bibinfo  {journal} {Accounts of Chemical Research}\ }\textbf {\bibinfo
  {volume} {49}},\ \bibinfo {pages} {2403} (\bibinfo {year} {2016})},\ \bibinfo
  {note} {pMID: 27779846}\BibitemShut {NoStop}%
\bibitem [{\citenamefont {Flick}\ \emph {et~al.}(2017)\citenamefont {Flick},
  \citenamefont {Ruggenthaler}, \citenamefont {Appel},\ and\ \citenamefont
  {Rubio}}]{flick2017}%
  \BibitemOpen
  \bibfield  {author} {\bibinfo {author} {\bibfnamefont {J.}~\bibnamefont
  {Flick}}, \bibinfo {author} {\bibfnamefont {M.}~\bibnamefont {Ruggenthaler}},
  \bibinfo {author} {\bibfnamefont {H.}~\bibnamefont {Appel}},\ and\ \bibinfo
  {author} {\bibfnamefont {A.}~\bibnamefont {Rubio}},\ }\bibfield  {title}
  {\bibinfo {title} {Atoms and molecules in cavities, from weak to strong
  coupling in quantum-electrodynamics (qed) chemistry},\ }\href
  {https://doi.org/10.1073/pnas.1615509114} {\bibfield  {journal} {\bibinfo
  {journal} {Proc. Nat. Acad. Sci. USA}\ }\textbf {\bibinfo {volume} {114}},\
  \bibinfo {pages} {3026} (\bibinfo {year} {2017})}\BibitemShut {NoStop}%
\bibitem [{\citenamefont {Herrera}\ and\ \citenamefont
  {Owrutsky}(2020)}]{Herrera2020}%
  \BibitemOpen
  \bibfield  {author} {\bibinfo {author} {\bibfnamefont {F.}~\bibnamefont
  {Herrera}}\ and\ \bibinfo {author} {\bibfnamefont {J.}~\bibnamefont
  {Owrutsky}},\ }\bibfield  {title} {\bibinfo {title} {{Molecular polaritons
  for controlling chemistry with quantum optics}},\ }\href
  {https://doi.org/10.1063/1.5136320} {\bibfield  {journal} {\bibinfo
  {journal} {J. Chem. Phys.}\ }\textbf {\bibinfo {volume} {152}},\ \bibinfo
  {pages} {100902} (\bibinfo {year} {2020})}\BibitemShut {NoStop}%
\bibitem [{\citenamefont {Rivera}\ \emph {et~al.}(2019)\citenamefont {Rivera},
  \citenamefont {Flick},\ and\ \citenamefont {Narang}}]{rivera2018}%
  \BibitemOpen
  \bibfield  {author} {\bibinfo {author} {\bibfnamefont {N.}~\bibnamefont
  {Rivera}}, \bibinfo {author} {\bibfnamefont {J.}~\bibnamefont {Flick}},\ and\
  \bibinfo {author} {\bibfnamefont {P.}~\bibnamefont {Narang}},\ }\bibfield
  {title} {\bibinfo {title} {Variational theory of nonrelativistic quantum
  electrodynamics},\ }\href {https://doi.org/10.1103/PhysRevLett.122.193603}
  {\bibfield  {journal} {\bibinfo  {journal} {Phys. Rev. Lett.}\ }\textbf
  {\bibinfo {volume} {122}},\ \bibinfo {pages} {193603} (\bibinfo {year}
  {2019})}\BibitemShut {NoStop}%
\bibitem [{\citenamefont {Flick}\ and\ \citenamefont
  {Narang}(2018)}]{flick2018b}%
  \BibitemOpen
  \bibfield  {author} {\bibinfo {author} {\bibfnamefont {J.}~\bibnamefont
  {Flick}}\ and\ \bibinfo {author} {\bibfnamefont {P.}~\bibnamefont {Narang}},\
  }\bibfield  {title} {\bibinfo {title} {Cavity-correlated electron-nuclear
  dynamics from first principles},\ }\href
  {https://doi.org/10.1103/PhysRevLett.121.113002} {\bibfield  {journal}
  {\bibinfo  {journal} {Phys. Rev. Lett.}\ }\textbf {\bibinfo {volume} {121}},\
  \bibinfo {pages} {113002} (\bibinfo {year} {2018})}\BibitemShut {NoStop}%
\bibitem [{\citenamefont {Flick}\ \emph {et~al.}(2018)\citenamefont {Flick},
  \citenamefont {Rivera},\ and\ \citenamefont {Narang}}]{flick2018strong}%
  \BibitemOpen
  \bibfield  {author} {\bibinfo {author} {\bibfnamefont {J.}~\bibnamefont
  {Flick}}, \bibinfo {author} {\bibfnamefont {N.}~\bibnamefont {Rivera}},\ and\
  \bibinfo {author} {\bibfnamefont {P.}~\bibnamefont {Narang}},\ }\bibfield
  {title} {\bibinfo {title} {Strong light-matter coupling in quantum chemistry
  and quantum photonics},\ }\href@noop {} {\bibfield  {journal} {\bibinfo
  {journal} {Nanophotonics}\ }\textbf {\bibinfo {volume} {7}},\ \bibinfo
  {pages} {1479} (\bibinfo {year} {2018})}\BibitemShut {NoStop}%
\bibitem [{\citenamefont {Flick}\ and\ \citenamefont
  {Narang}(2020)}]{flickexcited}%
  \BibitemOpen
  \bibfield  {author} {\bibinfo {author} {\bibfnamefont {J.}~\bibnamefont
  {Flick}}\ and\ \bibinfo {author} {\bibfnamefont {P.}~\bibnamefont {Narang}},\
  }\bibfield  {title} {\bibinfo {title} {ab initio polaritonic potential-energy
  surfaces for excited-state nanophotonics and polaritonic chemistry},\ }\href
  {https://aip.scitation.org/doi/10.1063/5.0021033} {\bibfield  {journal}
  {\bibinfo  {journal} {J. Chem. Phys.}\ }\textbf {\bibinfo {volume} {153}},\
  \bibinfo {pages} {094116} (\bibinfo {year} {2020})}\BibitemShut {NoStop}%
\bibitem [{\citenamefont {Galego}\ \emph {et~al.}(2015)\citenamefont {Galego},
  \citenamefont {Garcia-Vidal},\ and\ \citenamefont {Feist}}]{galego2015}%
  \BibitemOpen
  \bibfield  {author} {\bibinfo {author} {\bibfnamefont {J.}~\bibnamefont
  {Galego}}, \bibinfo {author} {\bibfnamefont {F.~J.}\ \bibnamefont
  {Garcia-Vidal}},\ and\ \bibinfo {author} {\bibfnamefont {J.}~\bibnamefont
  {Feist}},\ }\bibfield  {title} {\bibinfo {title} {Cavity-induced
  modifications of molecular structure in the strong-coupling regime},\ }\href
  {https://doi.org/10.1103/PhysRevX.5.041022} {\bibfield  {journal} {\bibinfo
  {journal} {Phys. Rev. X}\ }\textbf {\bibinfo {volume} {5}},\ \bibinfo {pages}
  {041022} (\bibinfo {year} {2015})}\BibitemShut {NoStop}%
\bibitem [{\citenamefont {Galego}\ \emph {et~al.}(2016)\citenamefont {Galego},
  \citenamefont {Garcia-Vidal},\ and\ \citenamefont {Feist}}]{galego2016}%
  \BibitemOpen
  \bibfield  {author} {\bibinfo {author} {\bibfnamefont {J.}~\bibnamefont
  {Galego}}, \bibinfo {author} {\bibfnamefont {F.~J.}\ \bibnamefont
  {Garcia-Vidal}},\ and\ \bibinfo {author} {\bibfnamefont {J.}~\bibnamefont
  {Feist}},\ }\bibfield  {title} {\bibinfo {title} {Suppressing photochemical
  reactions with quantized light fields},\ }\bibfield  {journal} {\bibinfo
  {journal} {Nat. Commun.}\ }\textbf {\bibinfo {volume} {7}},\ \href
  {https://doi.org/10.1038/ncomms13841} {10.1038/ncomms13841} (\bibinfo {year}
  {2016})\BibitemShut {NoStop}%
\bibitem [{\citenamefont {Thomas}\ \emph {et~al.}(2016)\citenamefont {Thomas},
  \citenamefont {George}, \citenamefont {Shalabney}, \citenamefont {Dryzhakov},
  \citenamefont {Varma}, \citenamefont {Moran}, \citenamefont {Chervy},
  \citenamefont {Zhong}, \citenamefont {Devaux}, \citenamefont {Genet},
  \citenamefont {Hutchison},\ and\ \citenamefont
  {Ebbesen}}]{anoop2016vibreactivity}%
  \BibitemOpen
  \bibfield  {author} {\bibinfo {author} {\bibfnamefont {A.}~\bibnamefont
  {Thomas}}, \bibinfo {author} {\bibfnamefont {J.}~\bibnamefont {George}},
  \bibinfo {author} {\bibfnamefont {A.}~\bibnamefont {Shalabney}}, \bibinfo
  {author} {\bibfnamefont {M.}~\bibnamefont {Dryzhakov}}, \bibinfo {author}
  {\bibfnamefont {S.~J.}\ \bibnamefont {Varma}}, \bibinfo {author}
  {\bibfnamefont {J.}~\bibnamefont {Moran}}, \bibinfo {author} {\bibfnamefont
  {T.}~\bibnamefont {Chervy}}, \bibinfo {author} {\bibfnamefont
  {X.}~\bibnamefont {Zhong}}, \bibinfo {author} {\bibfnamefont
  {E.}~\bibnamefont {Devaux}}, \bibinfo {author} {\bibfnamefont
  {C.}~\bibnamefont {Genet}}, \bibinfo {author} {\bibfnamefont {J.~A.}\
  \bibnamefont {Hutchison}},\ and\ \bibinfo {author} {\bibfnamefont {T.~W.}\
  \bibnamefont {Ebbesen}},\ }\bibfield  {title} {\bibinfo {title} {Ground-state
  chemical reactivity under vibrational coupling to the vacuum electromagnetic
  field},\ }\href {https://doi.org/10.1002/anie.201605504} {\bibfield
  {journal} {\bibinfo  {journal} {Angew. Chem. Int. Ed.}\ }\textbf {\bibinfo
  {volume} {55}},\ \bibinfo {pages} {11462} (\bibinfo {year}
  {2016})}\BibitemShut {NoStop}%
\bibitem [{\citenamefont {Herrera}\ and\ \citenamefont
  {Spano}(2016)}]{herrera2016chemistry}%
  \BibitemOpen
  \bibfield  {author} {\bibinfo {author} {\bibfnamefont {F.}~\bibnamefont
  {Herrera}}\ and\ \bibinfo {author} {\bibfnamefont {F.~C.}\ \bibnamefont
  {Spano}},\ }\bibfield  {title} {\bibinfo {title} {Cavity-controlled chemistry
  in molecular ensembles},\ }\href
  {https://doi.org/10.1103/PhysRevLett.116.238301} {\bibfield  {journal}
  {\bibinfo  {journal} {Phys. Rev. Lett.}\ }\textbf {\bibinfo {volume} {116}},\
  \bibinfo {pages} {238301} (\bibinfo {year} {2016})}\BibitemShut {NoStop}%
\bibitem [{\citenamefont {Galego}\ \emph {et~al.}(2019)\citenamefont {Galego},
  \citenamefont {Climent}, \citenamefont {Garcia-Vidal},\ and\ \citenamefont
  {Feist}}]{galego2019}%
  \BibitemOpen
  \bibfield  {author} {\bibinfo {author} {\bibfnamefont {J.}~\bibnamefont
  {Galego}}, \bibinfo {author} {\bibfnamefont {C.}~\bibnamefont {Climent}},
  \bibinfo {author} {\bibfnamefont {F.~J.}\ \bibnamefont {Garcia-Vidal}},\ and\
  \bibinfo {author} {\bibfnamefont {J.}~\bibnamefont {Feist}},\ }\bibfield
  {title} {\bibinfo {title} {Cavity casimir-polder forces and their effects in
  ground-state chemical reactivity},\ }\href
  {https://doi.org/10.1103/PhysRevX.9.021057} {\bibfield  {journal} {\bibinfo
  {journal} {Phys. Rev. X}\ }\textbf {\bibinfo {volume} {9}},\ \bibinfo {pages}
  {021057} (\bibinfo {year} {2019})}\BibitemShut {NoStop}%
\bibitem [{\citenamefont {Groenhof}\ \emph {et~al.}(2019)\citenamefont
  {Groenhof}, \citenamefont {Climent}, \citenamefont {Feist}, \citenamefont
  {Morozov},\ and\ \citenamefont {Toppari}}]{groenhof2019relaxation}%
  \BibitemOpen
  \bibfield  {author} {\bibinfo {author} {\bibfnamefont {G.}~\bibnamefont
  {Groenhof}}, \bibinfo {author} {\bibfnamefont {C.}~\bibnamefont {Climent}},
  \bibinfo {author} {\bibfnamefont {J.}~\bibnamefont {Feist}}, \bibinfo
  {author} {\bibfnamefont {D.}~\bibnamefont {Morozov}},\ and\ \bibinfo {author}
  {\bibfnamefont {J.~J.}\ \bibnamefont {Toppari}},\ }\bibfield  {title}
  {\bibinfo {title} {Tracking polariton relaxation with multiscale molecular
  dynamics simulations},\ }\href {https://doi.org/10.1021/acs.jpclett.9b02192}
  {\bibfield  {journal} {\bibinfo  {journal} {J. Phys. Chem. Lett.}\ }\textbf
  {\bibinfo {volume} {10}},\ \bibinfo {pages} {5476} (\bibinfo {year}
  {2019})},\ \bibinfo {note} {pMID: 31453696}\BibitemShut {NoStop}%
\bibitem [{\citenamefont {Thomas}\ \emph {et~al.}(2019)\citenamefont {Thomas},
  \citenamefont {Lethuillier-Karl}, \citenamefont {Nagarajan}, \citenamefont
  {Vergauwe}, \citenamefont {J.~George}, \citenamefont {Shalabney},
  \citenamefont {Devaux}, \citenamefont {Genet}, \citenamefont {Moran},\ and\
  \citenamefont {Ebbesen}}]{ebbesenTilting}%
  \BibitemOpen
  \bibfield  {author} {\bibinfo {author} {\bibfnamefont {A.}~\bibnamefont
  {Thomas}}, \bibinfo {author} {\bibfnamefont {L.}~\bibnamefont
  {Lethuillier-Karl}}, \bibinfo {author} {\bibfnamefont {K.}~\bibnamefont
  {Nagarajan}}, \bibinfo {author} {\bibfnamefont {R.~M.~A.}\ \bibnamefont
  {Vergauwe}}, \bibinfo {author} {\bibfnamefont {T.~C.}\ \bibnamefont
  {J.~George}}, \bibinfo {author} {\bibfnamefont {A.}~\bibnamefont
  {Shalabney}}, \bibinfo {author} {\bibfnamefont {E.}~\bibnamefont {Devaux}},
  \bibinfo {author} {\bibfnamefont {C.}~\bibnamefont {Genet}}, \bibinfo
  {author} {\bibfnamefont {J.}~\bibnamefont {Moran}},\ and\ \bibinfo {author}
  {\bibfnamefont {T.~W.}\ \bibnamefont {Ebbesen}},\ }\bibfield  {title}
  {\bibinfo {title} {Tilting a ground-state reactivity landscape by vibrational
  strong coupling},\ }\href
  {https://science.sciencemag.org/content/363/6427/615.abstract} {\bibfield
  {journal} {\bibinfo  {journal} {Science}\ }\textbf {\bibinfo {volume}
  {364}},\ \bibinfo {pages} {615} (\bibinfo {year} {2019})}\BibitemShut
  {NoStop}%
\bibitem [{\citenamefont {Lidzey}\ \emph {et~al.}(1999)\citenamefont {Lidzey},
  \citenamefont {Bradley}, \citenamefont {Virgili}, \citenamefont {Armitage},
  \citenamefont {Skolnick},\ and\ \citenamefont
  {Walker}}]{lidzey1999polemission}%
  \BibitemOpen
  \bibfield  {author} {\bibinfo {author} {\bibfnamefont {D.~G.}\ \bibnamefont
  {Lidzey}}, \bibinfo {author} {\bibfnamefont {D.~D.~C.}\ \bibnamefont
  {Bradley}}, \bibinfo {author} {\bibfnamefont {T.}~\bibnamefont {Virgili}},
  \bibinfo {author} {\bibfnamefont {A.}~\bibnamefont {Armitage}}, \bibinfo
  {author} {\bibfnamefont {M.~S.}\ \bibnamefont {Skolnick}},\ and\ \bibinfo
  {author} {\bibfnamefont {S.}~\bibnamefont {Walker}},\ }\bibfield  {title}
  {\bibinfo {title} {Room temperature polariton emission from strongly coupled
  organic semiconductor microcavities},\ }\href
  {https://doi.org/10.1103/PhysRevLett.82.3316} {\bibfield  {journal} {\bibinfo
   {journal} {Phys. Rev. Lett.}\ }\textbf {\bibinfo {volume} {82}},\ \bibinfo
  {pages} {3316} (\bibinfo {year} {1999})}\BibitemShut {NoStop}%
\bibitem [{\citenamefont {del Pino}\ \emph {et~al.}(2015)\citenamefont {del
  Pino}, \citenamefont {Feist},\ and\ \citenamefont
  {Garcia-Vidal}}]{delpino2015quantum}%
  \BibitemOpen
  \bibfield  {author} {\bibinfo {author} {\bibfnamefont {J.}~\bibnamefont {del
  Pino}}, \bibinfo {author} {\bibfnamefont {J.}~\bibnamefont {Feist}},\ and\
  \bibinfo {author} {\bibfnamefont {F.~J.}\ \bibnamefont {Garcia-Vidal}},\
  }\bibfield  {title} {\bibinfo {title} {Quantum theory of collective strong
  coupling of molecular vibrations with a microcavity mode},\ }\href
  {https://doi.org/10.1088/1367-2630/17/5/053040} {\bibfield  {journal}
  {\bibinfo  {journal} {New J. Phys.}\ }\textbf {\bibinfo {volume} {17}},\
  \bibinfo {pages} {053040} (\bibinfo {year} {2015})}\BibitemShut {NoStop}%
\bibitem [{\citenamefont {George}\ \emph {et~al.}(2015)\citenamefont {George},
  \citenamefont {Wang}, \citenamefont {Chervy}, \citenamefont
  {Canaguier-Durand}, \citenamefont {Schaeffer}, \citenamefont {Lehn},
  \citenamefont {Hutchison}, \citenamefont {Genet},\ and\ \citenamefont
  {Ebbesen}}]{george2015ultra}%
  \BibitemOpen
  \bibfield  {author} {\bibinfo {author} {\bibfnamefont {J.}~\bibnamefont
  {George}}, \bibinfo {author} {\bibfnamefont {S.}~\bibnamefont {Wang}},
  \bibinfo {author} {\bibfnamefont {T.}~\bibnamefont {Chervy}}, \bibinfo
  {author} {\bibfnamefont {A.}~\bibnamefont {Canaguier-Durand}}, \bibinfo
  {author} {\bibfnamefont {G.}~\bibnamefont {Schaeffer}}, \bibinfo {author}
  {\bibfnamefont {J.-M.}\ \bibnamefont {Lehn}}, \bibinfo {author}
  {\bibfnamefont {J.~A.}\ \bibnamefont {Hutchison}}, \bibinfo {author}
  {\bibfnamefont {C.}~\bibnamefont {Genet}},\ and\ \bibinfo {author}
  {\bibfnamefont {T.~W.}\ \bibnamefont {Ebbesen}},\ }\bibfield  {title}
  {\bibinfo {title} {Ultra-strong coupling of molecular materials: spectroscopy
  and dynamics},\ }\href {https://doi.org/10.1039/C4FD00197D} {\bibfield
  {journal} {\bibinfo  {journal} {Faraday Discuss.}\ }\textbf {\bibinfo
  {volume} {178}},\ \bibinfo {pages} {281} (\bibinfo {year}
  {2015})}\BibitemShut {NoStop}%
\bibitem [{\citenamefont {Herrera}\ and\ \citenamefont
  {Spano}(2018)}]{herrera2018review}%
  \BibitemOpen
  \bibfield  {author} {\bibinfo {author} {\bibfnamefont {F.}~\bibnamefont
  {Herrera}}\ and\ \bibinfo {author} {\bibfnamefont {F.~C.}\ \bibnamefont
  {Spano}},\ }\bibfield  {title} {\bibinfo {title} {Theory of nanoscale organic
  cavities: The essential role of vibration-photon dressed states},\ }\href
  {https://doi.org/10.1021/acsphotonics.7b00728} {\bibfield  {journal}
  {\bibinfo  {journal} {ACS Photonics}\ }\textbf {\bibinfo {volume} {5}},\
  \bibinfo {pages} {65} (\bibinfo {year} {2018})}\BibitemShut {NoStop}%
\bibitem [{\citenamefont {Zeb}\ \emph {et~al.}(2018)\citenamefont {Zeb},
  \citenamefont {Kirton},\ and\ \citenamefont
  {Keeling}}]{zeb2018exactvibdressed}%
  \BibitemOpen
  \bibfield  {author} {\bibinfo {author} {\bibfnamefont {M.~A.}\ \bibnamefont
  {Zeb}}, \bibinfo {author} {\bibfnamefont {P.~G.}\ \bibnamefont {Kirton}},\
  and\ \bibinfo {author} {\bibfnamefont {J.}~\bibnamefont {Keeling}},\
  }\bibfield  {title} {\bibinfo {title} {Exact states and spectra of
  vibrationally dressed polaritons},\ }\href
  {https://doi.org/10.1021/acsphotonics.7b00916} {\bibfield  {journal}
  {\bibinfo  {journal} {ACS Photonics}\ }\textbf {\bibinfo {volume} {5}},\
  \bibinfo {pages} {249} (\bibinfo {year} {2018})}\BibitemShut {NoStop}%
\bibitem [{\citenamefont {Herrera}\ and\ \citenamefont
  {Spano}(2017)}]{herrera2017dark}%
  \BibitemOpen
  \bibfield  {author} {\bibinfo {author} {\bibfnamefont {F.}~\bibnamefont
  {Herrera}}\ and\ \bibinfo {author} {\bibfnamefont {F.~C.}\ \bibnamefont
  {Spano}},\ }\bibfield  {title} {\bibinfo {title} {Dark vibronic polaritons
  and the spectroscopy of organic microcavities},\ }\href
  {https://doi.org/10.1103/PhysRevLett.118.223601} {\bibfield  {journal}
  {\bibinfo  {journal} {Phys. Rev. Lett.}\ }\textbf {\bibinfo {volume} {118}},\
  \bibinfo {pages} {223601} (\bibinfo {year} {2017})}\BibitemShut {NoStop}%
\bibitem [{\citenamefont {Xiang}\ \emph {et~al.}(2018)\citenamefont {Xiang},
  \citenamefont {Ribeiro}, \citenamefont {Dunkelberger}, \citenamefont {Wang},
  \citenamefont {Li}, \citenamefont {Blake S.~Simpkins}, \citenamefont
  {Yuen-Zhou},\ and\ \citenamefont {Xiong}}]{Owrutzsky2DIR}%
  \BibitemOpen
  \bibfield  {author} {\bibinfo {author} {\bibfnamefont {B.}~\bibnamefont
  {Xiang}}, \bibinfo {author} {\bibfnamefont {R.~F.}\ \bibnamefont {Ribeiro}},
  \bibinfo {author} {\bibfnamefont {A.~D.}\ \bibnamefont {Dunkelberger}},
  \bibinfo {author} {\bibfnamefont {J.}~\bibnamefont {Wang}}, \bibinfo {author}
  {\bibfnamefont {Y.}~\bibnamefont {Li}}, \bibinfo {author} {\bibfnamefont
  {J.~C.~O.}\ \bibnamefont {Blake S.~Simpkins}}, \bibinfo {author}
  {\bibfnamefont {J.}~\bibnamefont {Yuen-Zhou}},\ and\ \bibinfo {author}
  {\bibfnamefont {W.}~\bibnamefont {Xiong}},\ }\bibfield  {title} {\bibinfo
  {title} {Two-dimensional infrared spectroscopy of vibrational polaritons},\
  }\href {https://www.pnas.org/content/115/19/4845} {\bibfield  {journal}
  {\bibinfo  {journal} {PNAS}\ }\textbf {\bibinfo {volume} {115}},\ \bibinfo
  {pages} {4845} (\bibinfo {year} {2018})}\BibitemShut {NoStop}%
\bibitem [{\citenamefont {Coles}\ \emph {et~al.}(2014)\citenamefont {Coles},
  \citenamefont {Somaschi}, \citenamefont {Michetti}, \citenamefont {Clark},
  \citenamefont {Lagoudakis}, \citenamefont {Savvidis},\ and\ \citenamefont
  {Lidzey}}]{coles2014polariton}%
  \BibitemOpen
  \bibfield  {author} {\bibinfo {author} {\bibfnamefont {D.~M.}\ \bibnamefont
  {Coles}}, \bibinfo {author} {\bibfnamefont {N.}~\bibnamefont {Somaschi}},
  \bibinfo {author} {\bibfnamefont {P.}~\bibnamefont {Michetti}}, \bibinfo
  {author} {\bibfnamefont {C.}~\bibnamefont {Clark}}, \bibinfo {author}
  {\bibfnamefont {P.~G.}\ \bibnamefont {Lagoudakis}}, \bibinfo {author}
  {\bibfnamefont {P.~G.}\ \bibnamefont {Savvidis}},\ and\ \bibinfo {author}
  {\bibfnamefont {D.~G.}\ \bibnamefont {Lidzey}},\ }\bibfield  {title}
  {\bibinfo {title} {Polariton-mediated energy transfer between organic dyes in
  a strongly coupled optical microcavity},\ }\href
  {https://doi.org/10.1038/nmat3950} {\bibfield  {journal} {\bibinfo  {journal}
  {Nat. Mater.}\ }\textbf {\bibinfo {volume} {13}},\ \bibinfo {pages} {712}
  (\bibinfo {year} {2014})}\BibitemShut {NoStop}%
\bibitem [{\citenamefont {Zhong}\ \emph {et~al.}(2017)\citenamefont {Zhong},
  \citenamefont {Chervy}, \citenamefont {Zhang}, \citenamefont {Thomas},
  \citenamefont {George}, \citenamefont {Genet}, \citenamefont {Hutchison},\
  and\ \citenamefont {Ebbesen}}]{zhong2017entangled}%
  \BibitemOpen
  \bibfield  {author} {\bibinfo {author} {\bibfnamefont {X.}~\bibnamefont
  {Zhong}}, \bibinfo {author} {\bibfnamefont {T.}~\bibnamefont {Chervy}},
  \bibinfo {author} {\bibfnamefont {L.}~\bibnamefont {Zhang}}, \bibinfo
  {author} {\bibfnamefont {A.}~\bibnamefont {Thomas}}, \bibinfo {author}
  {\bibfnamefont {J.}~\bibnamefont {George}}, \bibinfo {author} {\bibfnamefont
  {C.}~\bibnamefont {Genet}}, \bibinfo {author} {\bibfnamefont {J.~A.}\
  \bibnamefont {Hutchison}},\ and\ \bibinfo {author} {\bibfnamefont {T.~W.}\
  \bibnamefont {Ebbesen}},\ }\bibfield  {title} {\bibinfo {title} {Energy
  transfer between spatially separated entangled molecules},\ }\href
  {https://doi.org/10.1002/anie.201703539} {\bibfield  {journal} {\bibinfo
  {journal} {Angew. Chem. Int. Ed.}\ }\textbf {\bibinfo {volume} {56}},\
  \bibinfo {pages} {9034} (\bibinfo {year} {2017})}\BibitemShut {NoStop}%
\bibitem [{\citenamefont {Juraschek}\ \emph {et~al.}(2019)\citenamefont
  {Juraschek}, \citenamefont {Neuman}, \citenamefont {Flick},\ and\
  \citenamefont {Narang}}]{juraschek2019cavity}%
  \BibitemOpen
  \bibfield  {author} {\bibinfo {author} {\bibfnamefont {D.~M.}\ \bibnamefont
  {Juraschek}}, \bibinfo {author} {\bibfnamefont {T.}~\bibnamefont {Neuman}},
  \bibinfo {author} {\bibfnamefont {J.}~\bibnamefont {Flick}},\ and\ \bibinfo
  {author} {\bibfnamefont {P.}~\bibnamefont {Narang}},\ }\bibfield  {title}
  {\bibinfo {title} {Cavity control of nonlinear phononics},\ }\href@noop {}
  {\bibfield  {journal} {\bibinfo  {journal} {arXiv preprint arXiv:1912.00122}\
  } (\bibinfo {year} {2019})}\BibitemShut {NoStop}%
\bibitem [{\citenamefont {Du}\ \emph {et~al.}(2018)\citenamefont {Du},
  \citenamefont {Martinez-Martinez}, \citenamefont {Ribeiro}, \citenamefont
  {Hu}, \citenamefont {Menon},\ and\ \citenamefont
  {Yuen-Zhou}}]{DuEnergyTransfer}%
  \BibitemOpen
  \bibfield  {author} {\bibinfo {author} {\bibfnamefont {M.}~\bibnamefont
  {Du}}, \bibinfo {author} {\bibfnamefont {L.~A.}\ \bibnamefont
  {Martinez-Martinez}}, \bibinfo {author} {\bibfnamefont {R.~F.}\ \bibnamefont
  {Ribeiro}}, \bibinfo {author} {\bibfnamefont {Z.}~\bibnamefont {Hu}},
  \bibinfo {author} {\bibfnamefont {V.~M.}\ \bibnamefont {Menon}},\ and\
  \bibinfo {author} {\bibfnamefont {J.}~\bibnamefont {Yuen-Zhou}},\ }\bibfield
  {title} {\bibinfo {title} {Theory for polariton-assisted remote energy
  transfer},\ }\href
  {https://pubs.rsc.org/en/content/articlelanding/2018/sc/c8sc00171e#!divAbstract}
  {\bibfield  {journal} {\bibinfo  {journal} {Chemical Science}\ }\textbf
  {\bibinfo {volume} {9}},\ \bibinfo {pages} {6659} (\bibinfo {year}
  {2018})}\BibitemShut {NoStop}%
\bibitem [{\citenamefont {Wang}\ \emph
  {et~al.}(2021{\natexlab{c}})\citenamefont {Wang}, \citenamefont {Neuman},
  \citenamefont {Flick},\ and\ \citenamefont {Narang}}]{Wang2020LossQEDFT}%
  \BibitemOpen
  \bibfield  {author} {\bibinfo {author} {\bibfnamefont {D.~S.}\ \bibnamefont
  {Wang}}, \bibinfo {author} {\bibfnamefont {T.}~\bibnamefont {Neuman}},
  \bibinfo {author} {\bibfnamefont {J.}~\bibnamefont {Flick}},\ and\ \bibinfo
  {author} {\bibfnamefont {P.}~\bibnamefont {Narang}},\ }\bibfield  {title}
  {\bibinfo {title} {{Light–matter interaction of a molecule in a dissipative
  cavity from first principles}},\ }\href
  {https://aip.scitation.org/doi/full/10.1063/5.0036283} {\bibfield  {journal}
  {\bibinfo  {journal} {J. Chem. Phys.}\ }\textbf {\bibinfo {volume} {154}},\
  \bibinfo {pages} {104109} (\bibinfo {year} {2021}{\natexlab{c}})}\BibitemShut
  {NoStop}%
\bibitem [{\citenamefont {Schuster}\ \emph {et~al.}(2010)\citenamefont
  {Schuster}, \citenamefont {Sears}, \citenamefont {Ginossar}, \citenamefont
  {Dicarlo}, \citenamefont {Frunzio}, \citenamefont {Morton}, \citenamefont
  {Wu}, \citenamefont {Briggs}, \citenamefont {Buckley}, \citenamefont
  {Awschalom},\ and\ \citenamefont {Schoelkopf}}]{Schuster2010}%
  \BibitemOpen
  \bibfield  {author} {\bibinfo {author} {\bibfnamefont {D.~I.}\ \bibnamefont
  {Schuster}}, \bibinfo {author} {\bibfnamefont {A.~P.}\ \bibnamefont {Sears}},
  \bibinfo {author} {\bibfnamefont {E.}~\bibnamefont {Ginossar}}, \bibinfo
  {author} {\bibfnamefont {L.}~\bibnamefont {Dicarlo}}, \bibinfo {author}
  {\bibfnamefont {L.}~\bibnamefont {Frunzio}}, \bibinfo {author} {\bibfnamefont
  {J.~J.}\ \bibnamefont {Morton}}, \bibinfo {author} {\bibfnamefont
  {H.}~\bibnamefont {Wu}}, \bibinfo {author} {\bibfnamefont {G.~A.}\
  \bibnamefont {Briggs}}, \bibinfo {author} {\bibfnamefont {B.~B.}\
  \bibnamefont {Buckley}}, \bibinfo {author} {\bibfnamefont {D.~D.}\
  \bibnamefont {Awschalom}},\ and\ \bibinfo {author} {\bibfnamefont {R.~J.}\
  \bibnamefont {Schoelkopf}},\ }\bibfield  {title} {\bibinfo {title}
  {{High-cooperativity coupling of electron-spin ensembles to superconducting
  cavities}},\ }\href {https://doi.org/10.1103/PhysRevLett.105.140501}
  {\bibfield  {journal} {\bibinfo  {journal} {Phys. Rev. Lett.}\ }\textbf
  {\bibinfo {volume} {105}},\ \bibinfo {pages} {140501} (\bibinfo {year}
  {2010})}\BibitemShut {NoStop}%
\bibitem [{\citenamefont {Sandner}\ \emph {et~al.}(2012)\citenamefont
  {Sandner}, \citenamefont {Ritsch}, \citenamefont {Ams{\"{u}}ss},
  \citenamefont {Koller}, \citenamefont {N{\"{o}}bauer}, \citenamefont {Putz},
  \citenamefont {Schmiedmayer},\ and\ \citenamefont {Majer}}]{Sandner2012}%
  \BibitemOpen
  \bibfield  {author} {\bibinfo {author} {\bibfnamefont {K.}~\bibnamefont
  {Sandner}}, \bibinfo {author} {\bibfnamefont {H.}~\bibnamefont {Ritsch}},
  \bibinfo {author} {\bibfnamefont {R.}~\bibnamefont {Ams{\"{u}}ss}}, \bibinfo
  {author} {\bibfnamefont {C.}~\bibnamefont {Koller}}, \bibinfo {author}
  {\bibfnamefont {T.}~\bibnamefont {N{\"{o}}bauer}}, \bibinfo {author}
  {\bibfnamefont {S.}~\bibnamefont {Putz}}, \bibinfo {author} {\bibfnamefont
  {J.}~\bibnamefont {Schmiedmayer}},\ and\ \bibinfo {author} {\bibfnamefont
  {J.}~\bibnamefont {Majer}},\ }\bibfield  {title} {\bibinfo {title} {{Strong
  magnetic coupling of an inhomogeneous nitrogen-vacancy ensemble to a
  cavity}},\ }\href {https://doi.org/10.1103/PhysRevA.85.053806} {\bibfield
  {journal} {\bibinfo  {journal} {Phys. Rev. A}\ }\textbf {\bibinfo {volume}
  {85}},\ \bibinfo {pages} {053806} (\bibinfo {year} {2012})}\BibitemShut
  {NoStop}%
\bibitem [{\citenamefont {Wolters}\ \emph {et~al.}(2010)\citenamefont
  {Wolters}, \citenamefont {Schell}, \citenamefont {Kewes}, \citenamefont
  {N{\"{u}}sse}, \citenamefont {Schoengen}, \citenamefont {D{\"{o}}scher},
  \citenamefont {Hannappel}, \citenamefont {L{\"{o}}chel}, \citenamefont
  {Barth},\ and\ \citenamefont {Benson}}]{Wolters2010}%
  \BibitemOpen
  \bibfield  {author} {\bibinfo {author} {\bibfnamefont {J.}~\bibnamefont
  {Wolters}}, \bibinfo {author} {\bibfnamefont {A.~W.}\ \bibnamefont {Schell}},
  \bibinfo {author} {\bibfnamefont {G.}~\bibnamefont {Kewes}}, \bibinfo
  {author} {\bibfnamefont {N.}~\bibnamefont {N{\"{u}}sse}}, \bibinfo {author}
  {\bibfnamefont {M.}~\bibnamefont {Schoengen}}, \bibinfo {author}
  {\bibfnamefont {H.}~\bibnamefont {D{\"{o}}scher}}, \bibinfo {author}
  {\bibfnamefont {T.}~\bibnamefont {Hannappel}}, \bibinfo {author}
  {\bibfnamefont {B.}~\bibnamefont {L{\"{o}}chel}}, \bibinfo {author}
  {\bibfnamefont {M.}~\bibnamefont {Barth}},\ and\ \bibinfo {author}
  {\bibfnamefont {O.}~\bibnamefont {Benson}},\ }\bibfield  {title} {\bibinfo
  {title} {{Enhancement of the zero phonon line emission from a single nitrogen
  vacancy center in a nanodiamond via coupling to a photonic crystal cavity}},\
  }\href {https://doi.org/10.1063/1.3499300} {\bibfield  {journal} {\bibinfo
  {journal} {Appl. Phys. Lett.}\ }\textbf {\bibinfo {volume} {97}},\ \bibinfo
  {pages} {95} (\bibinfo {year} {2010})}\BibitemShut {NoStop}%
\bibitem [{\citenamefont {Englund}\ \emph {et~al.}(2010)\citenamefont
  {Englund}, \citenamefont {Shields}, \citenamefont {Rivoire}, \citenamefont
  {Hatami}, \citenamefont {Vu{\v{c}}kovi{\'{c}}}, \citenamefont {Park},\ and\
  \citenamefont {Lukin}}]{Englund2010}%
  \BibitemOpen
  \bibfield  {author} {\bibinfo {author} {\bibfnamefont {D.}~\bibnamefont
  {Englund}}, \bibinfo {author} {\bibfnamefont {B.}~\bibnamefont {Shields}},
  \bibinfo {author} {\bibfnamefont {K.}~\bibnamefont {Rivoire}}, \bibinfo
  {author} {\bibfnamefont {F.}~\bibnamefont {Hatami}}, \bibinfo {author}
  {\bibfnamefont {J.}~\bibnamefont {Vu{\v{c}}kovi{\'{c}}}}, \bibinfo {author}
  {\bibfnamefont {H.}~\bibnamefont {Park}},\ and\ \bibinfo {author}
  {\bibfnamefont {M.~D.}\ \bibnamefont {Lukin}},\ }\bibfield  {title} {\bibinfo
  {title} {{Deterministic coupling of a single nitrogen vacancy center to a
  photonic crystal cavity}},\ }\href {https://doi.org/10.1021/nl101662v}
  {\bibfield  {journal} {\bibinfo  {journal} {Nano Lett.}\ }\textbf {\bibinfo
  {volume} {10}},\ \bibinfo {pages} {3922} (\bibinfo {year}
  {2010})}\BibitemShut {NoStop}%
\bibitem [{\citenamefont {Schr{\"{o}}der}\ \emph {et~al.}(2017)\citenamefont
  {Schr{\"{o}}der}, \citenamefont {Walsh}, \citenamefont {Zheng}, \citenamefont
  {Mouradian}, \citenamefont {Li}, \citenamefont {Malladi}, \citenamefont
  {Bakhru}, \citenamefont {Lu}, \citenamefont {Stein}, \citenamefont {Heuck},\
  and\ \citenamefont {Englund}}]{Schroder2017}%
  \BibitemOpen
  \bibfield  {author} {\bibinfo {author} {\bibfnamefont {T.}~\bibnamefont
  {Schr{\"{o}}der}}, \bibinfo {author} {\bibfnamefont {M.}~\bibnamefont
  {Walsh}}, \bibinfo {author} {\bibfnamefont {J.}~\bibnamefont {Zheng}},
  \bibinfo {author} {\bibfnamefont {S.}~\bibnamefont {Mouradian}}, \bibinfo
  {author} {\bibfnamefont {L.}~\bibnamefont {Li}}, \bibinfo {author}
  {\bibfnamefont {G.}~\bibnamefont {Malladi}}, \bibinfo {author} {\bibfnamefont
  {H.}~\bibnamefont {Bakhru}}, \bibinfo {author} {\bibfnamefont
  {M.}~\bibnamefont {Lu}}, \bibinfo {author} {\bibfnamefont {A.}~\bibnamefont
  {Stein}}, \bibinfo {author} {\bibfnamefont {M.}~\bibnamefont {Heuck}},\ and\
  \bibinfo {author} {\bibfnamefont {D.}~\bibnamefont {Englund}},\ }\bibfield
  {title} {\bibinfo {title} {{Scalable fabrication of coupled NV center -
  photonic crystal cavity systems by self-aligned N ion implantation}},\ }\href
  {https://doi.org/10.1364/ome.7.001514} {\bibfield  {journal} {\bibinfo
  {journal} {Opt. Mater. Express}\ }\textbf {\bibinfo {volume} {7}},\ \bibinfo
  {pages} {1514} (\bibinfo {year} {2017})}\BibitemShut {NoStop}%
\bibitem [{\citenamefont {Vogl}\ \emph {et~al.}(2019)\citenamefont {Vogl},
  \citenamefont {Lecamwasam}, \citenamefont {Buchler}, \citenamefont {Lu},\
  and\ \citenamefont {Lam}}]{Vogl2019}%
  \BibitemOpen
  \bibfield  {author} {\bibinfo {author} {\bibfnamefont {T.}~\bibnamefont
  {Vogl}}, \bibinfo {author} {\bibfnamefont {R.}~\bibnamefont {Lecamwasam}},
  \bibinfo {author} {\bibfnamefont {B.~C.}\ \bibnamefont {Buchler}}, \bibinfo
  {author} {\bibfnamefont {Y.}~\bibnamefont {Lu}},\ and\ \bibinfo {author}
  {\bibfnamefont {P.~K.}\ \bibnamefont {Lam}},\ }\bibfield  {title} {\bibinfo
  {title} {{Compact Cavity-Enhanced Single-Photon Generation with Hexagonal
  Boron Nitride}},\ }\href {https://doi.org/10.1021/acsphotonics.9b00314}
  {\bibfield  {journal} {\bibinfo  {journal} {ACS Photonics}\ }\textbf
  {\bibinfo {volume} {6}},\ \bibinfo {pages} {1955} (\bibinfo {year}
  {2019})}\BibitemShut {NoStop}%
\bibitem [{\citenamefont {Caldwell}\ \emph {et~al.}(2019)\citenamefont
  {Caldwell}, \citenamefont {Aharonovich}, \citenamefont {Cassabois},
  \citenamefont {Edgar}, \citenamefont {Gil},\ and\ \citenamefont
  {Basov}}]{Caldwell2019}%
  \BibitemOpen
  \bibfield  {author} {\bibinfo {author} {\bibfnamefont {J.~D.}\ \bibnamefont
  {Caldwell}}, \bibinfo {author} {\bibfnamefont {I.}~\bibnamefont
  {Aharonovich}}, \bibinfo {author} {\bibfnamefont {G.}~\bibnamefont
  {Cassabois}}, \bibinfo {author} {\bibfnamefont {J.~H.}\ \bibnamefont
  {Edgar}}, \bibinfo {author} {\bibfnamefont {B.}~\bibnamefont {Gil}},\ and\
  \bibinfo {author} {\bibfnamefont {D.~N.}\ \bibnamefont {Basov}},\ }\bibfield
  {title} {\bibinfo {title} {{Photonics with hexagonal boron nitride}},\ }\href
  {https://doi.org/10.1038/s41578-019-0124-1} {\bibfield  {journal} {\bibinfo
  {journal} {Nat. Rev. Mater.}\ }\textbf {\bibinfo {volume} {4}},\ \bibinfo
  {pages} {552} (\bibinfo {year} {2019})}\BibitemShut {NoStop}%
\bibitem [{\citenamefont {Janitz}\ \emph {et~al.}(2020)\citenamefont {Janitz},
  \citenamefont {Bhaskar},\ and\ \citenamefont {Childress}}]{Janitz2020}%
  \BibitemOpen
  \bibfield  {author} {\bibinfo {author} {\bibfnamefont {E.}~\bibnamefont
  {Janitz}}, \bibinfo {author} {\bibfnamefont {M.~K.}\ \bibnamefont
  {Bhaskar}},\ and\ \bibinfo {author} {\bibfnamefont {L.}~\bibnamefont
  {Childress}},\ }\bibfield  {title} {\bibinfo {title} {{Cavity quantum
  electrodynamics with color centers in diamond}},\ }\href
  {https://doi.org/10.1364/optica.398628} {\bibfield  {journal} {\bibinfo
  {journal} {Optica}\ }\textbf {\bibinfo {volume} {7}},\ \bibinfo {pages}
  {1232} (\bibinfo {year} {2020})},\ \Eprint {https://arxiv.org/abs/2101.02793}
  {arXiv:2101.02793} \BibitemShut {NoStop}%
\bibitem [{\citenamefont {Grosso}\ \emph {et~al.}(2017)\citenamefont {Grosso},
  \citenamefont {Moon}, \citenamefont {Lienhard}, \citenamefont {Ali},
  \citenamefont {Efetov}, \citenamefont {Furchi}, \citenamefont
  {Jarillo-Herrero}, \citenamefont {Ford}, \citenamefont {Aharonovich},\ and\
  \citenamefont {Englund}}]{Grosso2017}%
  \BibitemOpen
  \bibfield  {author} {\bibinfo {author} {\bibfnamefont {G.}~\bibnamefont
  {Grosso}}, \bibinfo {author} {\bibfnamefont {H.}~\bibnamefont {Moon}},
  \bibinfo {author} {\bibfnamefont {B.}~\bibnamefont {Lienhard}}, \bibinfo
  {author} {\bibfnamefont {S.}~\bibnamefont {Ali}}, \bibinfo {author}
  {\bibfnamefont {D.~K.}\ \bibnamefont {Efetov}}, \bibinfo {author}
  {\bibfnamefont {M.~M.}\ \bibnamefont {Furchi}}, \bibinfo {author}
  {\bibfnamefont {P.}~\bibnamefont {Jarillo-Herrero}}, \bibinfo {author}
  {\bibfnamefont {M.~J.}\ \bibnamefont {Ford}}, \bibinfo {author}
  {\bibfnamefont {I.}~\bibnamefont {Aharonovich}},\ and\ \bibinfo {author}
  {\bibfnamefont {D.}~\bibnamefont {Englund}},\ }\bibfield  {title} {\bibinfo
  {title} {{Tunable and high-purity room temperature single-photon emission
  from atomic defects in hexagonal boron nitride}},\ }\href
  {https://doi.org/10.1038/s41467-017-00810-2} {\bibfield  {journal} {\bibinfo
  {journal} {Nat. Commun.}\ }\textbf {\bibinfo {volume} {8}},\ \bibinfo {pages}
  {1} (\bibinfo {year} {2017})}\BibitemShut {NoStop}%
\bibitem [{\citenamefont {Mendelson}\ \emph {et~al.}(2020)\citenamefont
  {Mendelson}, \citenamefont {Chugh}, \citenamefont {Reimers}, \citenamefont
  {Cheng}, \citenamefont {Gottscholl}, \citenamefont {Long}, \citenamefont
  {Mellor}, \citenamefont {Zettl}, \citenamefont {Dyakonov}, \citenamefont
  {Beton}, \citenamefont {Novikov}, \citenamefont {Jagadish}, \citenamefont
  {Tan}, \citenamefont {Ford},\ and\ \citenamefont {Toth}}]{Mendelson}%
  \BibitemOpen
  \bibfield  {author} {\bibinfo {author} {\bibfnamefont {N.}~\bibnamefont
  {Mendelson}}, \bibinfo {author} {\bibfnamefont {D.}~\bibnamefont {Chugh}},
  \bibinfo {author} {\bibfnamefont {J.~R.}\ \bibnamefont {Reimers}}, \bibinfo
  {author} {\bibfnamefont {T.~S.}\ \bibnamefont {Cheng}}, \bibinfo {author}
  {\bibfnamefont {A.}~\bibnamefont {Gottscholl}}, \bibinfo {author}
  {\bibfnamefont {H.}~\bibnamefont {Long}}, \bibinfo {author} {\bibfnamefont
  {C.~J.}\ \bibnamefont {Mellor}}, \bibinfo {author} {\bibfnamefont
  {A.}~\bibnamefont {Zettl}}, \bibinfo {author} {\bibfnamefont
  {V.}~\bibnamefont {Dyakonov}}, \bibinfo {author} {\bibfnamefont {P.~H.}\
  \bibnamefont {Beton}}, \bibinfo {author} {\bibfnamefont {S.~V.}\ \bibnamefont
  {Novikov}}, \bibinfo {author} {\bibfnamefont {C.}~\bibnamefont {Jagadish}},
  \bibinfo {author} {\bibfnamefont {H.~H.}\ \bibnamefont {Tan}}, \bibinfo
  {author} {\bibfnamefont {M.~J.}\ \bibnamefont {Ford}},\ and\ \bibinfo
  {author} {\bibfnamefont {M.}~\bibnamefont {Toth}},\ }\bibfield  {title}
  {\bibinfo {title} {Identifying carbon as the source of visible single-photon
  emission from hexagonal boron nitride},\ }\href
  {http://dx.doi.org/10.1038/s41563-020-00850-y} {\bibfield  {journal}
  {\bibinfo  {journal} {Nat. Mater.}\ } (\bibinfo {year} {2020})}\BibitemShut
  {NoStop}%
\bibitem [{\citenamefont {Tokatly}(2013)}]{tokatly2013}%
  \BibitemOpen
  \bibfield  {author} {\bibinfo {author} {\bibfnamefont {I.~V.}\ \bibnamefont
  {Tokatly}},\ }\bibfield  {title} {\bibinfo {title} {Time-dependent density
  functional theory for many-electron systems interacting with cavity
  photons},\ }\href {https://doi.org/10.1103/PhysRevLett.110.233001} {\bibfield
   {journal} {\bibinfo  {journal} {Phys. Rev. Lett.}\ }\textbf {\bibinfo
  {volume} {110}},\ \bibinfo {pages} {233001} (\bibinfo {year}
  {2013})}\BibitemShut {NoStop}%
\bibitem [{\citenamefont {Ruggenthaler}\ \emph {et~al.}(2014)\citenamefont
  {Ruggenthaler}, \citenamefont {Flick}, \citenamefont {Pellegrini},
  \citenamefont {Appel}, \citenamefont {Tokatly},\ and\ \citenamefont
  {Rubio}}]{ruggenthaler2014}%
  \BibitemOpen
  \bibfield  {author} {\bibinfo {author} {\bibfnamefont {M.}~\bibnamefont
  {Ruggenthaler}}, \bibinfo {author} {\bibfnamefont {J.}~\bibnamefont {Flick}},
  \bibinfo {author} {\bibfnamefont {C.}~\bibnamefont {Pellegrini}}, \bibinfo
  {author} {\bibfnamefont {H.}~\bibnamefont {Appel}}, \bibinfo {author}
  {\bibfnamefont {I.~V.}\ \bibnamefont {Tokatly}},\ and\ \bibinfo {author}
  {\bibfnamefont {A.}~\bibnamefont {Rubio}},\ }\bibfield  {title} {\bibinfo
  {title} {Quantum-electrodynamical density-functional theory: Bridging quantum
  optics and electronic-structure theory},\ }\href
  {https://doi.org/10.1103/PhysRevA.90.012508} {\bibfield  {journal} {\bibinfo
  {journal} {Phys. Rev. A}\ }\textbf {\bibinfo {volume} {90}},\ \bibinfo
  {pages} {012508} (\bibinfo {year} {2014})}\BibitemShut {NoStop}%
\bibitem [{\citenamefont {Flick}\ \emph {et~al.}(2019)\citenamefont {Flick},
  \citenamefont {Welakuh}, \citenamefont {Ruggenthaler}, \citenamefont
  {Appel},\ and\ \citenamefont {Rubio}}]{flick2019lmrnqe}%
  \BibitemOpen
  \bibfield  {author} {\bibinfo {author} {\bibfnamefont {J.}~\bibnamefont
  {Flick}}, \bibinfo {author} {\bibfnamefont {D.~M.}\ \bibnamefont {Welakuh}},
  \bibinfo {author} {\bibfnamefont {M.}~\bibnamefont {Ruggenthaler}}, \bibinfo
  {author} {\bibfnamefont {H.}~\bibnamefont {Appel}},\ and\ \bibinfo {author}
  {\bibfnamefont {A.}~\bibnamefont {Rubio}},\ }\bibfield  {title} {\bibinfo
  {title} {Light–matter response in nonrelativistic quantum
  electrodynamics},\ }\href {https://doi.org/10.1021/acsphotonics.9b00768}
  {\bibfield  {journal} {\bibinfo  {journal} {ACS Photonics}\ }\textbf
  {\bibinfo {volume} {6}},\ \bibinfo {pages} {2757} (\bibinfo {year}
  {2019})}\BibitemShut {NoStop}%
\bibitem [{\citenamefont {Momma}\ and\ \citenamefont
  {Izumi}(2008)}]{Momma2008}%
  \BibitemOpen
  \bibfield  {author} {\bibinfo {author} {\bibfnamefont {K.}~\bibnamefont
  {Momma}}\ and\ \bibinfo {author} {\bibfnamefont {F.}~\bibnamefont {Izumi}},\
  }\bibfield  {title} {\bibinfo {title} {{VESTA: A three-dimensional
  visualization system for electronic and structural analysis}},\ }\href
  {https://doi.org/10.1107/S0021889808012016} {\bibfield  {journal} {\bibinfo
  {journal} {J. Appl. Crystallogr.}\ }\textbf {\bibinfo {volume} {41}},\
  \bibinfo {pages} {653} (\bibinfo {year} {2008})}\BibitemShut {NoStop}%
\bibitem [{\citenamefont {Marques}\ \emph {et~al.}(2003)\citenamefont
  {Marques}, \citenamefont {Castro}, \citenamefont {Bertsch},\ and\
  \citenamefont {Rubio}}]{octopus1}%
  \BibitemOpen
  \bibfield  {author} {\bibinfo {author} {\bibfnamefont {M.~A.~L.}\
  \bibnamefont {Marques}}, \bibinfo {author} {\bibfnamefont {A.}~\bibnamefont
  {Castro}}, \bibinfo {author} {\bibfnamefont {G.~F.}\ \bibnamefont
  {Bertsch}},\ and\ \bibinfo {author} {\bibfnamefont {A.}~\bibnamefont
  {Rubio}},\ }\bibfield  {title} {\bibinfo {title} {{Octopus: a
  First-Principles Tool for Excited Electron-Ion Dynamics}},\ }\href
  {https://www.sciencedirect.com/science/article/pii/S0010465502006860}
  {\bibfield  {journal} {\bibinfo  {journal} {Comput. Phys. Commun.}\ }\textbf
  {\bibinfo {volume} {151}},\ \bibinfo {pages} {60} (\bibinfo {year}
  {2003})}\BibitemShut {NoStop}%
\bibitem [{\citenamefont {Andrade}\ \emph {et~al.}(2015)\citenamefont
  {Andrade}, \citenamefont {A.}, \citenamefont {De~Giovannini}, \citenamefont
  {Larsen}, \citenamefont {Oliveira}, \citenamefont {Alberdi-Rodriguez},
  \citenamefont {Varas}, \citenamefont {Theophilou}, \citenamefont {Helbig},
  \citenamefont {Verstraete}, \citenamefont {Stella}, \citenamefont {Nogueira},
  \citenamefont {Aspuru-Guzik}, \citenamefont {Castro}, \citenamefont
  {Marques},\ and\ \citenamefont {Rubio}}]{octopus2}%
  \BibitemOpen
  \bibfield  {author} {\bibinfo {author} {\bibfnamefont {X.}~\bibnamefont
  {Andrade}}, \bibinfo {author} {\bibfnamefont {S.~D.}\ \bibnamefont {A.}},
  \bibinfo {author} {\bibfnamefont {U.}~\bibnamefont {De~Giovannini}}, \bibinfo
  {author} {\bibfnamefont {A.~H.}\ \bibnamefont {Larsen}}, \bibinfo {author}
  {\bibfnamefont {M.~J.~T.}\ \bibnamefont {Oliveira}}, \bibinfo {author}
  {\bibfnamefont {J.}~\bibnamefont {Alberdi-Rodriguez}}, \bibinfo {author}
  {\bibfnamefont {A.}~\bibnamefont {Varas}}, \bibinfo {author} {\bibfnamefont
  {I.}~\bibnamefont {Theophilou}}, \bibinfo {author} {\bibfnamefont
  {N.}~\bibnamefont {Helbig}}, \bibinfo {author} {\bibfnamefont {M.~J.}\
  \bibnamefont {Verstraete}}, \bibinfo {author} {\bibfnamefont
  {L.}~\bibnamefont {Stella}}, \bibinfo {author} {\bibfnamefont
  {F.}~\bibnamefont {Nogueira}}, \bibinfo {author} {\bibfnamefont
  {A.}~\bibnamefont {Aspuru-Guzik}}, \bibinfo {author} {\bibfnamefont
  {A.}~\bibnamefont {Castro}}, \bibinfo {author} {\bibfnamefont {M.~A.~L.}\
  \bibnamefont {Marques}},\ and\ \bibinfo {author} {\bibfnamefont
  {A.}~\bibnamefont {Rubio}},\ }\bibfield  {title} {\bibinfo {title}
  {{Real-Space Grids and the Octopus Code as Tools for the Development of New
  Simulation Approaches for Electronic Systems}},\ }\href
  {https://pubs.rsc.org/en/content/articlelanding/2015/cp/c5cp00351b#!divAbstract}
  {\bibfield  {journal} {\bibinfo  {journal} {Phys. Chem. Chem. Phys.}\
  }\textbf {\bibinfo {volume} {17}},\ \bibinfo {pages} {31371} (\bibinfo {year}
  {2015})}\BibitemShut {NoStop}%
\bibitem [{\citenamefont {Tancogne-Dejean}\ \emph {et~al.}(2020)\citenamefont
  {Tancogne-Dejean}, \citenamefont {Oliveira}, \citenamefont {Andrade},
  \citenamefont {Appel}, \citenamefont {Borca}, \citenamefont {Le~Breton},
  \citenamefont {Buchholz}, \citenamefont {Castro}, \citenamefont {Corni},
  \citenamefont {Correa}, \citenamefont {De~Giovannini}, \citenamefont
  {Delgado}, \citenamefont {Eich}, \citenamefont {Flick}, \citenamefont {Gil},
  \citenamefont {Gomez}, \citenamefont {Helbig}, \citenamefont {Hübener},
  \citenamefont {Jestädt} \emph {et~al.}}]{octopus3}%
  \BibitemOpen
  \bibfield  {author} {\bibinfo {author} {\bibfnamefont {N.}~\bibnamefont
  {Tancogne-Dejean}}, \bibinfo {author} {\bibfnamefont {M.~J.~T.}\ \bibnamefont
  {Oliveira}}, \bibinfo {author} {\bibfnamefont {X.}~\bibnamefont {Andrade}},
  \bibinfo {author} {\bibfnamefont {H.}~\bibnamefont {Appel}}, \bibinfo
  {author} {\bibfnamefont {C.~H.}\ \bibnamefont {Borca}}, \bibinfo {author}
  {\bibfnamefont {G.}~\bibnamefont {Le~Breton}}, \bibinfo {author}
  {\bibfnamefont {F.}~\bibnamefont {Buchholz}}, \bibinfo {author}
  {\bibfnamefont {A.}~\bibnamefont {Castro}}, \bibinfo {author} {\bibfnamefont
  {S.}~\bibnamefont {Corni}}, \bibinfo {author} {\bibfnamefont {A.~A.}\
  \bibnamefont {Correa}}, \bibinfo {author} {\bibfnamefont {U.}~\bibnamefont
  {De~Giovannini}}, \bibinfo {author} {\bibfnamefont {A.}~\bibnamefont
  {Delgado}}, \bibinfo {author} {\bibfnamefont {F.~G.}\ \bibnamefont {Eich}},
  \bibinfo {author} {\bibfnamefont {J.}~\bibnamefont {Flick}}, \bibinfo
  {author} {\bibfnamefont {G.}~\bibnamefont {Gil}}, \bibinfo {author}
  {\bibfnamefont {A.}~\bibnamefont {Gomez}}, \bibinfo {author} {\bibfnamefont
  {N.}~\bibnamefont {Helbig}}, \bibinfo {author} {\bibfnamefont
  {H.}~\bibnamefont {Hübener}}, \bibinfo {author} {\bibfnamefont
  {R.}~\bibnamefont {Jestädt}}, \emph {et~al.},\ }\bibfield  {title} {\bibinfo
  {title} {Octopus, a computational framework for exploring light-driven
  phenomena and quantum dynamics in extended and finite systems},\ }\href
  {https://doi.org/10.1063/1.5142502} {\bibfield  {journal} {\bibinfo
  {journal} {J. Chem. Phys.}\ }\textbf {\bibinfo {volume} {152}},\ \bibinfo
  {pages} {124119} (\bibinfo {year} {2020})}\BibitemShut {NoStop}%
\bibitem [{\citenamefont {Barcza}\ \emph {et~al.}(2020)\citenamefont {Barcza},
  \citenamefont {Iv{\'{a}}dy}, \citenamefont {Szilv{\'{a}}si}, \citenamefont
  {V{\"{o}}r{\"{o}}s}, \citenamefont {Veis}, \citenamefont {Gali},\ and\
  \citenamefont {Legeza}}]{Barcza}%
  \BibitemOpen
  \bibfield  {author} {\bibinfo {author} {\bibfnamefont {G.}~\bibnamefont
  {Barcza}}, \bibinfo {author} {\bibfnamefont {V.}~\bibnamefont {Iv{\'{a}}dy}},
  \bibinfo {author} {\bibfnamefont {T.}~\bibnamefont {Szilv{\'{a}}si}},
  \bibinfo {author} {\bibfnamefont {M.}~\bibnamefont {V{\"{o}}r{\"{o}}s}},
  \bibinfo {author} {\bibfnamefont {L.}~\bibnamefont {Veis}}, \bibinfo {author}
  {\bibfnamefont {A.}~\bibnamefont {Gali}},\ and\ \bibinfo {author}
  {\bibfnamefont {O.}~\bibnamefont {Legeza}},\ }\bibfield  {title} {\bibinfo
  {title} {{DMRG on Top of Plane-Wave Kohn-Sham Orbitals: Case Study of
  Defected Boron Nitride}},\ }\href {https://arxiv.org/abs/2006.04557}
  {\bibfield  {journal} {\bibinfo  {journal} {arXiv:2006.04557}\ } (\bibinfo
  {year} {2020})}\BibitemShut {NoStop}%
\bibitem [{\citenamefont {Reimers}\ \emph {et~al.}(2020)\citenamefont
  {Reimers}, \citenamefont {Shen}, \citenamefont {Kianinia}, \citenamefont
  {Bradac}, \citenamefont {Aharonovich}, \citenamefont {Ford},\ and\
  \citenamefont {Piecuch}}]{Reimers2020}%
  \BibitemOpen
  \bibfield  {author} {\bibinfo {author} {\bibfnamefont {J.~R.}\ \bibnamefont
  {Reimers}}, \bibinfo {author} {\bibfnamefont {J.}~\bibnamefont {Shen}},
  \bibinfo {author} {\bibfnamefont {M.}~\bibnamefont {Kianinia}}, \bibinfo
  {author} {\bibfnamefont {C.}~\bibnamefont {Bradac}}, \bibinfo {author}
  {\bibfnamefont {I.}~\bibnamefont {Aharonovich}}, \bibinfo {author}
  {\bibfnamefont {M.~J.}\ \bibnamefont {Ford}},\ and\ \bibinfo {author}
  {\bibfnamefont {P.}~\bibnamefont {Piecuch}},\ }\bibfield  {title} {\bibinfo
  {title} {Photoluminescence, photophysics, and photochemistry of the
  ${{\mathrm{V}}_{\mathrm{B}}}^{\ensuremath{-}}$ defect in hexagonal boron
  nitride},\ }\href {https://doi.org/10.1103/PhysRevB.102.144105} {\bibfield
  {journal} {\bibinfo  {journal} {Phys. Rev. B}\ }\textbf {\bibinfo {volume}
  {102}},\ \bibinfo {pages} {144105} (\bibinfo {year} {2020})}\BibitemShut
  {NoStop}%
\bibitem [{\citenamefont {Perdew}\ \emph {et~al.}(1996)\citenamefont {Perdew},
  \citenamefont {Burke},\ and\ \citenamefont {Ernzerhof}}]{Perdew1996}%
  \BibitemOpen
  \bibfield  {author} {\bibinfo {author} {\bibfnamefont {J.~P.}\ \bibnamefont
  {Perdew}}, \bibinfo {author} {\bibfnamefont {K.}~\bibnamefont {Burke}},\ and\
  \bibinfo {author} {\bibfnamefont {M.}~\bibnamefont {Ernzerhof}},\ }\bibfield
  {title} {\bibinfo {title} {{Generalized Gradient Approximation Made
  Simple}},\ }\href {https://doi.org/10.1103/PhysRevLett.77.3865} {\bibfield
  {journal} {\bibinfo  {journal} {Phys. Rev. Lett.}\ }\textbf {\bibinfo
  {volume} {77}},\ \bibinfo {pages} {3865} (\bibinfo {year}
  {1996})}\BibitemShut {NoStop}%
\bibitem [{\citenamefont {Perdew}\ and\ \citenamefont
  {Zunger}(1981)}]{perdew1981}%
  \BibitemOpen
  \bibfield  {author} {\bibinfo {author} {\bibfnamefont {J.~P.}\ \bibnamefont
  {Perdew}}\ and\ \bibinfo {author} {\bibfnamefont {A.}~\bibnamefont
  {Zunger}},\ }\bibfield  {title} {\bibinfo {title} {Self-interaction
  correction to density-functional approximations for many-electron systems},\
  }\href {https://doi.org/10.1103/PhysRevB.23.5048} {\bibfield  {journal}
  {\bibinfo  {journal} {Phys. Rev. B}\ }\textbf {\bibinfo {volume} {23}},\
  \bibinfo {pages} {5048} (\bibinfo {year} {1981})}\BibitemShut {NoStop}%
\bibitem [{\citenamefont {Perdew}\ and\ \citenamefont
  {Wang}(1992)}]{Perdew1992}%
  \BibitemOpen
  \bibfield  {author} {\bibinfo {author} {\bibfnamefont {J.~P.}\ \bibnamefont
  {Perdew}}\ and\ \bibinfo {author} {\bibfnamefont {Y.}~\bibnamefont {Wang}},\
  }\bibfield  {title} {\bibinfo {title} {{Accurate and simple analytic
  representation of the electron-gas correlation energy}},\ }\href
  {https://doi.org/10.1103/PhysRevB.98.079904} {\bibfield  {journal} {\bibinfo
  {journal} {Phys. Rev. B}\ }\textbf {\bibinfo {volume} {45}},\ \bibinfo
  {pages} {13224} (\bibinfo {year} {1992})}\BibitemShut {NoStop}%
\bibitem [{\citenamefont {Casida}(1995)}]{casida1995time}%
  \BibitemOpen
  \bibfield  {author} {\bibinfo {author} {\bibfnamefont {M.~E.}\ \bibnamefont
  {Casida}},\ }\bibfield  {title} {\bibinfo {title} {Time-dependent density
  functional response theory for molecules},\ }in\ \href@noop {} {\emph
  {\bibinfo {booktitle} {Recent Advances In Density Functional Methods: (Part
  I)}}}\ (\bibinfo  {publisher} {World Scientific},\ \bibinfo {address}
  {Singapore},\ \bibinfo {year} {1995})\ pp.\ \bibinfo {pages}
  {155--192}\BibitemShut {NoStop}%
\bibitem [{\citenamefont {Craig}\ and\ \citenamefont
  {Thirunamachandran}(2012)}]{craig2012molecular}%
  \BibitemOpen
  \bibfield  {author} {\bibinfo {author} {\bibfnamefont {D.}~\bibnamefont
  {Craig}}\ and\ \bibinfo {author} {\bibfnamefont {T.}~\bibnamefont
  {Thirunamachandran}},\ }\href
  {https://books.google.com/books?id=S6DDAgAAQBAJ} {\emph {\bibinfo {title}
  {Molecular Quantum Electrodynamics}}},\ Dover Books on Chemistry\ (\bibinfo
  {publisher} {Dover Publications},\ \bibinfo {year} {2012})\BibitemShut
  {NoStop}%
\bibitem [{\citenamefont {{Jaynes}}\ and\ \citenamefont
  {{Cummings}}(1963)}]{JaynesCummings1963}%
  \BibitemOpen
  \bibfield  {author} {\bibinfo {author} {\bibfnamefont {E.~T.}\ \bibnamefont
  {{Jaynes}}}\ and\ \bibinfo {author} {\bibfnamefont {F.~W.}\ \bibnamefont
  {{Cummings}}},\ }\bibfield  {title} {\bibinfo {title} {Comparison of quantum
  and semiclassical radiation theories with application to the beam maser},\
  }\href {https://doi.org/10.1109/PROC.1963.1664} {\bibfield  {journal}
  {\bibinfo  {journal} {Proceedings of the IEEE}\ }\textbf {\bibinfo {volume}
  {51}},\ \bibinfo {pages} {89} (\bibinfo {year} {1963})}\BibitemShut {NoStop}%
\bibitem [{\citenamefont {Kokalj}(1999)}]{Kokalj1999}%
  \BibitemOpen
  \bibfield  {author} {\bibinfo {author} {\bibfnamefont {A.}~\bibnamefont
  {Kokalj}},\ }\bibfield  {title} {\bibinfo {title} {{XCrySDen: A New Program
  for Displaying Crystalline Structures and Electron Densities}},\ }\href
  {https://doi.org/10.1016/S1093-3263(99)00028-5} {\bibfield  {journal}
  {\bibinfo  {journal} {J. Mol. Graph. Model.}\ }\textbf {\bibinfo {volume}
  {17}},\ \bibinfo {pages} {176} (\bibinfo {year} {1999})}\BibitemShut
  {NoStop}%
\bibitem [{\citenamefont {Tawfik}\ \emph {et~al.}(2017)\citenamefont {Tawfik},
  \citenamefont {Ali}, \citenamefont {Fronzi}, \citenamefont {Kianinia},
  \citenamefont {Tran}, \citenamefont {Stampfl}, \citenamefont {Aharonovich},
  \citenamefont {Toth},\ and\ \citenamefont {Ford}}]{Tawfik2017}%
  \BibitemOpen
  \bibfield  {author} {\bibinfo {author} {\bibfnamefont {S.~A.}\ \bibnamefont
  {Tawfik}}, \bibinfo {author} {\bibfnamefont {S.}~\bibnamefont {Ali}},
  \bibinfo {author} {\bibfnamefont {M.}~\bibnamefont {Fronzi}}, \bibinfo
  {author} {\bibfnamefont {M.}~\bibnamefont {Kianinia}}, \bibinfo {author}
  {\bibfnamefont {T.~T.}\ \bibnamefont {Tran}}, \bibinfo {author}
  {\bibfnamefont {C.}~\bibnamefont {Stampfl}}, \bibinfo {author} {\bibfnamefont
  {I.}~\bibnamefont {Aharonovich}}, \bibinfo {author} {\bibfnamefont
  {M.}~\bibnamefont {Toth}},\ and\ \bibinfo {author} {\bibfnamefont {M.~J.}\
  \bibnamefont {Ford}},\ }\bibfield  {title} {\bibinfo {title}
  {{First-Principles Ivestigation of Quantum Emission from hBN Defects}},\
  }\href {https://doi.org/10.1039/c7nr04270a} {\bibfield  {journal} {\bibinfo
  {journal} {Nanoscale}\ }\textbf {\bibinfo {volume} {9}},\ \bibinfo {pages}
  {13575} (\bibinfo {year} {2017})}\BibitemShut {NoStop}%
\bibitem [{\citenamefont {Grosso}\ \emph {et~al.}(2020)\citenamefont {Grosso},
  \citenamefont {Moon}, \citenamefont {Ciccarino}, \citenamefont {Flick},
  \citenamefont {Mendelson}, \citenamefont {Mennel}, \citenamefont {Toth},
  \citenamefont {Aharonovich}, \citenamefont {Narang},\ and\ \citenamefont
  {Englund}}]{Grosso2020}%
  \BibitemOpen
  \bibfield  {author} {\bibinfo {author} {\bibfnamefont {G.}~\bibnamefont
  {Grosso}}, \bibinfo {author} {\bibfnamefont {H.}~\bibnamefont {Moon}},
  \bibinfo {author} {\bibfnamefont {C.~J.}\ \bibnamefont {Ciccarino}}, \bibinfo
  {author} {\bibfnamefont {J.}~\bibnamefont {Flick}}, \bibinfo {author}
  {\bibfnamefont {N.}~\bibnamefont {Mendelson}}, \bibinfo {author}
  {\bibfnamefont {L.}~\bibnamefont {Mennel}}, \bibinfo {author} {\bibfnamefont
  {M.}~\bibnamefont {Toth}}, \bibinfo {author} {\bibfnamefont {I.}~\bibnamefont
  {Aharonovich}}, \bibinfo {author} {\bibfnamefont {P.}~\bibnamefont
  {Narang}},\ and\ \bibinfo {author} {\bibfnamefont {D.~R.}\ \bibnamefont
  {Englund}},\ }\bibfield  {title} {\bibinfo {title} {{Low-Temperature
  Electron-Phonon Interaction of Quantum Emitters in Hexagonal Boron
  Nitride}},\ }\href {https://doi.org/10.1021/acsphotonics.9b01789} {\bibfield
  {journal} {\bibinfo  {journal} {ACS Photonics}\ }\textbf {\bibinfo {volume}
  {7}},\ \bibinfo {pages} {1410} (\bibinfo {year} {2020})}\BibitemShut
  {NoStop}%
\bibitem [{\citenamefont {Towns}\ \emph {et~al.}(2014)\citenamefont {Towns},
  \citenamefont {Cockerill}, \citenamefont {Dahan}, \citenamefont {Foster},
  \citenamefont {Gaither}, \citenamefont {Grimshaw}, \citenamefont {Hazlewood},
  \citenamefont {Lathrop}, \citenamefont {Lifka}, \citenamefont {Peterson},
  \citenamefont {Roskies}, \citenamefont {Scott},\ and\ \citenamefont
  {Wilkins-Diehr}}]{xsede}%
  \BibitemOpen
  \bibfield  {author} {\bibinfo {author} {\bibfnamefont {J.}~\bibnamefont
  {Towns}}, \bibinfo {author} {\bibfnamefont {T.}~\bibnamefont {Cockerill}},
  \bibinfo {author} {\bibfnamefont {M.}~\bibnamefont {Dahan}}, \bibinfo
  {author} {\bibfnamefont {I.}~\bibnamefont {Foster}}, \bibinfo {author}
  {\bibfnamefont {K.}~\bibnamefont {Gaither}}, \bibinfo {author} {\bibfnamefont
  {A.}~\bibnamefont {Grimshaw}}, \bibinfo {author} {\bibfnamefont
  {V.}~\bibnamefont {Hazlewood}}, \bibinfo {author} {\bibfnamefont
  {S.}~\bibnamefont {Lathrop}}, \bibinfo {author} {\bibfnamefont
  {D.}~\bibnamefont {Lifka}}, \bibinfo {author} {\bibfnamefont {G.~D.}\
  \bibnamefont {Peterson}}, \bibinfo {author} {\bibfnamefont {R.}~\bibnamefont
  {Roskies}}, \bibinfo {author} {\bibfnamefont {J.~R.}\ \bibnamefont {Scott}},\
  and\ \bibinfo {author} {\bibfnamefont {N.}~\bibnamefont {Wilkins-Diehr}},\
  }\bibfield  {title} {\bibinfo {title} {Xsede: Accelerating scientific
  discovery},\ }\href {https://doi.org/10.1109/MCSE.2014.80} {\bibfield
  {journal} {\bibinfo  {journal} {Computing in Science \& Engineering}\
  }\textbf {\bibinfo {volume} {16}},\ \bibinfo {pages} {62} (\bibinfo {year}
  {2014})}\BibitemShut {NoStop}%
\bibitem [{\citenamefont {Schlipf}\ and\ \citenamefont
  {Gygi}(2015)}]{Schlipf2015}%
  \BibitemOpen
  \bibfield  {author} {\bibinfo {author} {\bibfnamefont {M.}~\bibnamefont
  {Schlipf}}\ and\ \bibinfo {author} {\bibfnamefont {F.}~\bibnamefont {Gygi}},\
  }\bibfield  {title} {\bibinfo {title} {{Optimization Algorithm for the
  gGneration of ONCV Pseudopotentials}},\ }\href
  {https://doi.org/10.1016/j.cpc.2015.05.011} {\bibfield  {journal} {\bibinfo
  {journal} {Comput. Phys. Commun.}\ }\textbf {\bibinfo {volume} {196}},\
  \bibinfo {pages} {36} (\bibinfo {year} {2015})}\BibitemShut {NoStop}%
\bibitem [{\citenamefont {Hamann}(2013)}]{Hamann2013}%
  \BibitemOpen
  \bibfield  {author} {\bibinfo {author} {\bibfnamefont {D.~R.}\ \bibnamefont
  {Hamann}},\ }\bibfield  {title} {\bibinfo {title} {{Optimized Norm-conserving
  Vanderbilt Pseudopotentials}},\ }\href
  {https://doi.org/10.1103/PhysRevB.88.085117} {\bibfield  {journal} {\bibinfo
  {journal} {Phys. Rev. B}\ }\textbf {\bibinfo {volume} {88}},\ \bibinfo
  {pages} {085117} (\bibinfo {year} {2013})}\BibitemShut {NoStop}%
\bibitem [{\citenamefont {Tran}\ \emph {et~al.}(2016)\citenamefont {Tran},
  \citenamefont {Bray}, \citenamefont {Ford}, \citenamefont {Toth},\ and\
  \citenamefont {Aharonovich}}]{Tran2016a}%
  \BibitemOpen
  \bibfield  {author} {\bibinfo {author} {\bibfnamefont {T.~T.}\ \bibnamefont
  {Tran}}, \bibinfo {author} {\bibfnamefont {K.}~\bibnamefont {Bray}}, \bibinfo
  {author} {\bibfnamefont {M.~J.}\ \bibnamefont {Ford}}, \bibinfo {author}
  {\bibfnamefont {M.}~\bibnamefont {Toth}},\ and\ \bibinfo {author}
  {\bibfnamefont {I.}~\bibnamefont {Aharonovich}},\ }\bibfield  {title}
  {\bibinfo {title} {{Quantum emission from Hexagonal Boron Nitride
  Monolayers}},\ }\href {https://doi.org/10.1038/nnano.2015.242} {\bibfield
  {journal} {\bibinfo  {journal} {Nat. Nanotechnol.}\ }\textbf {\bibinfo
  {volume} {11}},\ \bibinfo {pages} {37} (\bibinfo {year} {2016})}\BibitemShut
  {NoStop}%
\bibitem [{\citenamefont {Wu}\ \emph {et~al.}(2017)\citenamefont {Wu},
  \citenamefont {Galatas}, \citenamefont {Sundararaman}, \citenamefont
  {Rocca},\ and\ \citenamefont {Ping}}]{Wu2017}%
  \BibitemOpen
  \bibfield  {author} {\bibinfo {author} {\bibfnamefont {F.}~\bibnamefont
  {Wu}}, \bibinfo {author} {\bibfnamefont {A.}~\bibnamefont {Galatas}},
  \bibinfo {author} {\bibfnamefont {R.}~\bibnamefont {Sundararaman}}, \bibinfo
  {author} {\bibfnamefont {D.}~\bibnamefont {Rocca}},\ and\ \bibinfo {author}
  {\bibfnamefont {Y.}~\bibnamefont {Ping}},\ }\bibfield  {title} {\bibinfo
  {title} {{First-principles Engineering of Charged Defects for Two-Dimensional
  Quantum Technologies}},\ }\href
  {https://doi.org/10.1103/PhysRevMaterials.1.071001} {\bibfield  {journal}
  {\bibinfo  {journal} {Phys. Rev. Mater.}\ }\textbf {\bibinfo {volume} {1}},\
  \bibinfo {pages} {071001} (\bibinfo {year} {2017})}\BibitemShut {NoStop}%
\bibitem [{\citenamefont {Tancogne-Dejean}\ and\ \citenamefont
  {Rubio}(2018)}]{Tancogne-dejean2018}%
  \BibitemOpen
  \bibfield  {author} {\bibinfo {author} {\bibfnamefont {N.}~\bibnamefont
  {Tancogne-Dejean}}\ and\ \bibinfo {author} {\bibfnamefont {A.}~\bibnamefont
  {Rubio}},\ }\bibfield  {title} {\bibinfo {title} {{Atomic-Like High-Harmonic
  Generation from Two-Dimensional Materials}},\ }\href
  {https://advances.sciencemag.org/content/4/2/eaao5207/tab-pdf} {\bibfield
  {journal} {\bibinfo  {journal} {Sci. Adv.}\ }\textbf {\bibinfo {volume}
  {4}},\ \bibinfo {pages} {1} (\bibinfo {year} {2018})}\BibitemShut {NoStop}%
\bibitem [{\citenamefont {Abdi}\ \emph {et~al.}(2018)\citenamefont {Abdi},
  \citenamefont {Chou}, \citenamefont {Gali},\ and\ \citenamefont
  {Plenio}}]{Abdi2018}%
  \BibitemOpen
  \bibfield  {author} {\bibinfo {author} {\bibfnamefont {M.}~\bibnamefont
  {Abdi}}, \bibinfo {author} {\bibfnamefont {J.~P.}\ \bibnamefont {Chou}},
  \bibinfo {author} {\bibfnamefont {A.}~\bibnamefont {Gali}},\ and\ \bibinfo
  {author} {\bibfnamefont {M.~B.}\ \bibnamefont {Plenio}},\ }\bibfield  {title}
  {\bibinfo {title} {{Color Centers in Hexagonal Boron Nitride Monolayers: A
  Group Theory and ab Initio Analysis}},\ }\href
  {https://doi.org/10.1021/acsphotonics.7b01442} {\bibfield  {journal}
  {\bibinfo  {journal} {ACS Photonics}\ }\textbf {\bibinfo {volume} {5}},\
  \bibinfo {pages} {1967} (\bibinfo {year} {2018})}\BibitemShut {NoStop}%
\bibitem [{\citenamefont {Ernzerhof}(2003)}]{Ernzerhof2003}%
  \BibitemOpen
  \bibfield  {author} {\bibinfo {author} {\bibfnamefont {M.}~\bibnamefont
  {Ernzerhof}},\ }\bibfield  {title} {\bibinfo {title} {Hybrid functionals
  based on a screened coulomb potential},\ }\href
  {https://doi.org/10.1063/1.1564060} {\bibfield  {journal} {\bibinfo
  {journal} {J. Chem. Phys.}\ }\textbf {\bibinfo {volume} {118}},\ \bibinfo
  {pages} {8207} (\bibinfo {year} {2003})}\BibitemShut {NoStop}%
\bibitem [{\citenamefont {Maruyama}\ and\ \citenamefont
  {Okada}(2018)}]{Maruyama2018}%
  \BibitemOpen
  \bibfield  {author} {\bibinfo {author} {\bibfnamefont {M.}~\bibnamefont
  {Maruyama}}\ and\ \bibinfo {author} {\bibfnamefont {S.}~\bibnamefont
  {Okada}},\ }\bibfield  {title} {\bibinfo {title} {{Energetics and Electronic
  Structure of Triangular Hexagonal Boron Nitride Nanoflakes}},\ }\href
  {https://doi.org/10.1038/s41598-018-34874-x} {\bibfield  {journal} {\bibinfo
  {journal} {Sci. Rep.}\ }\textbf {\bibinfo {volume} {8}},\ \bibinfo {pages}
  {1} (\bibinfo {year} {2018})}\BibitemShut {NoStop}%
\bibitem [{\citenamefont {Ruggenthaler}\ \emph {et~al.}(2018)\citenamefont
  {Ruggenthaler}, \citenamefont {Tancogne-Dejean}, \citenamefont {Flick},
  \citenamefont {Appel},\ and\ \citenamefont {Rubio}}]{Ruggenthaler2018}%
  \BibitemOpen
  \bibfield  {author} {\bibinfo {author} {\bibfnamefont {M.}~\bibnamefont
  {Ruggenthaler}}, \bibinfo {author} {\bibfnamefont {N.}~\bibnamefont
  {Tancogne-Dejean}}, \bibinfo {author} {\bibfnamefont {J.}~\bibnamefont
  {Flick}}, \bibinfo {author} {\bibfnamefont {H.}~\bibnamefont {Appel}},\ and\
  \bibinfo {author} {\bibfnamefont {A.}~\bibnamefont {Rubio}},\ }\bibfield
  {title} {\bibinfo {title} {{From a quantum-electrodynamical light–matter
  description to novel spectroscopies}},\ }\href
  {https://doi.org/10.1038/s41570-018-0118} {\bibfield  {journal} {\bibinfo
  {journal} {Nat. Rev. Chem.}\ }\textbf {\bibinfo {volume} {2}},\ \bibinfo
  {pages} {0118} (\bibinfo {year} {2018})}\BibitemShut {NoStop}%
\bibitem [{\citenamefont {Sidler}\ \emph {et~al.}(2021)\citenamefont {Sidler},
  \citenamefont {Sch{\"{a}}fer}, \citenamefont {Ruggenthaler},\ and\
  \citenamefont {Rubio}}]{Sidler2021}%
  \BibitemOpen
  \bibfield  {author} {\bibinfo {author} {\bibfnamefont {D.}~\bibnamefont
  {Sidler}}, \bibinfo {author} {\bibfnamefont {C.}~\bibnamefont
  {Sch{\"{a}}fer}}, \bibinfo {author} {\bibfnamefont {M.}~\bibnamefont
  {Ruggenthaler}},\ and\ \bibinfo {author} {\bibfnamefont {A.}~\bibnamefont
  {Rubio}},\ }\bibfield  {title} {\bibinfo {title} {{Polaritonic Chemistry:
  Collective Strong Coupling Implies Strong Local Modification of Chemical
  Properties}},\ }\href {https://doi.org/10.1021/acs.jpclett.0c03436}
  {\bibfield  {journal} {\bibinfo  {journal} {J. Phys. Chem. Lett.}\ }\textbf
  {\bibinfo {volume} {12}},\ \bibinfo {pages} {508} (\bibinfo {year} {2021})},\
  \Eprint {https://arxiv.org/abs/2011.03284} {2011.03284} \BibitemShut
  {NoStop}%
\end{thebibliography}
\end{document}